\documentclass[10pt,twocolumn,letterpaper,pagenumbers]{article}

\usepackage{algorithm}
\usepackage{algorithmic}

\usepackage[pagenumbers]{iccv} 
\usepackage{makecell}
\usepackage{multirow}
\usepackage{hhline}
\usepackage{adjustbox}
\usepackage{booktabs, tabularx}
\usepackage{gensymb}
\usepackage{comment}
\usepackage{listings}
\usepackage{color}

\definecolor{dkgreen}{rgb}{0,0.6,0}
\definecolor{gray}{rgb}{0.5,0.5,0.5}
\definecolor{mauve}{rgb}{0.58,0,0.82}

\newcolumntype{C}{>{\centering\arraybackslash}X}

\definecolor{iccvblue}{rgb}{0.21,0.49,0.74}
\usepackage[pagebackref,breaklinks,colorlinks,allcolors=iccvblue]{hyperref}

\def\paperID{6628} 
\def\confName{ICCV}
\def\confYear{2025}

\title{Advancing Multimodal LLMs by Large-Scale 3D Visual Instruction Dataset Generation}

\author{Liu He\\
Purdue University\\
{\tt\small he425@purdue.edu}
\and
Xiao Zeng\\
Amazon\\
{\tt\small zenxiao@amazon.com}
\and
Yizhi Song\\
Purdue University\\
{\tt\small song630@purdue.edu}
\and
Albert Y. C. Chen\\
Amazon\\
{\tt\small aycchen@amazon.com}
\and
Lu Xia\\
Amazon\\
{\tt\small luxial@amazon.com}
\and
Shashwat Verma\\
Amazon\\
{\tt\small shashwv@amazon.com}
\and
Sankalp Dayal\\
Amazon\\
{\tt\small sankalpd@amazon.com}
\and
Min Sun\\
Amazon\\
{\tt\small minnsun@amazon.com}
\and
Cheng-Hao Kuo\\
Amazon\\
{\tt\small chkuo@amazon.com}
\and
Daniel Aliaga\\
Purdue University\\
{\tt\small aliaga@cs.purdue.edu}
}

\begin{document}

\twocolumn[{
\renewcommand\twocolumn[1][]{#1}
\maketitle
\vspace{-25pt}
\begin{center}
    \vspace{-5pt}
    \includegraphics[width=1.0\linewidth]{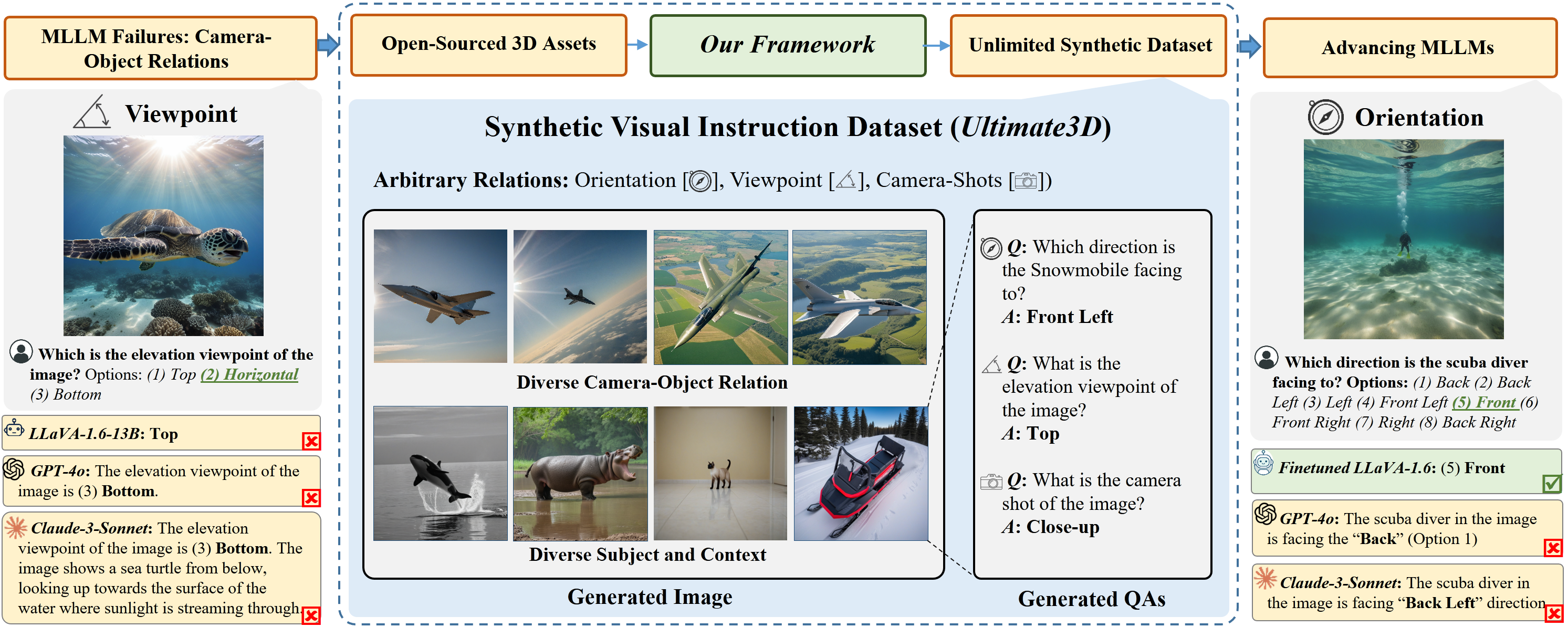}
    \vspace{-15pt}
    \captionof{figure}{\textbf{Generating Synthetic Visual Instruction Dataset}. Our framework uses open-sourced 3D assets to generate photo-realistic images with precisely controlled camera-object relation. Corresponding text instructions are also generated by Large Language Model (LLM). The generated \textit{Ultimate3D} dataset (240K) and benchmark (8K) advance baseline MLLM models (LLaVA-1.6, Llama-3.2-Vision, etc.) to outperform commercial MLLMs (GPT-4o, Claude-3-Sonnet, etc.) on camera-object relation recognition tasks.}
    \label{fig:teaser}
\end{center}
}]

\begin{abstract}
Multimodal Large Language Models (MLLMs) struggle with accurately capturing camera-object relations, especially for object orientation, camera viewpoint, and camera shots. This stems from the fact that existing MLLMs are trained on images with limited diverse camera-object relations and corresponding textual descriptions. To address this, we propose a synthetic generation pipeline to create large-scale 3D visual instruction datasets. Our framework takes 3D assets as input and uses rendering and diffusion-based image generation models to create photorealistic images preserving precise camera-object relations. Additionally, large language models (LLMs) are used to generate text prompts for guiding visual instruction tuning and controlling image generation. We create Ultimate3D, a dataset of 240K VQAs with precise camera-object annotations, and corresponding benchmark. MLLMs fine-tuned on our proposed dataset outperform commercial models by a large margin, achieving an average accuracy improvement of \textbf{33.4\%} on camera-object relation recognition tasks. Our code, dataset, and benchmark will contribute to broad MLLM applications.

\end{abstract}
    
\vspace{-15pt}
\section{Introduction}
\label{sec:intro}

The advent of Multimodal Large Language Models (MLLMs)~\cite{dai2023instructblip, li2023blip, liu2024visual, liu2024improved, li2024llava, chen2024internvl, lin2024vila, tong2024cambrian, llamavision2024} have demonstrated remarkable success on tasks including image captioning, embodied robotic manipulation, and visual question answering (VQA). MLLMs bridge visual understanding from vision encoders~\cite{dosovitskiy2020image, oquab2023dinov2} with powerful verbal capability from large language models (LLMs)~\cite{vicuna2023, touvron2023llama}. Collectively this provides end-to-end inferencing across image and text modalities. In fact, the recently released GPT-4o~\cite{openai2023gpt4} and Claude-3.5-Sonnet~\cite{anthropic2024claudev3} have extended their performances to be comparable to human intelligence in some of the experimented tasks~\cite{yang2023dawn}.

However, research works have found needs and weaknesses of MLLMs~\cite{tong2024eyes, fu2024blink, yuksekgonul2022and}. Tong~\etal~\cite{tong2024eyes} point out the deficiency in visual representations learnt by ViT~\cite{dosovitskiy2020image}. In particular, two weakness categories (i.e., orientation/direction, viewpoint/perspective) focus on camera-object relations such as distinguishing simple concepts about object orientation ("right", "front", "back", etc.), camera viewpoint location ("top", "bottom", etc.), and camera-shot styles ("close-up", "long-shot", etc.). As two examples in Fig.~\ref{fig:teaser} show, current SOTA model (i.e., GPT-4o) struggle with understanding such camera-object relations.
The challenges to improving 3D-aware MLLM abilities on recognizing camera-object relations include: (1) capturing the complexity of the task, and (2) having a dataset containing millions of image-text pairs as is a common requirement for training MLLM on spatial perception tasks~\cite{cheng2024spatialrgpt, chen2024spatialvlm, guo2024regiongpt}. 

Nevertheless, to the best of our knowledge, none of the existing MLLMs or training datasets focus on the aforementioned camera-object relations. Prior methods for MLLM training data generation can be divided into the following two groups:
(1) \textbf{Relabeling of existing real images}. These works focus on generating textual captions by deep learning models given existing images~\cite{chen2024allava, li2024if, zhang2024video, sharifzadeh2024synth, zhang2023llavar, du2023makes}. Specifically, SpatialRGPT~\cite{cheng2024spatialrgpt} and SpatialVLM~\cite{chen2024spatialvlm} address spatial perception and reasoning weakness by providing a comprehensive data annotation pipeline leveraging multiple image segmentation, depth estimation, and camera calibration models to produce dense predictions for real images. But, in practice the accuracy of the predicted labels is not high and the dataset size and distribution is constrained by existing image datasets which predominantly only contain "front" view objects.
%
(2) \textbf{Generating synthetic images}. Typically text-to-image DMs are utilized for synthetic image generation based on provided text prompts~\cite{liu2024synthvlm, du2023training, li2023stablellava}. But, their score-based data curation is not robust to camera-object relations, or generation failures (See Supp Sec. 4). Alternatively, simulation-based generation pipelines for visual reasoning~\cite{johnson2017clevr, liu2019clevr, greff2022kubric} may provide ground truth camera-object relations, but the generated image quality, complexity, and diversity is still far from realistic. In summary, existing synthetic dataset generation pipelines fail to provide image-text pairs with ground truth camera-object relation, realistic image quality, and diverse image subject and context.

In this work, we propose a synthetic generation pipeline to create unlimited image-text pairs focusing on camera-object relations. With 3D assets (e.g., 3D models with texture, materials, and animation) as input, our framework utilizes a renderer (e.g., Blender) with arbitrary camera-object relation parameters to render a series of 3D ground truth priors (i.e., RGB, depth, segmentation). Those 3D priors are used as comprehensive conditions for stacked diffusion-based generation models (e.g., ControlNet) to generate photorealistic images preserving precise camera-object relations. Simultaneously, LLMs are used to generate versatile text prompts for the controlling of image context and object appearance, and to generate diverse VQA prompts regarding camera-object relations. With our framework we have created the \textit{Ultimate3D} dataset as a future benchmark for camera-object relations. MLLMs fine-tuned by our dataset explicitly outperform commercial models by a large margin (i.e., an average 33.4\% accuracy improvement) on recognizing camera-object relations.

Our contributions include:

\begin{itemize}
\item A synthetic generation pipeline for unlimited image-text pair creation of diverse camera-object relations.
\item A dataset and benchmark for further improving and evaluating camera-object relations in MLLMs.
\item Improved open-sourced MLLMs (e.g., LLaVA) that outperform commercial SOTA (e.g., GPT-4o) in camera-object relation tasks.

\end{itemize}

\begin{figure*}[t]
    \vspace{-20pt}
  \includegraphics[width=\linewidth]{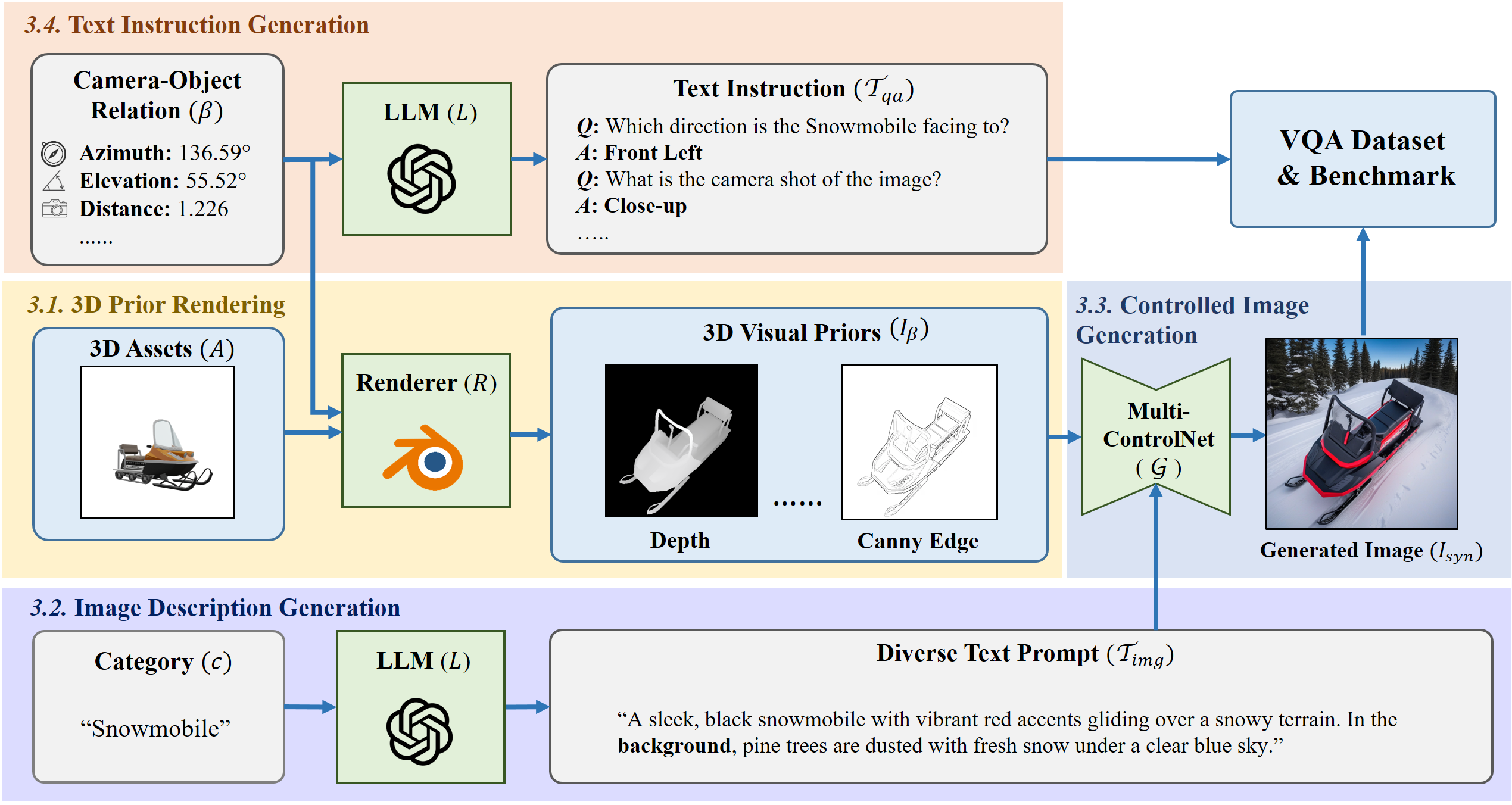}
    \vspace{-17pt}
  \caption{\textbf{Our Framework.} Each shade box corresponds to a section in Sec.~\ref{sec:method}. Given open-sourced 3D assets, our approach leverages a 3D Renderer to generate 3D visual priors ($I_{\beta}$) preserving ground truth camera-object relation ($\beta$). Meanwhile, LLMs take 3D asset category to generate diverse image descriptions ($\mathcal{T}_{img}$) as conditional guidance. Both the 3D visual priors and diverse text prompts are used to generate synthetic images ($I_{syn}$) by multiple ControlNet-based networks. Corresponding text QA instructions ($\mathcal{T}_{qa}$) are generated by LLMs given the ground truth camera-object relation. Our \textit{Ultimate3D} dataset and benchmark (i.e. pairs of $\mathcal{T}_{qa}$ and $I_{syn}$) contribute to fine-tuning and evaluation of MLLMs.}
  \vspace{-13pt}
  \label{fig:framework}
\end{figure*}

\section{Related Works}
\label{sec:relatedworks}

\subsection{Datasets and Benchmarks for MLLMs}
MLLMs use various dataset recipes for visual instruction tuning~\cite{dai2023instructblip, li2023blip, liu2024visual, liu2024improved, li2024llava, chen2024internvl, lin2024vila, tong2024cambrian}. Typically, dozens of datasets, focusing on different areas, are combined to enable the comprehensive abilities of MLLMs. Datasets include image captioning~\cite{lin2014microsoft, young2014image, yu2016modeling}, visual reasoning~\cite{hudson2019gqa, liu2023visual}, general VQA data~\cite{gurari2018vizwiz, wang2023see, liu2024visual} on knowledge grounding~\cite{marino2019ok, lu2022learn}, image understanding~\cite{goyal2017making}, and OCR~\cite{mishra2019ocr,singh2019towards}. Further, most visual instruction datasets are relabeled or created from other existing image-based datasets (e.g., COCO~\cite{lin2014microsoft}, LAION~\cite{schuhmann2022laion}, LVIS~\cite{gupta2019lvis}, Visual Genome~\cite{krishna2017visual}, ADE20K~\cite{zhou2019semantic}). As far as we know, none of the datasets focus on camera-object relations. Thus, the missing ground truth camera-object relation data exacerbates the task for deep learning models, such as SpatialRGPT~\cite{cheng2024spatialrgpt} and SpatialVLM~\cite{chen2024spatialvlm}. In fact, multiple papers have indicated MLLMs' poor distinguishing ability on camera-object relations~\cite{tong2024eyes, fu2024blink}. Our paper fills this gap by providing a generation pipeline, dataset, and benchmark.

In particular for the visual instruction data generation, LLMs, like GPT-4o~\cite{openai2023gpt4}, have been successfully utilized in LLaVA~\cite{liu2024visual, liu2023visual} for dialogue text instruction generation, and used in MMVP~\cite{tong2024eyes} for answer grading. However, generating high-quality images corresponding to proposed text prompts has not been well developed. In this project, we leverage the diverse text generation capability of GPT-4o for the text prompt generation and we provide an image generation pipeline regarding 3D-awareness.

\subsection{Simulation-based Image Generation}
CLEVR~\cite{johnson2017clevr} and Kubric~\cite{greff2022kubric} use 3D engines to generate synthetic images by rendering 3D primitives at multiple arbitrary camera-object relations. Generally, simulation-based image rendering easily preserve ground truth camera-object relations in arbitrary viewpoint settings. However, the generated images are neither photorealistic nor diverse. Realism is limited by the assets' level of detail and textures, and diversity is restricted by the small number of assets. These limitations constrain the usage of synthetic images in MLLM training to only object counting or simple spatial reasoning tasks~\cite{tong2024cambrian}. In this project, we leverage the 3D geometry prior rendered by a 3D renderer to precisely guide the image generation pipeline for preserving ground truth camera-object relation.

\subsection{Generative Images}
Diffusion models provide advanced image generation by iteratively restoring the original data by denoising from Gaussian noise~\cite{ho2020denoising,song2020denoising}. Open-sourced Stable Diffusion~\cite{rombach2022high}, and SDXL~\cite{podell2023sdxl} enable high-resolution and photo-realistic image generation and editing. 
While generative models provide versatile context and subject, representative works (e.g., AURORA~\cite{krojer2024learning}, EditBench~\cite{wang2023imagen}, Prompt-to-Prompt~\cite{hertz2022prompt}, and MagicBrush~\cite{zhang2024magicbrush}) fail to control camera-object relations in the generated content, and quality guarantees rely on human checking. 
ControlNet~\cite{zhang2023adding}, built on versatile backbone DMs (e.g., SDXL), provides spatial control in the generated images by using various image priors (e.g., Canny edges, semantic segmentation, relative depth).
%

In particular, Ma et al.~\cite{ma2023generating} explore using synthetic image generation for 3D pose estimation by conditioning diffusion models on pose angles. However, their approach does not address advancing MLLMs 3D spatial understanding, nor does it consider VQA instruction generation. Moreover, their reliance on 2D Canny priors (i.e., silhouette) brings severe artifacts for images with backward viewpoint and complex depth variations. In contrast, our method incorporates depth priors to enhance geometric accuracy and leverages SDXL’s superior fidelity to improve image realism. Further discussions and user studies are in Sec.~\ref{sec:ablation} and Sec.~\ref{sec:dataset_quality}.

\section{Method}
\label{sec:method}
Our framework is described in Fig.~\ref{fig:framework} and Supp Sec.11. Given input 3D asset $A$, its category $c$, and arbitrary camera-object relation $\beta$, our framework generates photorealistic image $I_{syn}$ preserving $\beta$, and corresponding text QA instruction $\mathcal{T}_{qa}$. Our pipeline consists of 3D visual prior rendering (Sec.~\ref{sec:prior_gen}), diverse image description generation (Sec.~\ref{sec:text_gen}), controlled image generation (Sec.~\ref{sec:img_gen}), and 3D-aware text instruction generation (Sec.~\ref{sec:vqagen}).

\begin{figure}[h]
    \vspace{-10pt}
  \includegraphics[width=\linewidth]{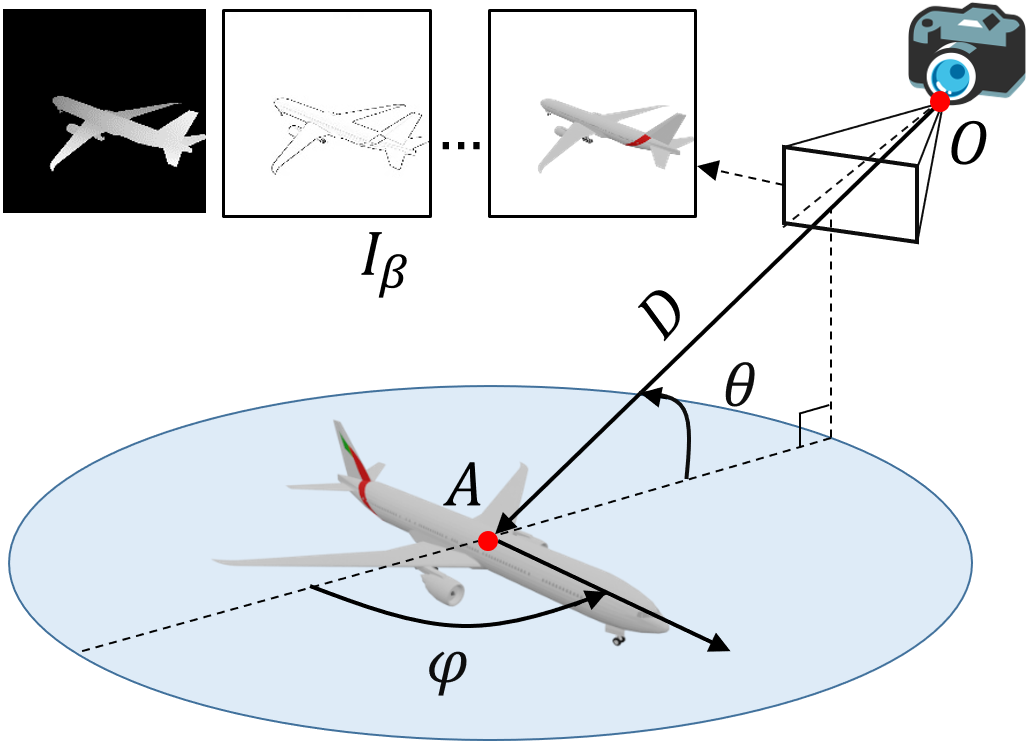}
  \caption{\textbf{3D Visual Prior Rendering.} With a general camera model, our method utilizes a 3D Renderer (e.g., Blender) to render multiple 3D visual priors given arbitrary camera-object relations.}
    \vspace{-8pt}
  \label{fig:CAD}
\end{figure}

\subsection{3D Visual Prior Rendering}

Our system renders several 3D visual priors including 2D depth images, Canny-edge images, and segmentation masks from arbitrary viewpoints. These 3D visual priors contain the scene's geometry and will be used together with LLM-generated image descriptions to collectively control the DM-based image generation in Sec.~\ref{sec:img_gen}.

\label{sec:prior_gen}
\textbf{Camera-Object Relation Definition.} As Fig.~\ref{fig:CAD} shows, we propose a simple camera model to define camera-object relations -- a component which is missing in~\cite{ma2023generating}. Given a 3D asset represented by the centroid $A$ and a camera represented by aperture center $O$, we assume the camera is directly facing the asset. Any location/direction relation between camera and captured object can be represented by $\beta = \{\varphi, \theta, D\}$ consisting of three components:

\begin{itemize}
    \item The \textbf{object orientation angle $\varphi \in [0, 2\pi]$}, which is the counterclockwise azimuth of the 3D asset's facing direction with respect to the camera-facing direction $\overrightarrow{OA}$.
    \item The \textbf{viewpoint angle $\theta \in [-\frac{\pi}{2}, \frac{\pi}{2}]$}, which is the elevation angle of the camera over the horizontal plane of the asset. 
    \item The \textbf{camera-object distance $D \in [0, \infty)$}, which is the length of the line $\overline{OA}$.
\end{itemize}


\textbf{Visual Prior Rendering.} Given an arbitrary $\beta$, the system will render a corresponding set of 2D images $I_{\beta}$ (Fig.~\ref{fig:CAD}) of the 3D asset. The system renders $I_{\beta}$ using 3D renderer $R$. $R$ uses as the default camera intrinsic a perspective camera with focal length 35mm, and a default extrinsic provided by $\beta$. We keep the 3D asset in upright position and located at the center of the camera frame. Images captured at arbitrary $\beta$ will result in various viewpoints, object orientations, and camera-shot distances. Specifically, the renderer $R$ generates images of RGB data, semantic masks, and pixel-level depth maps. 
The rendered RGB images are further processed to provide Canny edges as additional visual priors.

\subsection{Diverse Image Description Generation}
\label{sec:text_gen}
Our approach uses GPT-4o as a LLM to generate diverse text prompts for image descriptions ($\mathcal{T}_{img}$) for DM-based image generation; an example is in Fig.~\ref{fig:framework}. Our framework uses system prompting and few-shot prompting tricks which are broadly utilized for image captioning and QA composing~\cite{liu2023visual, liu2024visual}.

For the generation of $\mathcal{T}_{img}$, the LLM is guided to generate diverse descriptions of background context and object details given the object category $c$. The diversity and rationality of the description directly influences the image generation quality (see Supp Sec. 9). More details are in Supplemental Sec. 1.

\subsection{Controlled Image Generation}
\label{sec:img_gen}

We utilize DMs to generate photorealistic images $I_{syn}$ given comprehensive visual prior $I_{\beta}$. A successful generation of $I_{syn}$ shall preserve ground truth camera-object relation information, and be of high fidelity. We first describe a traditional text-to-image DM pipeline, extend it to stacked multiple ControlNets using various images of $I_{\beta}$, and obtain precisely-controlled image generation.

\textbf{Text-to-Image Generation.}
We use text-to-image DM (i.e., SDXL-1.0~\cite{podell2023sdxl}) as our backbone. The denoising U-Net, $\mathcal{G}_{\theta}$ parameterized by $\theta$, conducts iterative denoising diffusion steps of Gaussian noise $\epsilon$ for a total of $T$ time steps. A text-encoder transforms the text prompt $\mathcal{T}_{img}$ into text embedding for guidance injection. The iterative denoising is defined as:

\begin{equation}
    z_{t-1} = \mathcal{G}_{\theta}(\mathcal{T}_{img}, z_{t}, t), t = T, ..., 1
    \label{eq:sdxl}
\end{equation}

where $z_{T} = \epsilon \sim \mathcal{N}(0, I)$, and $z_{0}$ is the latent variable of the generated image. It can be decoded to an RGB image by pretrained image decoder $\mathcal{D}$ as $I_{0} = \mathcal{D}(z_{0})$. 

\textbf{Precise Controlling via Multiple ControlNets.}
ControlNet~\cite{zhang2023adding} guides image generation by injecting additional controlling features derived from various 2D image visual priors. There are dozens of off-the-shelf ControlNets, each pretrained by a single type of visual prior. In our work, we leverage multiple ControlNets to capture comprehensive visual priors from $I_{\beta}$ (depth, Canny edge, etc.) where each one is weighted by coefficients $w$. The denoising process using multiple ControlNets ($\mathcal{C}$) is defined as:

\begin{equation}
    z_{t-1} = \mathcal{G}_{\theta}(\mathcal{T}_{img}, z_{t}, t, \sum_{k=1}^{|\beta|}w_{k} \cdot \mathcal{C}^{k}(I^{k}_{\beta})), t = T, ..., 1.~
    \label{eq:controlnet}
\end{equation}
Similarly, the output generated image $I_{syn} = \mathcal{D}(z_{0})$ and $I_{syn}$ is expected to have ground truth camera-object relations.

\subsection{Text Instruction Generation}
\label{sec:vqagen}
We utilize GPT-4o as a LLM to generate QA pairs on camera-object relation ($\mathcal{T}_{qa}$) for visual instruction tuning (more details in Supplemental Sec.2).

Specifically, the LLM is prompted to generate QA pairs given the camera-object relation $\beta$. We subdivide each component of $\beta$ into categorical types for effective referring through textual response by MLLMs:
\begin{itemize}
    \item Object orientation angle $\varphi$ is classified into eight object orientation types, each covers a bin of $\frac{\pi}{4}$ azimuth (i.e. "right", "front right", "front", "front left", "left", "back left", "back", "back right").
    \item The viewpoint angle $\theta$ is classified into three viewpoint types, each covering a range of $\frac{\pi}{3}$ elevation angle (i.e. "horizontal", "top", "bottom").
    \item The camera-object distance $D$ is in relative units of Blender. It reflects the ratio of object dimension over the full picture size of the camera ($D=1$ indicates fully covered). The three camera-shot types (i.e. "close-up", "medium-shot", "long-shot") is decided by two cutoff values 1.25 and 3.0.
\end{itemize}

Examples of generated multiple-choice style QA pairs as MMVP benchmark~\cite{tong2024eyes} are showed in Fig.~\ref{fig:framework}. For question generation, we prompt LLMs to generate template questions for each prediction task of object orientation, viewpoint type, and camera-shot type. We list all possible options, including the ground truth $\beta$, in random order. MLLMs are expected to choose the option according to $\beta$.

\begin{table*}[t!]
\vspace{-16pt}
\caption{\textbf{Quantitative Comparisons.} We fine-tune LLaVA models by \textit{Ultimate3D} dataset, then evaluate the MLLM response accuracy ($\%$) on \textit{Ultimate3D} and \textit{MMVP} benchmarks (\textit{MMVP} is a public benchmark showing our cross-dataset capability). Fine-tuned LLaVA models outperform SOTAs by an average of $33.4\%$ among all three tasks.}
\vspace{-1mm}
\centering
\small
\renewcommand{\arraystretch}{1.2}  
\begin{tabular}{l|cccc|cccc|cc}
\noalign{\hrule height 1.0pt}
\multirow{3}{*}{\textbf{MLLMs}} & \multicolumn{4}{c}{\textbf{Orientation} $(\%\uparrow)$} & \multicolumn{4}{|c|}{\textbf{Viewpoint} $(\%\uparrow)$} & \multicolumn{2}{c}{\textbf{Camera-Shots} $(\%\uparrow)$} \\
\cline{2-11}
 & \multicolumn{3}{c|}{\textit{Ultimate3D}} & \multirow{2}{*}{\textit{MMVP}} & \multicolumn{3}{c|}{\textit{Ultimate3D}} & \multirow{2}{*}{\textit{MMVP}} & \multicolumn{1}{c|}{\textit{Ultimate3D}} & \multirow{2}{*}{\textit{MMVP}} \\
\cline{2-4} \cline{6-8} \cline{10-10}
 & Both & Syn & \multicolumn{1}{c|}{Real} &  & Both & Syn & \multicolumn{1}{c|}{Real} &  & \multicolumn{1}{c|}{Syn} &  \\
\noalign{\hrule height 1.0pt}
\cellcolor{orange!15}LLaVA-1.5-7B & 18.3 & 11.1 & 22.5 & 25.0 & 31.8 & 38.4 & 25.3 & 29.2 & 46.0 & 66.7 \\ 
\cellcolor{orange!15}LLaVA-1.6-7B & 18.2 & 14.7 & 20.3 & 25.0 & 36.8 & 43.9 & 29.8 & 33.3 & 46.6 & 71.4 \\
\cellcolor{orange!15}LLaVA-1.5-13B & 17.9 & 18.7 & 17.4 & 22.7 & 38.9 & 35.0 & 42.8 & 41.7 & 44.6 & 61.0 \\
\cellcolor{orange!15}LLaVA-1.6-13B & 16.1 & 16.4 & 16.0 & 15.9 & 30.7 & 30.3 & 31.1 & 33.3 & 42.3 & 61.9 \\
\cellcolor{orange!15}Llama-3.2-V-11B & 5.6 & 5.3 & 5.7 & 6.8 & 19.8 & 26.2 & 13.6 & 38.1 & 29.6 & 25.0 \\
\noalign{\hrule height 1.0pt}
\cellcolor{green!12}Finetuned LLaVA-1.5-7B & 68.8 & 85.6 & 58.8 & 54.5 & 70.6 & 81.3 & 60.0 & 66.7 & 94.0 & 66.7 \\ 
\cellcolor{green!12}Finetuned LLaVA-1.6-7B & 71.5 & 86.9 & 62.4 & \underline{61.4} & \underline{71.3} & \textbf{83.3} & \underline{59.5} & 66.7 & 94.1 & 50.0 \\
\cellcolor{green!12}Finetuned LLaVA-1.5-13B & 70.0 & 85.6 & 60.8 & 50.0 & 69.7 & 81.9 & 57.7 & 66.7 & \underline{94.2} & \textbf{76.2} \\
\cellcolor{green!12}Finetuned LLaVA-1.6-13B & \underline{72.4} & \textbf{88.1} & \underline{63.1} & \textbf{65.9} & \textbf{72.3} & \underline{83.0} & \textbf{61.8} & \textbf{75.0} & \textbf{94.8} & \underline{71.4} \\
\cellcolor{green!12}Finetuned Llama-3.2-V-11B & \textbf{74.2} & \underline{87.3} & \textbf{66.4} & 50.0 & 69.1 & 79.6 & 58.7 & 58.3 & 93.1 & \textbf{76.2} \\

\noalign{\hrule height 1.0pt}
\cellcolor{cyan!10}GPT-4o & 43.1 & 51.0 & 38.5 & 40.9 & 54.1 & 63.5 & 44.9 & 
\underline{70.8} & 41.7 & 42.9 \\
\cellcolor{cyan!10}GPT-4o-mini & 43.5 & 52.3 & 38.3 & 38.6 & 54.5 & 63.9 & 45.1 & 66.7 & 41.5 & 42.9 \\
\cellcolor{cyan!10}Claude-3-Sonnet & 43.1 & 52.6 & 37.5 & 36.4 & 54.3 & 64.2 & 44.5 & 66.7 & 41.7 & 47.6 \\
\cellcolor{cyan!10}Claude-3.5-Sonnet & 40.3 & 47.7 & 35.8 & 54.5 &	55.6  & 65.9 & 45.3 & 66.7 & 41.8 & 42.9 \\
\noalign{\hrule height 1.0pt}
\end{tabular}
\vspace{-8pt}
\label{tbl:quan_ulti3d}
\end{table*}

\begin{figure*}[t]
  \vspace{-20pt}
  \includegraphics[width=\linewidth]{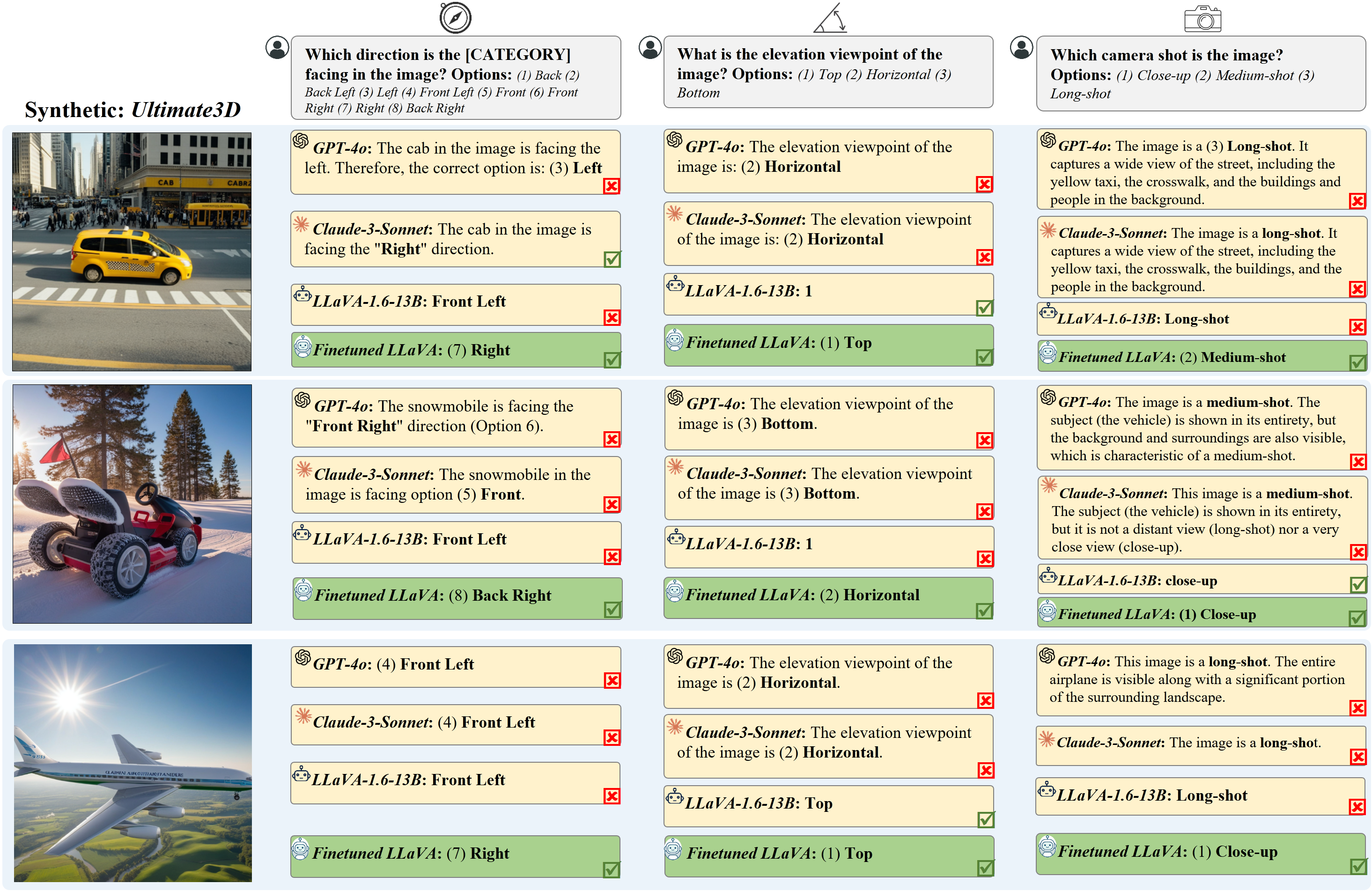}
  \includegraphics[width=\linewidth]{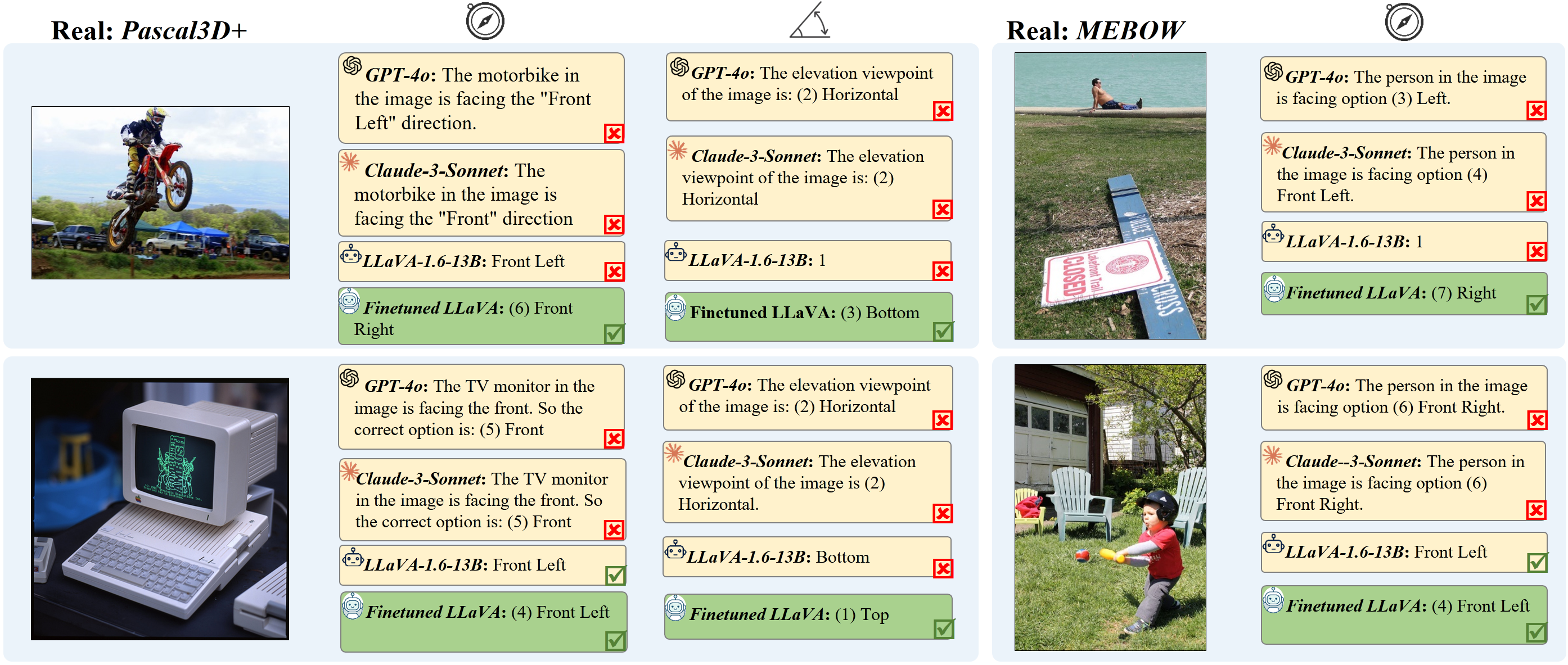}
    \vspace{-17pt}
  \caption{\textbf{Qualitative Results.} We show responses by GPT-4o, Claude-3-Sonnet, LLaVA-1.6-13B, and fine-tuned LLaVA-1.6-13B, on \textit{Ultimate3D} benchmark. Each model is asked the questions (in gray boxes) regarding object orientation, camera viewpoint, and camera-shots type, together with the input images on the left. The model responses illustrate that the LLaVA-1.6-13B model fine-tuned by \textit{Ultimate3D} can precisely recognize camera-object relations in both real and synthetic images. GPT-4o and Claude-3-Sonnet provide performance similar to random guess on object orientation, display a bias to underestimate viewpoint angle, and show a bias to overestimate camera-shot distance. The original LLaVA-1.6 model defaults to a few fixed answers regardless of the input images.}
  \label{fig:qualitatiev}
  \vspace{-12pt}
\end{figure*}

\section{Experiments}
\label{sec:experiments}

\subsection{Visual Instruction Generation}
\label{subsec:dataset}
\textbf{Data Source.}
Our framework uses open-sourced 3D assets datasets from ObjaverseXL~\cite{deitke2023objaverse, deitke2024objaverse} and ShapeNet~\cite{chang2015shapenet}. Partial 3D assets are labeled following the synset list from ImageNet-1K~\cite{russakovsky2015imagenet}. We leverage this subset and define the asset category by its synset label. In particular, we manually select 100 synsets with explicit "facing" orientation from the original list of ImageNet. In total, there are 1196 assets used for synthetic visual instruction generation.

\noindent\textbf{ControlNets.}
We utilize SDXL-1.0~\cite{podell2023sdxl} as the backbone of ControlNets. For selection of multiple visual priors, we found using both Canny edges and relative depth provides the best mix of camera-object relation preservation and image quality. The diffusion step $T=30$, and relative weights for depth and Canny edge ControlNets are 0.5 and 0.8, respectively. Several seconds are needed to generate an image by a single H100 GPU. Ablations and user studies are shown in Sec.~\ref{sec:ablation} and Sec.~\ref{sec:dataset_quality}.

\subsection{Ultimate3D Dataset and Benchmark.}
Our framework creates the \textit{Ultimate3D} dataset consisting of 85K synthetic images and 18K cropped images of a single human per photograph and corresponding human orientation labels from MEBOW~\cite{wu2020mebow}. For the synthetic images, we use 1180 3D assets covering 100 categories. For each 3D asset, there are 72 possible camera-object relations (8 orientations, 3 viewpoints, and 3 camera shot types) and we render a corresponding synthetic image for each relation. Then, each synthetic image is described by up to 3 textual QAs corresponding to object orientation, camera viewpoint, and camera-shot type, yielding a total of 240K corresponding QAs. Examples of synthetic images are in Supp Sec. 9. The images from MEBOW are useful because otherwise there is a strong lack of 3D assets of human which we find crucial to fine-tuning. The MEBOW images only have one QA corresponding to human orientation.

The \textit{Ultimate3D} benchmark (Fig.~\ref{fig:qualitatiev}) has two sets: (1) Synthetic Set. It contains 1200 synthetic images and 3600 QAs generated by our framework. Each image and QA has been manually reviewed to ensure correctness. This set covers all 100 categories with explicit facing orientation. The 3D assets utilized for benchmark generation do not overlap with those used for dataset generation. (2) Real Set. It contains two portions. The first portion is 2443 real images collected from Pascal3D+~\cite{xiang_wacv14} covering 10 categories (e.g. aeroplane, bicycle, etc.). Each image corresponds to a QA based on provided labels of object orientation or camera viewpoint. The second portion is 800 real images of humans collected from the MEBOW dataset~\cite{wu2020mebow}. Each image corresponds to a QA on human orientation based on provided labels. In total, there are 3243 VQAs in the Real set. We resample the real images to keep a uniform distribution for each orientation and viewpoint type. Since no camera-shot information is provided in the real image set, the QAs in the real set exclude camera-shot type questions.

Both the \textit{Ultimate3D} dataset and the benchmark have no camera-object relation bias, which is very uncommon in the real image datasets (e.g. most images capture objects facing straight to the camera~\cite{xiang_wacv14}). In contrast, in our dataset the number of images in each orientation, viewpoint, and camera-shot type are \textbf{evenly} distributed.

\subsection{Advancing 3D-Aware MLLMs}
We explore how much our synthetic \textit{Ultimate3D} dataset can improve the MLLMs recognition capability on camera-object relation. The evaluation results on both synthetic and real image VQA benchmarks indicate the fine-tuned baseline models are significantly improved from random guess and outperform commercial SOTAs.

\noindent\textbf{Training Details.} We choose LLaVA-1.5/1.6~\cite{liu2024improved} with both 7B and 13B parameters, and Llama-3.2-Vision-11B~\cite{llamavision2024} as baselines. Each model is fine-tuned on the \textit{Ultimate3D} dataset for one epoch. Specifically, both the MLP of vision-language connector and the LLM decoder are trained, while the vision encoder is frozen. All fine-tuning hyperparameters follow the official instructions from~\cite{liu2024improved, llamavision2024}; training is 12 hours on 4 H100 GPUs.

\noindent\textbf{Evaluation Settings.}
We evaluate the baseline MLLMs and their fine-tuned versions on \textit{Ultimate3D} benchmark and MMVP benchmark~\cite{tong2024eyes}. MMVP benchmark is an intentionally selected real image dataset which is claimed to be difficultly distinguished by MLLMs. We manually label 89 VQAs regarding camera-object relations. During evaluation, MLLMs' textual responses are evaluated by GPT-4o, which is prompted to judge if the responses correspond to the ground truth answers. We calculate the accuracy rate ($\%$) of the MLLMs responses as the evaluation metric. We also include commercial SOTA models (e.g. GPT-4o and Claude-3.5-Sonnet) in the evaluation.

\noindent\textbf{Quantitative Results.}
The quantitative results of response accuracy ($\%$) are shown in Tab.~\ref{tbl:quan_ulti3d}. Baseline LLaVA models only provide one of several fixed options ("front" orientation, "top" viewpoint, etc.) regardless of the content of the input images. The accuracy of baselines are near random guess level (e.g. 33\% or 12.5\% depending on the option type). This weakness may result from the bias on LLaVA's visual instruction datasets. Since these datasets are based on real image datasets, most images are captured from a front facing direction, and from an above or side viewpoint.

All MLLMs finetuned by our synthetic generated dataset significantly improve the accuracy of recognition on camera-object relation. On \textit{Ultimate3D} benchmark, fine-tuned LLaVA-1.6-13B model outperforms commercial SOTAs by average of $33.4\%$ percentage. Moreover, the performance shows plausible generalization on real and synthetic image benchmarks. In MMVP benchmark, fine-tuned LLaVA-1.6-13B model also provides average of $19.23\%$ higher accuracy than commercial SOTAs. Smaller LLaVA-1.6-7B model provides similarly competitive performances to its 13B version. This indicates the potential of using a compact model to distinguish the camera-object relations. 

Moreover, Llama-3.2-Vision-11B obtains the largest accuracy improvement (average 60.46$\%\uparrow$) on all three types of camera-object relation recognition tasks. The original released model refuses to answer most of the questions (e.g., response: "I'm not able to provide information about the individual in this image."). But our fine-tuned model provides the best performance on recognizing orientation compared to all alternative MLLMs.


\noindent\textbf{Qualitative Results.} We show the qualitative comparisons in Fig.~\ref{fig:qualitatiev}. Commercial SOTA models (e.g. GPT-4o) and open-sourced models (e.g. LLaVA-1.6-13B) struggle with precise recognition of all three types of camera-object relations. For the two commercial models, their responses of predicting object orientation are near random regardless of the input images. We find a bias that both commercial models tend to underestimate viewpoint angle (e.g. mistake top view as horizontal), and overestimate camera-shot distance (e.g. mistake medium-shot as long-shot). The original LLaVA model has severe bias in its responses to all types of questions. Most of its answers are "Front Left" orientation, "Top" viewpoint, and "Close-up" camera-shots. As comparison, the LLaVA model fine-tuned by our \textit{Ultimate3D} dataset provides plausible performance on all tasks across the synthetic and real sets of our benchmark. Additional comparisons are in Supplemental Sec. 8.

We hypothesize that both dataset bias and lack of camera-object visual instruction dataset cause the aforementioned behavior biases across various MLLMs. Without modifications to model structure or training scheme, our fine-tuned LLaVA-1.6-13B model significantly outperforms commercial models which have a large number of model parameters and consume huge amount of training data. This indicates that lack of dataset is the main reason of current MLLMs' weakness on camera-object relation recognition.

\noindent\textbf{Limitations.} 
The numerical prediction of camera-object relation $\beta$ is more challenging for MLLMs compared to the categorical prediction we report. Preliminary tests show MLLMs can handle numerical prediction of camera-object relation for a single category, but struggle with a hundred categories as in our dataset. Since our main contribution is the discovery of the dataset bottleneck and the generation pipeline, we leave addressing this limitation as future work.

\subsection{Generalization of \textit{Ultimate3D} to General VQA}
To explore the generalization of 3D camera-object relation capturing with general VQA tasks, we fine-tune LLaVA-1.6-13B models by a half-half mixture of \textit{Ultimate3D} (240K VQAs) and randomly sampled 240K VQAs from LLaVA-Instruct-665K~\cite{liu2024improved}. Tab.~\ref{tab:comparison_vqav2} presents the dataset mixture finetuning performance on VQAv2~\cite{goyal2017making}, our \textit{Ultimate3D}, and \textit{MMVP} benchmarks. The results demonstrate that incorporating \textit{Ultimate3D} significantly enhances camera-object relation reasoning while maintaining general VQA performance.

\begin{table}[t]
\vspace{-7mm}
\caption{\textbf{Generalization.} We finetune LLaVA-1.6-13B model using a mixture of both general VQA and camera-object relation VQA datasets. Without degradation on general VQA tasks, finetuned model (green) improves its ability on camera-object relation recognition and beat SOTA (i.e. GPT-4o) by 28.42\% accuracy on \textit{Ultimate3D} benchmark. $*$ denotes the average accuracy ($\%$) across all 3 types of camera-object relation VQAs in Tab.~\ref{tbl:quan_ulti3d}.}
\vspace{-2mm}
\renewcommand{\arraystretch}{1.2} 
\centering
\small
\resizebox{\columnwidth}{!}{
    \begin{tabular}{l|ccc}
        \toprule
          MLLMs & VQAv2~\cite{goyal2017making} & \textit{Ultimate3D\textsuperscript{*}} & \textit{MMVP\textsuperscript{*}} \\
        \midrule
       \cellcolor{orange!15} LLaVA-1.6-13B~\cite{liu2024improved} & \underline{80.00} & 29.70 & 37.03 \\
        \midrule
       \cellcolor{green!12}\textit{Ulti3D} (100\%) & 78.77 & \textbf{79.83} & \textbf{70.77}\\
       \cellcolor{green!12}\textit{Ulti3D} (50\%) + ~\cite{liu2024improved}(50\%)& \textbf{80.01} & \underline{74.72} & \underline{65.80} \\
        \midrule
       \cellcolor{cyan!10}GPT-4o & / & 46.30 & 51.57 \\
       \cellcolor{cyan!10}Claude-3.5-Sonnet & / & 45.90 & 54.70 \\
    
        \bottomrule
    
    \end{tabular}
}

    \vspace{-2mm}
\label{tab:comparison_vqav2}
\end{table}%

\subsection{Ablations on Image Generation Pipeline}
\label{sec:ablation}
\begin{figure}[t]
  \vspace{-7mm}
  \includegraphics[width=\linewidth]{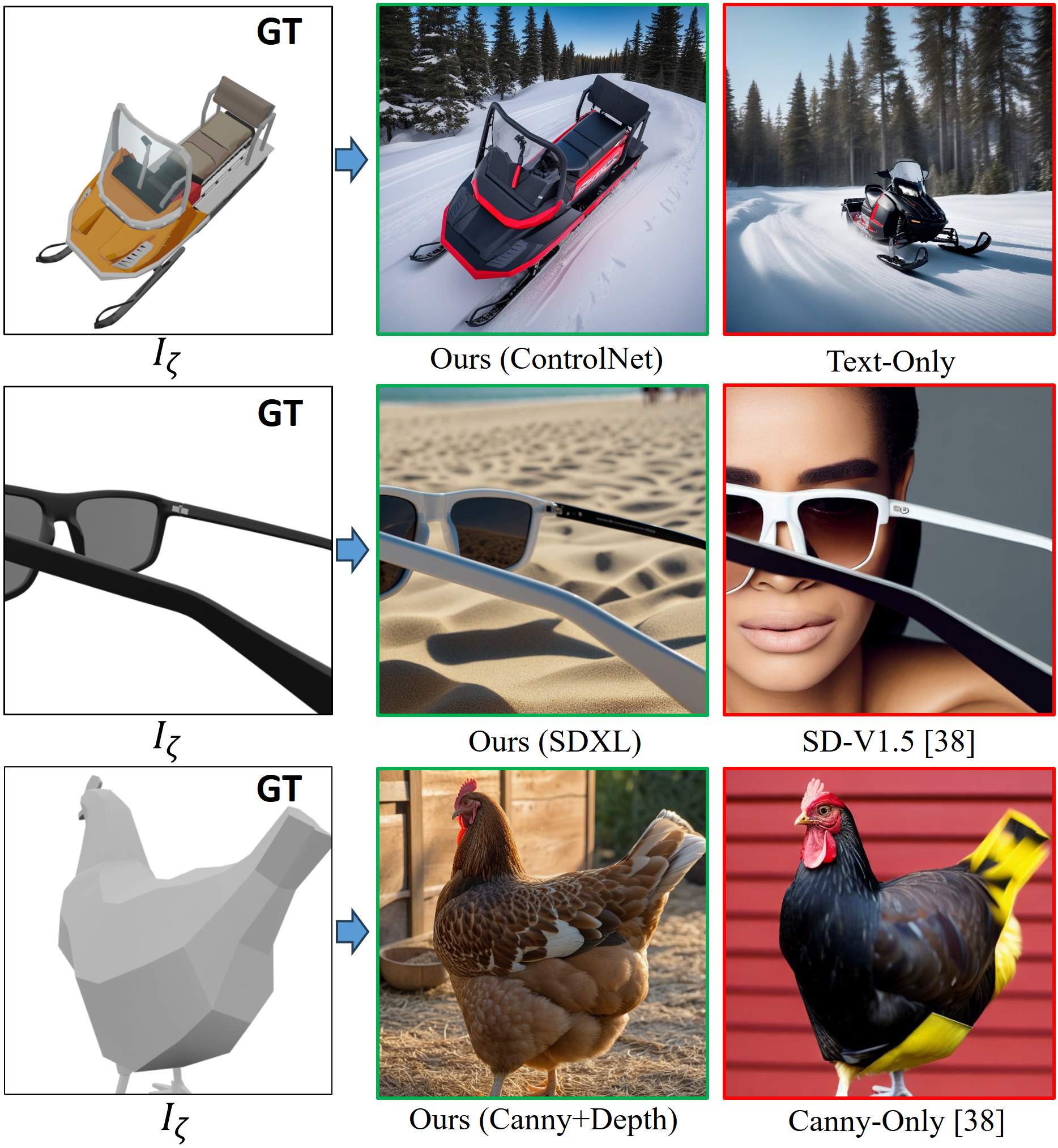}
  \vspace{-6mm}
  \caption{\textbf{Ablations on Image Generation Pipeline.}
  Our framework shows superior performance by using multi-ControlNet with SDXL backbones, and introducing both Canny edges and depth maps as visual priors. It consistently delivers better robustness and quality compared to text-only SDXL and~\cite{ma2023generating}. When these visual priors are removed (top-right: removing all visual priors; bottom-right: removing depth prior), the generator often fails to produce good results and brings lower generation success rate (see Sec~\ref{sec:dataset_quality},~\cite{ma2023generating} only achieves $55.11\%$ success rate v.s. ours $93.07\%$)
  }
  \vspace{-3mm}
  \label{fig:ablation}
\end{figure}

In Fig.~\ref{fig:ablation}, we ablate our image generation pipeline and illustrate better quality than~\cite{ma2023generating} by: (1) removing the visual priors; and (2) using alternative image generation backbones; (3) changing the group of visual priors given to the image generator. Compared to text-only SDXL and~\cite{ma2023generating}, we find that multiple visual priors may significantly improve the robustness of image generation under complex camera-object relation (especially for viewpoints of the object's back-side, and for 3D assets with low level-of-detail). In particular, the depth and Canny edge prior are crucial. However, we find that adding more priors (e.g., RGB, semantic masks, normal maps) does not explicitly improve quality. Moreover, our SDXL backbone outperforms other alternatives (i.e. SD-V1.5~\cite{rombach2022high}). Discussions in Suppl Sec. 5 and Sec. 6.

\subsection{User Study on Dataset Quality}
\label{sec:dataset_quality}
We investigate using text-image aligning metric and pose-estimation model prediction to evaluate quality of our Ultimate3D dataset but find those metrics to be deficient (Supp Sec. 4). Instead, we perform a comprehensive user study as a quality check of \textit{Ultimate3D} dataset. As Supp Sec.7 shows, the user will see two images side-by-side (Supp Fig.4): an RGB image rendered by Blender (representing $I_{\beta}$), and a synthetic image generated by our method using $I_{\beta}$. Preserving the RGB image's 3D geometry and structure in the synthetic image is regarded as success. Our study includes 225 randomly sampled image pairs where each image is independently reviewed by 3 users. Overall success rate is \textbf{93.07$\%$} (agreement rate among 3 users is 94.4\%).

In comparison, we perform another similar-objective user study on ~\cite{ma2023generating}. The overall success of this alternative method was only 55.11\% (we explain the cause of this reduced performance in Sec.~\ref{sec:ablation}).

\section{Conclusion}
\label{sec:conclusion}
In this work, we discover the dataset bottleneck of MLLMs for distinguishing camera-object relations. Given only 3D assets, our framework is able to generate unlimited image-text pairs of diverse camera-object relations with no distribution bias. We also provide the \textit{Ultimate3D} dataset and benchmark for improving future MLLMs. Moreover, fine-tuned open-sourced MLLMs using our dataset outperform SOTA commercial models by a significant margin.

As future work, we plan to extend our framework for multi-object image generation with additional viewpoints. We will further improve numerical prediction capability on camera-object relation for open-vocabulary scenarios.

\section{Acknowledgments}
This work was conducted during Liu He's internship at Amazon. The author is grateful to Jingjing Zheng, Tony Qi, Nan Qiao, Ke Zhang, and Yuyin Sun for their valuable discussions and contributions to the underlying ideas of this work.

{
    \small
    \bibliographystyle{ieeenat_fullname}
    \bibliography{main}
}

\clearpage
\setcounter{page}{1}
\setcounter{section}{0}
\maketitlesupplementary

\noindent \textbf{\Large Overview}

Below is a summary of the contents in each section of this supplemental material:
\begin{itemize}
\item Sec.~\ref{suppsec:textprompt}: Details of text prompt generation as image description utilized for controlled image generation.
\item Sec.~\ref{suppsec:vqagen}: Details of VQA prompt generation given the ground truth camera-object relations. 
\item Sec.~\ref{suppsec:LLMgrading}: Details of LLM-based grading for MLLM response evaluation.
\item Sec.~\ref{supp_sec: aligning}: Preliminary studies on data curating for camera-object relation VQA dataset by text-image aligning metrics, and by pose estimation models.
\item Sec.~\ref{suppsec:dm_backbone_ablation}: Ablations on selections of DM backbones. 
\item Sec.~\ref{suppsec:genfailure}: Factors influence generation quality of our image generation pipelines.
\item Sec.~\ref{suppsec:userstudy}: Details of user study on image quality evaluation.
\item Sec.~\ref{suppsec:addqual}: More qualitative comparisons between finetuned LLaVA models to commerical SOTA models.
\item Sec.~\ref{suppsec:ultimate3D}: 48 image examples illustrate the diversity of \textit{Ultimate3D} dataset and benchmarks on object categories, camera-object relations, and background context.
\item Sec.~\ref{suppsec:finetune_dicussion}: Additional discussions and insights on using synthetic dataset generated by our framework for MLLM finetuning.
\item Sec.~\ref{suppsec:algorithm}: Algorithm of our 3D visual instruction dataset generation pipeline.
\end{itemize}

\section{Image Description Generation}
\label{suppsec:textprompt}
We provide system prompts given to GPT-4oto generate context description for image generation. Corresponding section in main paper is Sec. 3.2. In the prompt, we provide one-shot example for better robustness of the text generation.

Specifically, generated text prompt will be merged with default positive prompt like \textit{"detailed, 4K, 35mm photograph, professional"}, and customized camera-object prompt \textit{"the image shows a [$\beta$] view of a [$c$]"}, in order to enhance image generation quality.

\begin{lstlisting}
    SYSTEM_PROMPT_FOR_IMAGE_GEN = "You are an expert in generating concise and diverse descriptions of an object to guide image generation model like DAll-E3. You should use your common sense to generate 1 sentence description of the given object. Additionally, you should also generate 1 sentence of a common real-world scene given the object. Both sentences of descriptions should not include more than one of the given object to avoid ambiguity.
    
    Example Input: Please generate the visual prompt of chicken.
    Example Output: A white and brown chicken standing in a cage made of metal bars. The chicken has a long, curved beak and its feathers are fluffy and white.
    "
\end{lstlisting}

\section{Text Instruction Generation}
\label{suppsec:vqagen}
We provide the system prompt to generate QA pairs given camera-object relation as below. Corresponding section in main paper is Sec. 3.4. In the prompt, we provide few-shot example for better instructions for diversity of the generated text.

\begin{lstlisting}
    SYSTEM_PROMPT_FOR_VQA_GEN = "You are an AI visual assistant, and you are seeing a single image. What you see are provided with several sentences, describing the same image you are looking at. Answer all questions as you are seeing the image.

    Design a conversation between you and a person asking about this photo. The answers should be in a tone that a visual AI assistant is seeing the image and answering the question.
    Ask diverse questions and give corresponding answers.
    
    Include questions asking about the visual content of the image, including the object orientation and its corresponding azimuth degree, the camera viewpoint of the image and corresponding camera elevation angle, the camera shot type of the image and the distance from camera to the target object. Only include questions that have definite answers:
    (1) one can see the content and its orientation degree in the image that the question asks about and can answer confidently;
    (2) one can recognize camera viewpoint and camera shot of the image that the question asks about and can answer confidently;
    The question can be multiple choice questions given the choice, or open-ended questions. For multiple choice, the option expression can be diverse but with the same meaning. For example, "front" also means "directly towards the camera"; 'back' also means 'away from the camera', etc. Use the common sense to make options diverse if possible.
    Provide at most 5 pairs of question and answer pairs. Prior the confident questions. Do not ask any question that cannot be answered confidently.
    
    You should ask questions from multiple QA templates as the below examples show.
    
    Example Input Description:
    The image shows a front view of police van. A blue and white police van with the word "POLICE" emblazoned on the side is parked on a city street. Nearby, a couple of officers are discussing a recent incident while pedestrians walk by, some glancing curiously at the vehicle.
    The police van is with an azimuth of 181.518 degree, facing Front direction.
    The elevation angle of the camera to the police van is 81.41 degree, the camera viewpoint is Horizontal view.
    The relative distance from the camera to the police van is 1.132 meters, the camera shot type is Close-up.
    
    Example Output Question and Answer Pairs:
    Question:
    From the camera's perspective, is the police van in the picture facing straight or oriented at an angle? Options: (a) Directly towards the camera (b) At an angle
    ===
    Answer:
    (a) Directly towards the camera
    ===
    Question:
    Is the police van in the picture facing the camera or away from the camera? Options: (a) Away from the camera (b) Facing the camera
    ===
    Answer:
    (b) Facing the camera
    ===
    Question:
    Which direction is the police van facing in the image? Options: (1) Back (2) Front
    ===
    Answer:
    (2) Front
    ===
    Question:
    Is the police van facing back or front from the camera's perspective? Options: (a) Back (b) Front
    ===
    Answer:
    (b) Front
    ===
    Question:
    Is the photo taken directly above the police van or from the side? Options: (a) Taken directly (b) From the side
    ===
    Answer:
    (b) From the side
    ===
    Question:
    Is the photo taken far away the police van or taken closely?
    ===
    Answer:
    The relative distance from the camera to the police van is 1.132 meters, thus it is with a close-up shot. This indicates the photo is taken closely.
    ===
    Question:
    What is the elevation viewpoint of the image? Options: (1) Top (2) Horizontal (3) Bottom
    ===
    Answer:
    (2) Horizontal
    ===
    Question:
    Which camera shots is the image? Options: (1) Close-up (2) Medium-shot (3) Long-shot
    ===
    Answer:
    (1) Close-up"
\end{lstlisting}

\section{LLM-based Response Grading}
\label{suppsec:LLMgrading}
We provide the system prompt for evaluation of the MLLM response to our camera-object relation multiple choice questions.

\begin{lstlisting}
SYSTEM_PROMPT_FOR_GRADING_MLLM_RESPONSE ='You are a helpful and precise assistant for checking the quality of the answer. You should review all listed choices and compared to the Answer content, and judge whether the Answer is correct (yes), or not (no). You should only focus on the major content of the answer, not the detailed number or symbol. Please answer in only yes or no'
\end{lstlisting}



\section{Metric Limitations for Evaluating Camera-Object Relation}
\label{supp_sec: aligning}

\begin{figure}[h]
  \includegraphics[width=\linewidth]{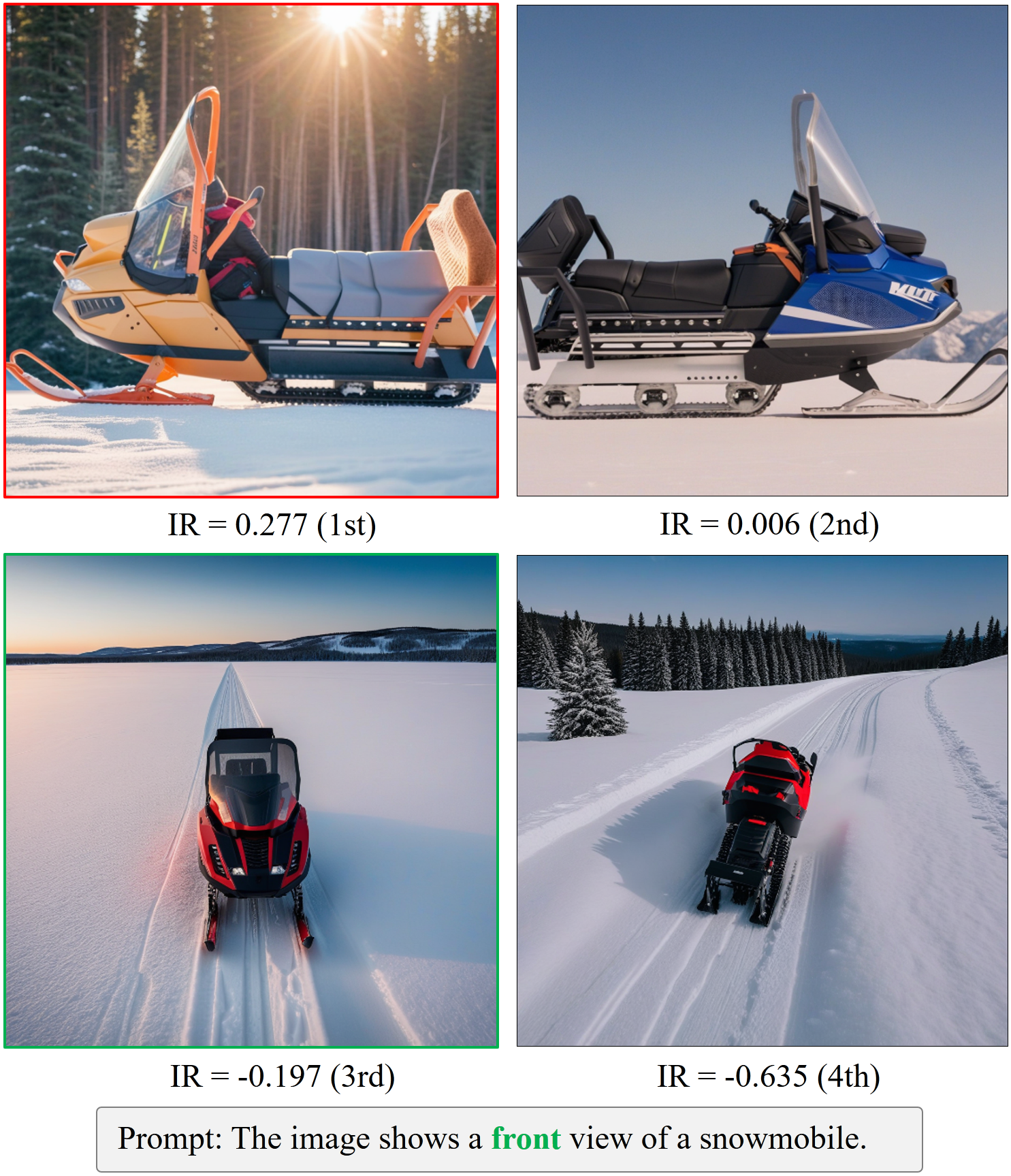}
  \caption{\textbf{Limitations of Text-Image Aligning Metrics.} We evaluate the ImageReward~\cite{xu2024imagereward} (IR) between the text promopt regarding object orientation with above 4 synthetic images. The image matches the correct camera-object relation text prompt ("front") may reach the highest score. However, image with the highest matching score (it faces to "left" direction) is not the correct one with expected orientation. This failure is prevailing across all categories.}
  \label{suppfig:aligning}
\end{figure}

In Fig.~\ref{suppfig:aligning}, we show an example of using ImageReward~\citep{xu2024imagereward} for the dataset curation by evaluating the alignment between generated image and text prompts. Preliminary test indicates that text-image aligning metric is not sensitive to camera-object relation text prompts. Thus we may not rely on those metrics for data curation purpose.

\begin{figure}[h]
  \includegraphics[width=\linewidth]{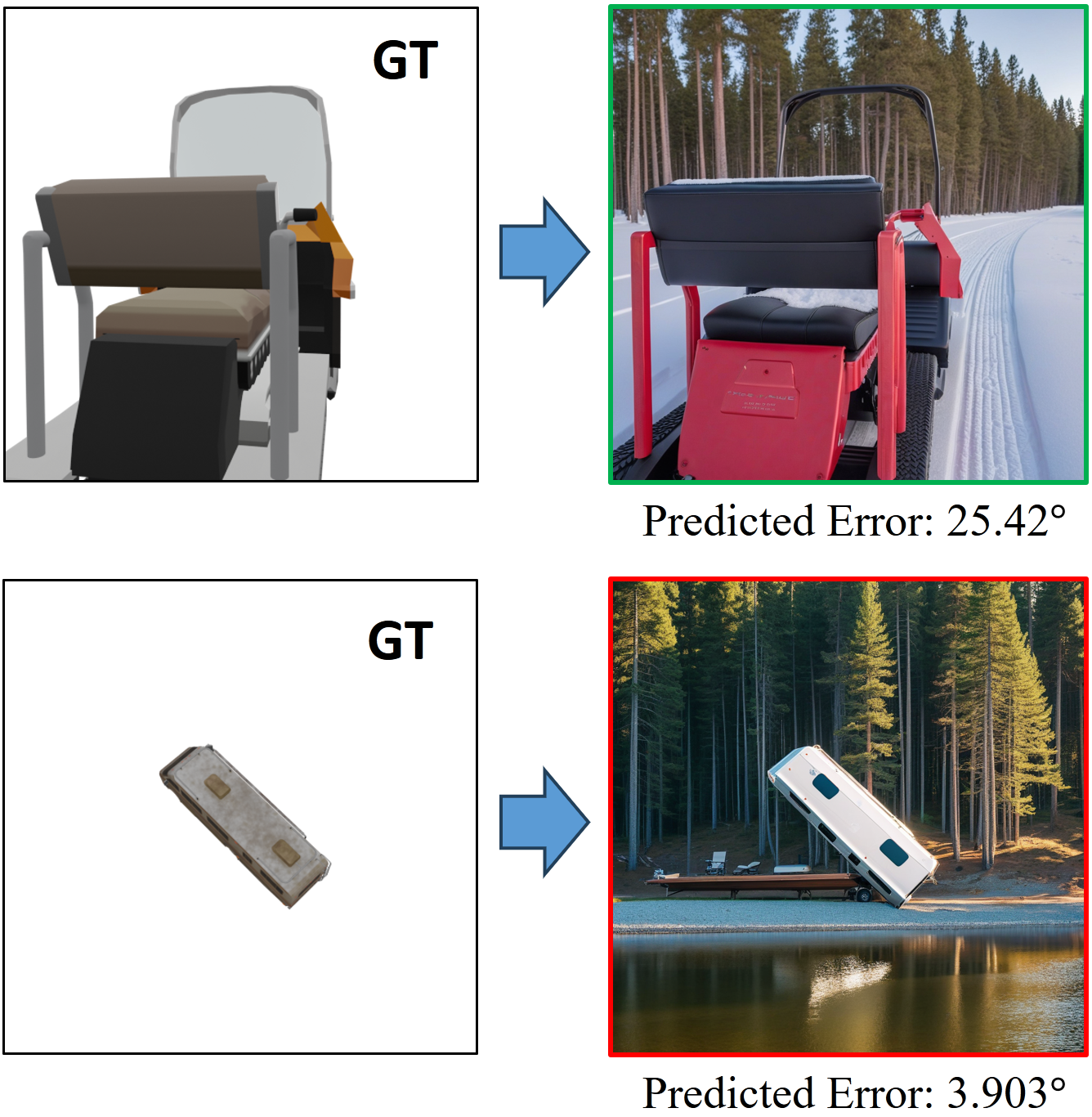}
  \caption{\textbf{Failures of Pose Estimation Model.} Enlighten by~\cite{ma2023generating}, we train a transformer-based pose estimation model on a 10-fold manner for generated images on each single category. The images with less prediction errors may indicate successful generation. However, the pose estimation model is not robust to complex objects and diverse camera-object relations in our dataset.}
  \label{suppfig:poseest}
\end{figure}

Alternatively, in Fig.~\ref{suppfig:poseest}, enlighten by~\cite{ma2023generating}, we train a transformer-based pose estimation model on a 10-fold manner for generated images on each single category. The images with less prediction errors may indicate successful generation. However, the pose estimation model is not robust to complex objects and diverse camera-object relations in our dataset. Using pose estimation model prediction error is not robust for data curation purpose.

\begin{table}[b]
  \footnotesize
  \centering
  \begin{tabular}{l|ccccc}
    \hline
    Ablations & CLIP-T $\uparrow$ & CLIP-I $\uparrow$ & DINO $\uparrow$ & IR $\uparrow$ & FID $\downarrow$ \\ \hline
    Ours (SD-V1.5) & 29.80 & 83.56 & 69.36 & -0.29 & 289.99 \\ 
    Ours (SDXL) & \textbf{34.40} & \textbf{87.44} & \textbf{70.79} & \textbf{0.23} & \textbf{256.94} \\
    \hline
  \end{tabular}
  \caption{\textbf{Ablations on DM Backbones}. SDXL provides better fidelity (CLIP-I, DINO), realism (FID), and text-image alignment (CLIP-T, IR) than other alternatives.}
  \label{tab:abltion_sd}
\end{table}

\section{Ablation of DM Backbones}
\label{suppsec:dm_backbone_ablation}
In Tab.~\ref{tab:abltion_sd}, we perform quantitative comparisons on general image visual quality between different DM backbones: SD V1.5 and SDXL. We randomly sample 1,000 generated images ($I_{syn}$), each image corresponds to: (1) the RGB visual prior rendered using Blender (one element of $I_{\beta}$); (2) the text prompts ($\mathcal{T}_{img}$) provided to the diffusion model for image generation. We evaluate the generated images on the following aspects:
\begin{itemize}
    \item Prompt following. CLIP-T (CLIP-Text Score \cite{hessel2021clipscore}) and IR (Image Reward \cite{xu2024imagereward}) metrics are employed to assess the alignment between the generated images and the input prompts. The results indicate that images produced by the SDXL backbone exhibit superior prompt alignment.
    \item Fidelity. CLIP-I (CLIP-Image Score \cite{hessel2021clipscore}) and DINO Score \cite{oquab2023dinov2} are used to measure the similarity between the visual prior images and the synthesized outputs. Findings demonstrate that SDXL more effectively preserves fidelity to the visual prior.
    \item Realism. FID \cite{heusel2017gans} is used to compare the distributions of generated images with a subset of LAION-aesthetic \cite{schuhmann2022laion} images, revealing that SDXL significantly surpasses SD V1.5 in terms of image quality and realism.
\end{itemize}

\section{Factors influencing Image Generation}
\label{suppsec:genfailure}
\begin{figure*}[t]
  \includegraphics[width=\linewidth]{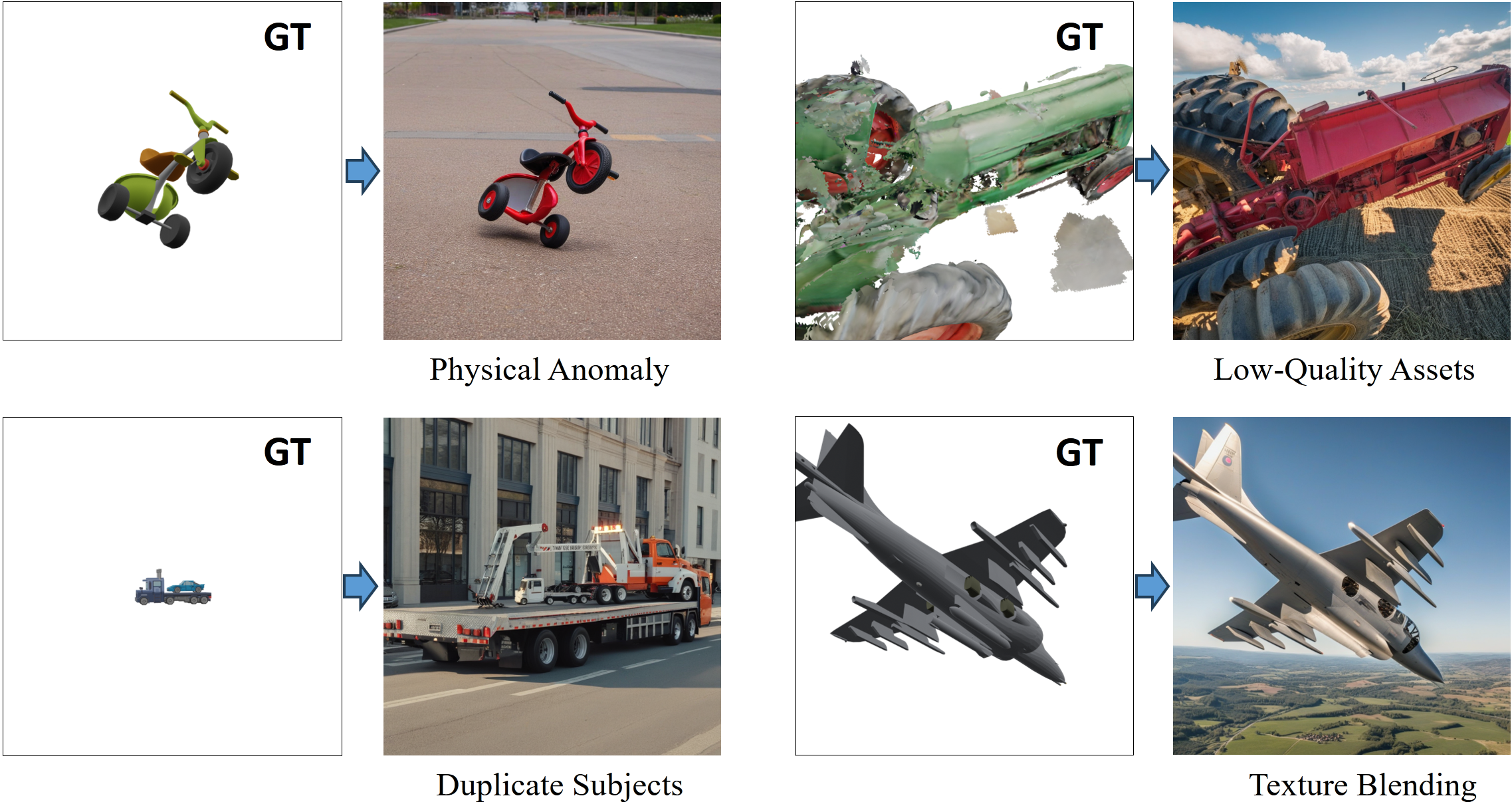}
  \caption{\textbf{Factors influencing Image Generation.} The upper-left example shows that bottom-view camera-object relations can introduce physical anomalies, such as the tricycle appearing with its front lifted unnaturally. The upper-right example highlights that low-quality 3D assets can produce fragmented depth maps or Canny edge visual priors, resulting in unrealistic generated images. The lower-left example shows that small subjects viewed from a long-shot perspective may lead to duplicate subjects in the generated image. The lower-right example illustrates how complex 3D asset structures can cause texture blending issues, particularly where intricate geometry is present.}
  \label{fig:supp_genfailure}
\end{figure*}

Several factors may lead to unreasonable artifacts in our image generation pipeline. In Fig.~\ref{fig:supp_genfailure}, the upper-left example shows that bottom-view camera-object relations can introduce physical anomalies, such as the tricycle appearing with its front lifted unnaturally. The upper-right example highlights that low-quality 3D assets can produce fragmented depth maps or Canny edge visual priors, resulting in unrealistic generated images. The lower-left example shows that small subjects viewed from a long-shot perspective may lead to duplicate subjects in the generated image. The lower-right example illustrates how complex 3D asset structures can cause texture blending issues, particularly where intricate geometry is present.

There are several ways to reduce the influences from those factors. First, choosing 3D assets with reasonable number of meshes, and with plausible textures may filter the low-quality and texture blending failures. Second, using simple and compact image description prompts to generate long-shot camera-shot images will significantly reduce the duplicate subject failures. Reduce the image generation with high elevation angle from bottom view will significantly reduce physical anomaly.

\section{User Study Details}
\label{suppsec:userstudy}
\begin{figure*}[h]
  \includegraphics[width=\linewidth]{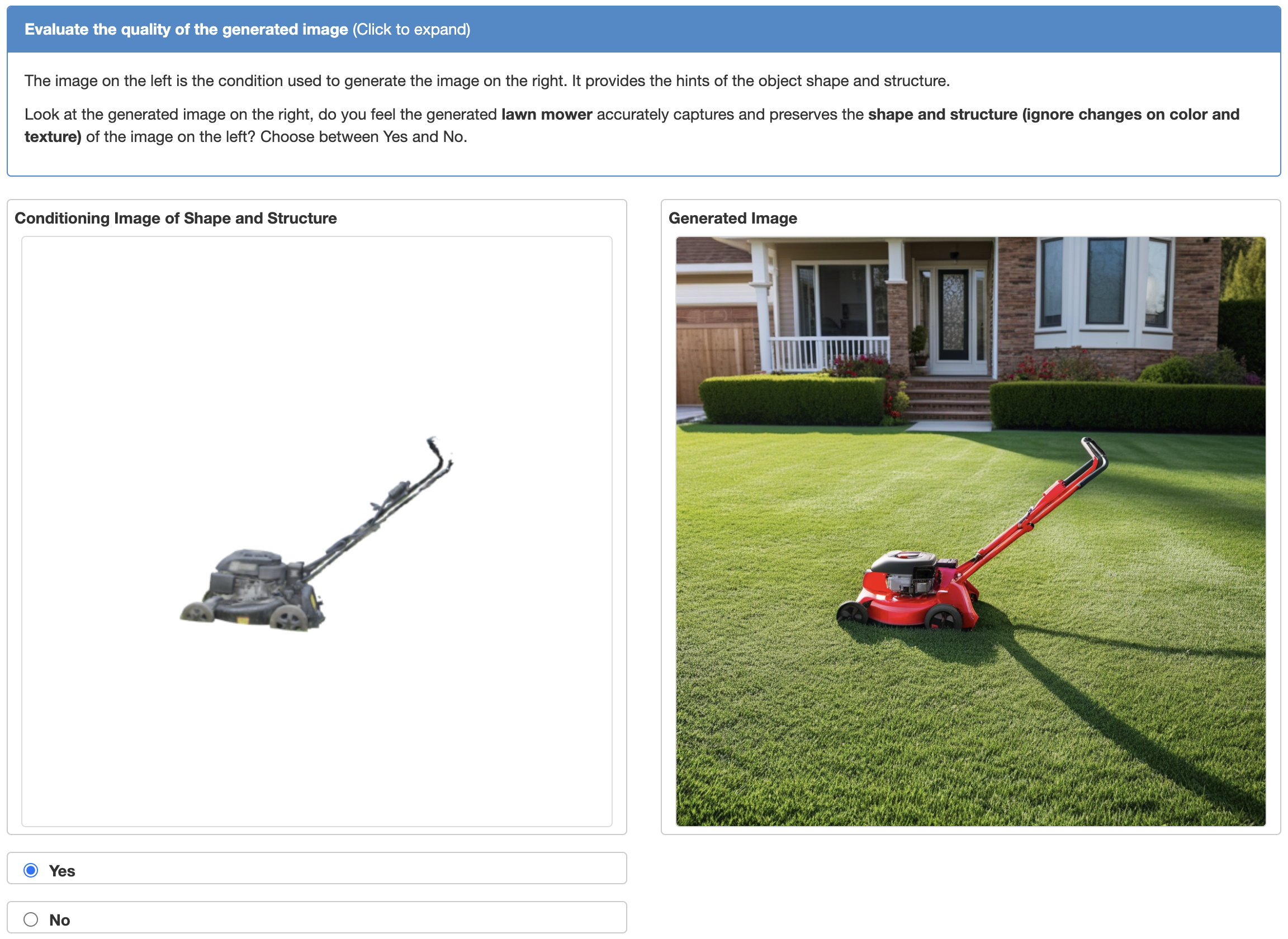}
  \caption{\textbf{User Study UI Page.} Users will see a pair of side-by-side images. The left hand side is an RGB image generated by Blender, and the right hand side is the synthetic image generated by DMs using that RGB image as guidance prior. A vivid preserving of 3D geometry and structure in the synthetic image is a success. Users are required to provide a binary answer regarding the success of the image generation.}
  \label{suppfig:userstudy}
\end{figure*}

The Fig.~\ref{suppfig:userstudy} shows the UI page of our user study. We randomly selected 200 images from the \textit{Ultimate3D} dataset. Each image corresponds to an RGB prior generated by Blender. During the user study, users will see a pair of side-by-side images. One is an RGB image generated by Blender, and the other is the synthetic image generated by DMs using that RGB image as guidance prior. A vivid preserving of 3D geometry and structure in the synthetic image is a success. Users are required to provide a binary answer regarding the success of the image generation. Each image pair will be manually reviewed by at least 3 users.

\section{Additional Qualitative Results}
\label{suppsec:addqual}
\begin{figure*}[t]
  \includegraphics[width=\linewidth]{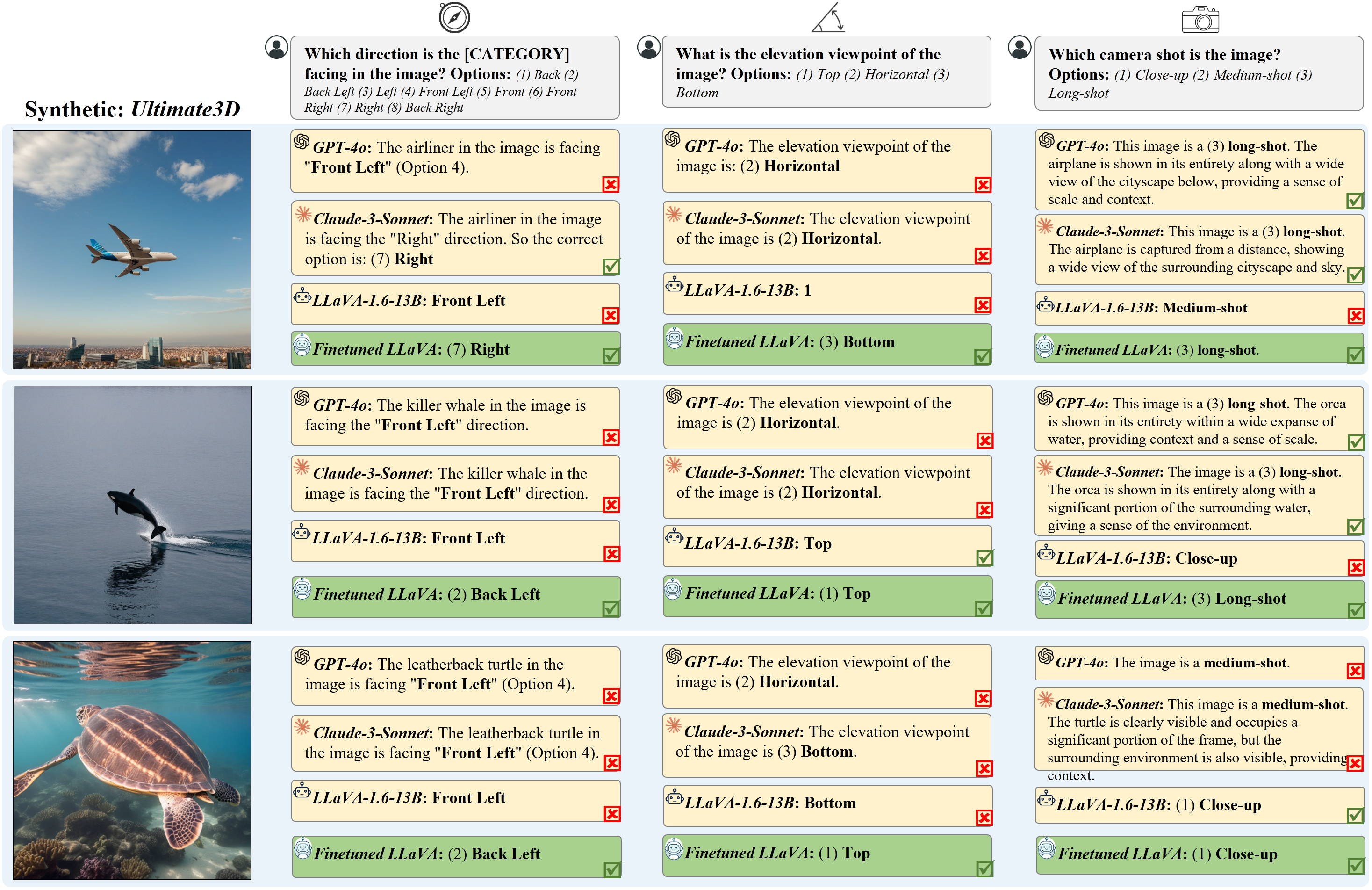}
  \includegraphics[width=\linewidth]{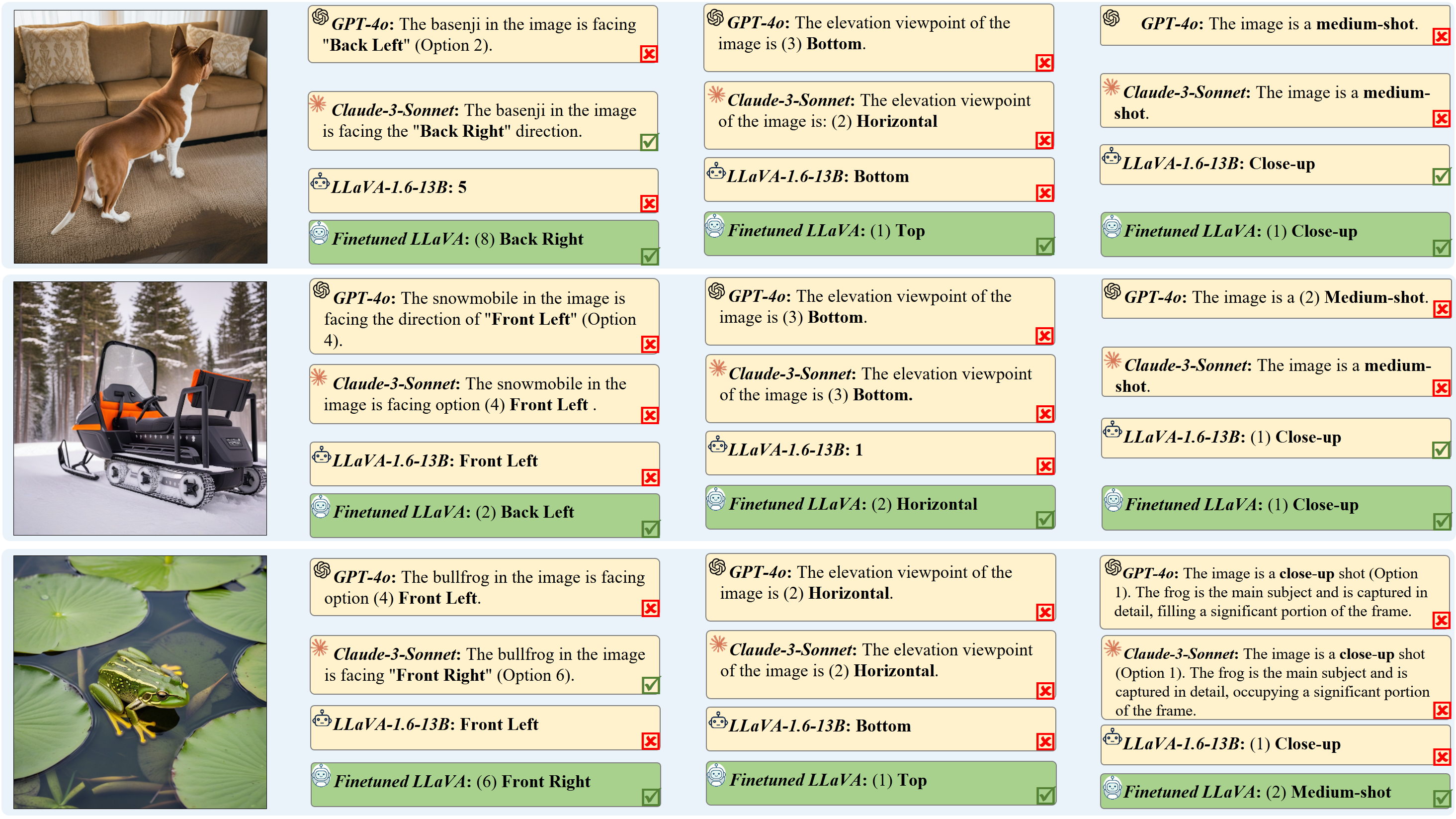}
  \caption{\textbf{Additional Qualitative Results.} We show the evaluations of camera-object relation recognition capability of GPT-4o, Claude-3-Sonnet, LLaVA-1.6-13B, and finetuned LLaVA-1.6-13B, on \textit{Ultimate3D} benchmark. Each model is asked the questions (in gray boxes) regarding object orientation, camera viewpoint, and camera-shots type, together with the input images on the left. The model responses illustrate that finetuned LLaVA-1.6-13B model by \textit{Ultimate3D} dataset significantly outperforms all compared models.}
  \label{fig:supp_qualitatiev}
\end{figure*}

Based on \textit{Ultimate3D} benchmark, we provide additional Fig.~\ref{fig:supp_qualitatiev} to show more qualitative comparisons between finetuned LLaVA model to commercial SOTAs. The finetuned LLaVA model outperforms other models.

\section{Additional Visuals of \textit{Ultimate3D} Dataset}
\label{suppsec:ultimate3D}
\begin{figure*}[hbt!]
\centering
\setlength{\tabcolsep}{3pt}
\begin{tabular}{p{0.5em}cccccc}
   \rotatebox{90}{\textbf{Front}} & 
   \includegraphics[width=0.15\textwidth]{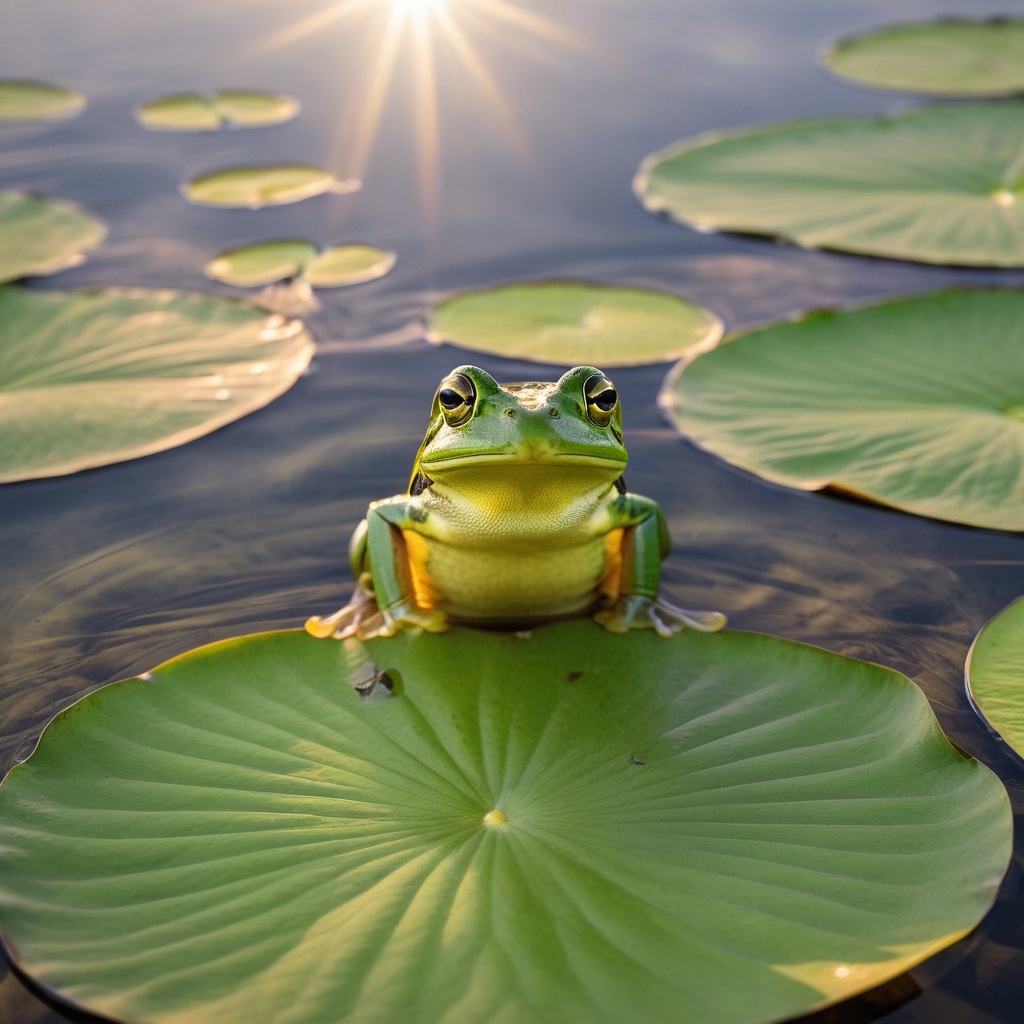} &    
    \includegraphics[width=0.15\textwidth]{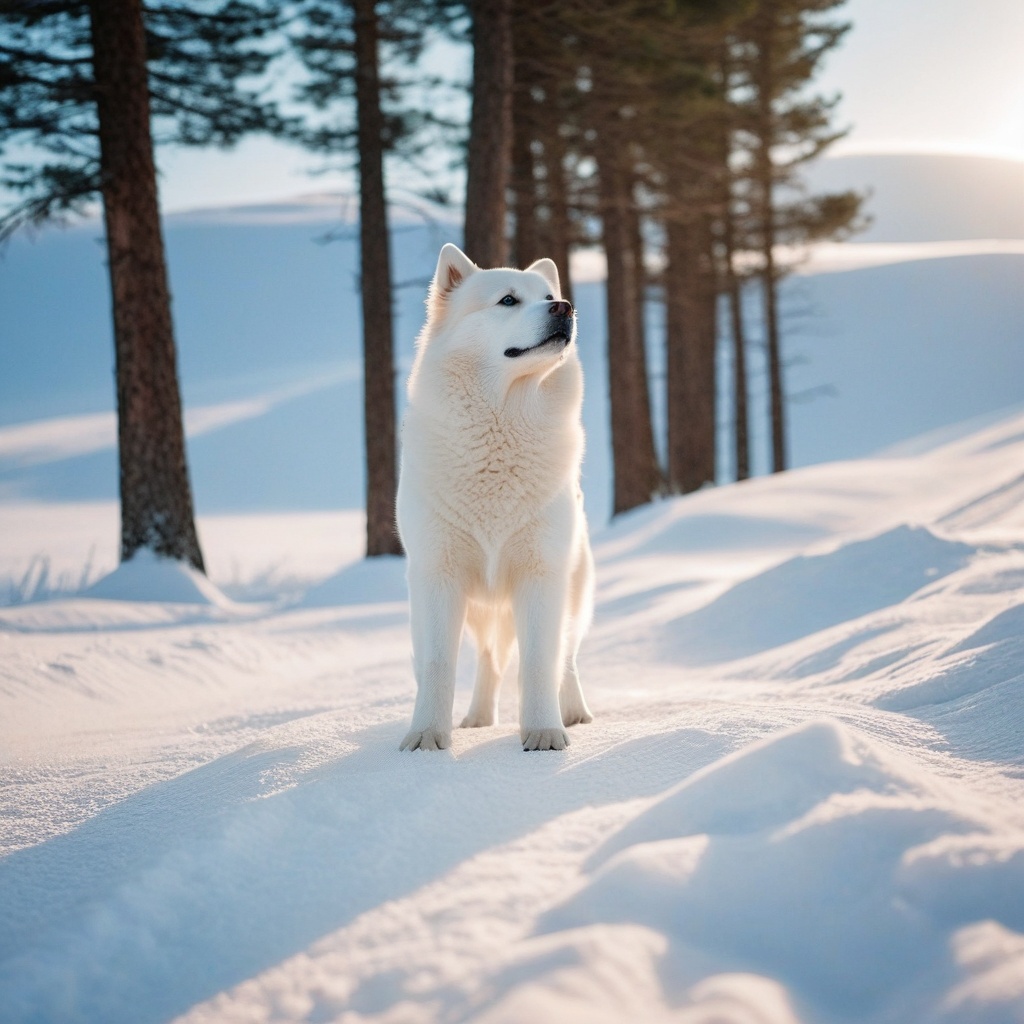} &
   \includegraphics[width=0.15\textwidth]{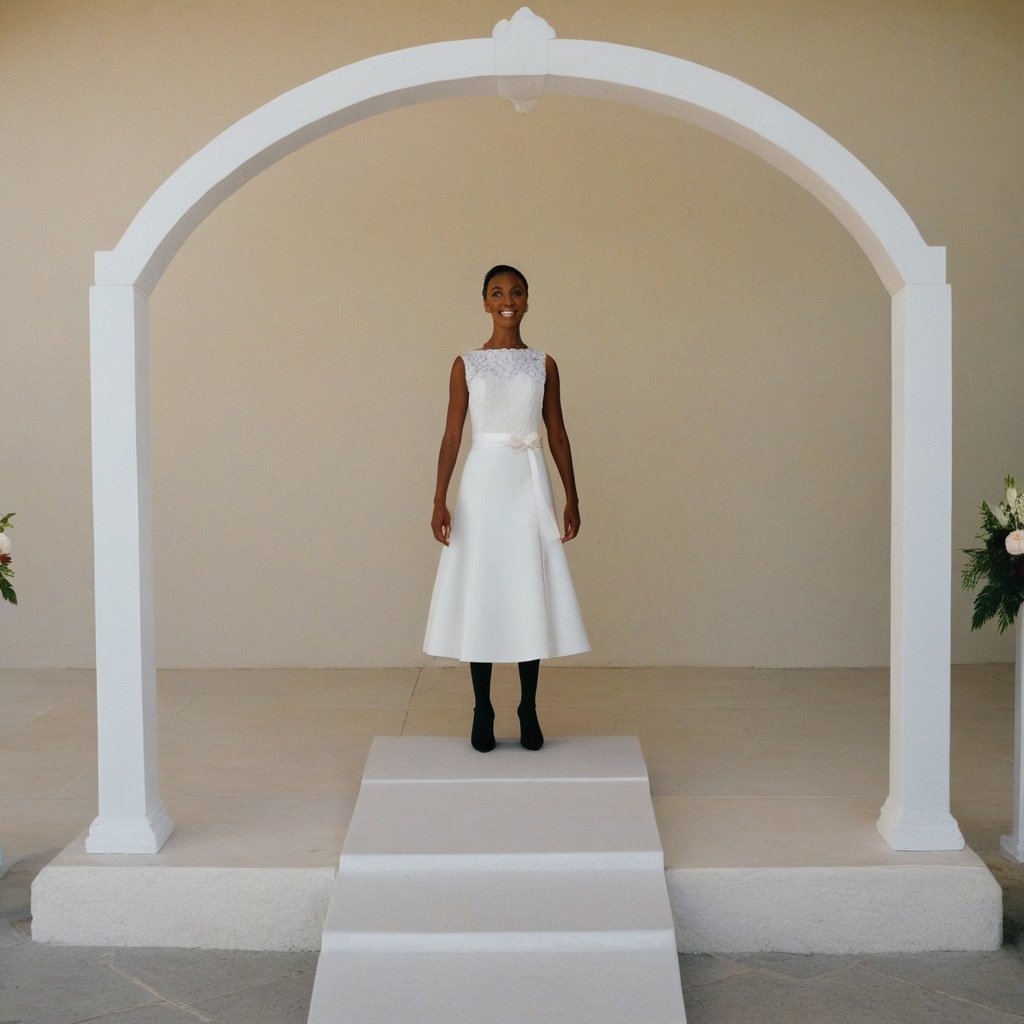} &
    \includegraphics[width=0.15\textwidth]{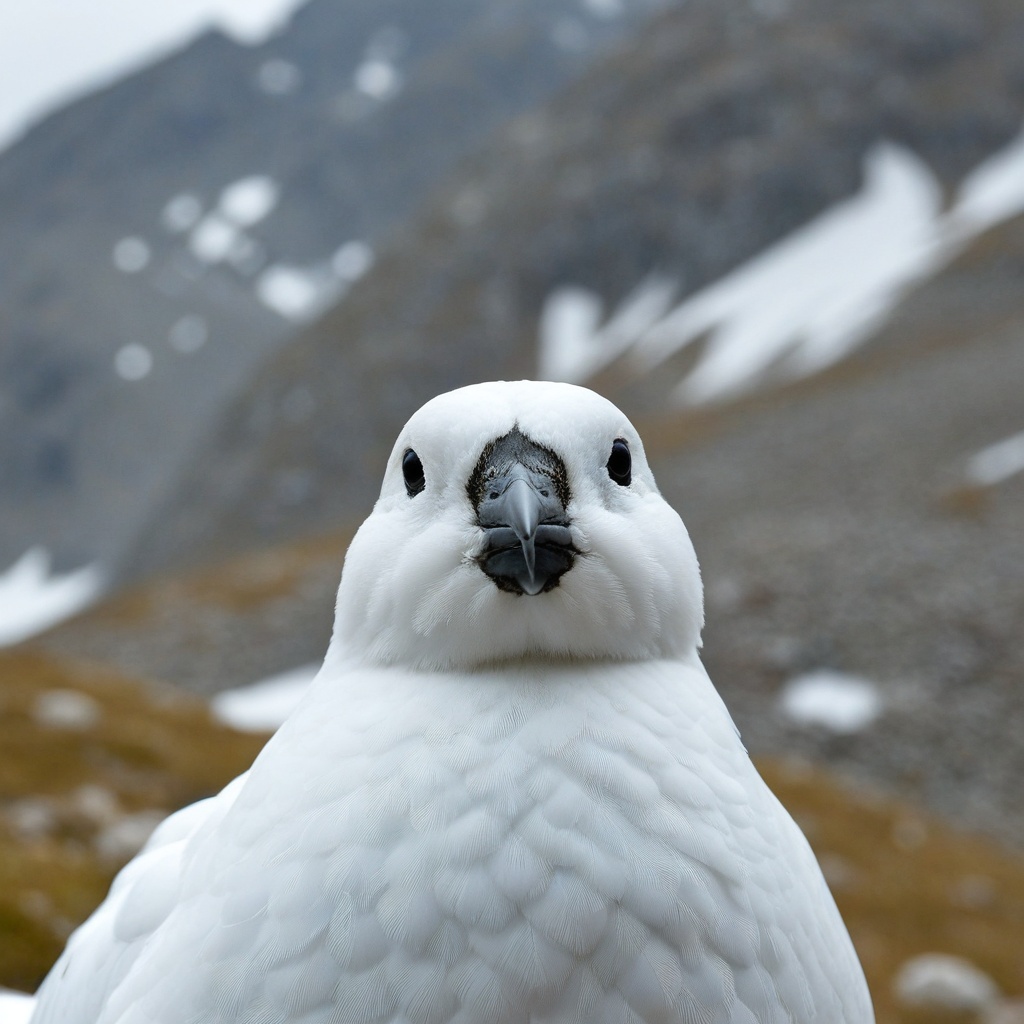} &
    \includegraphics[width=0.15\textwidth]{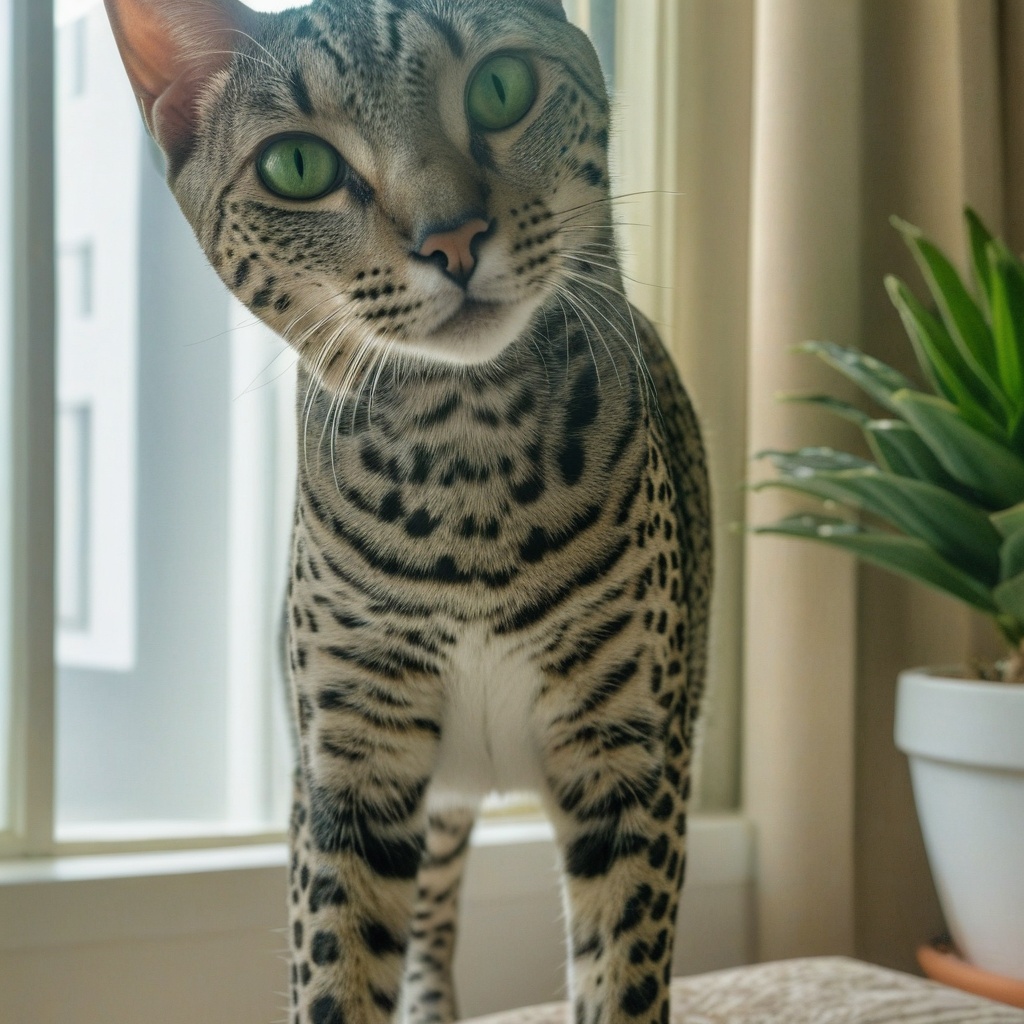} &
    \includegraphics[width=0.15\textwidth]{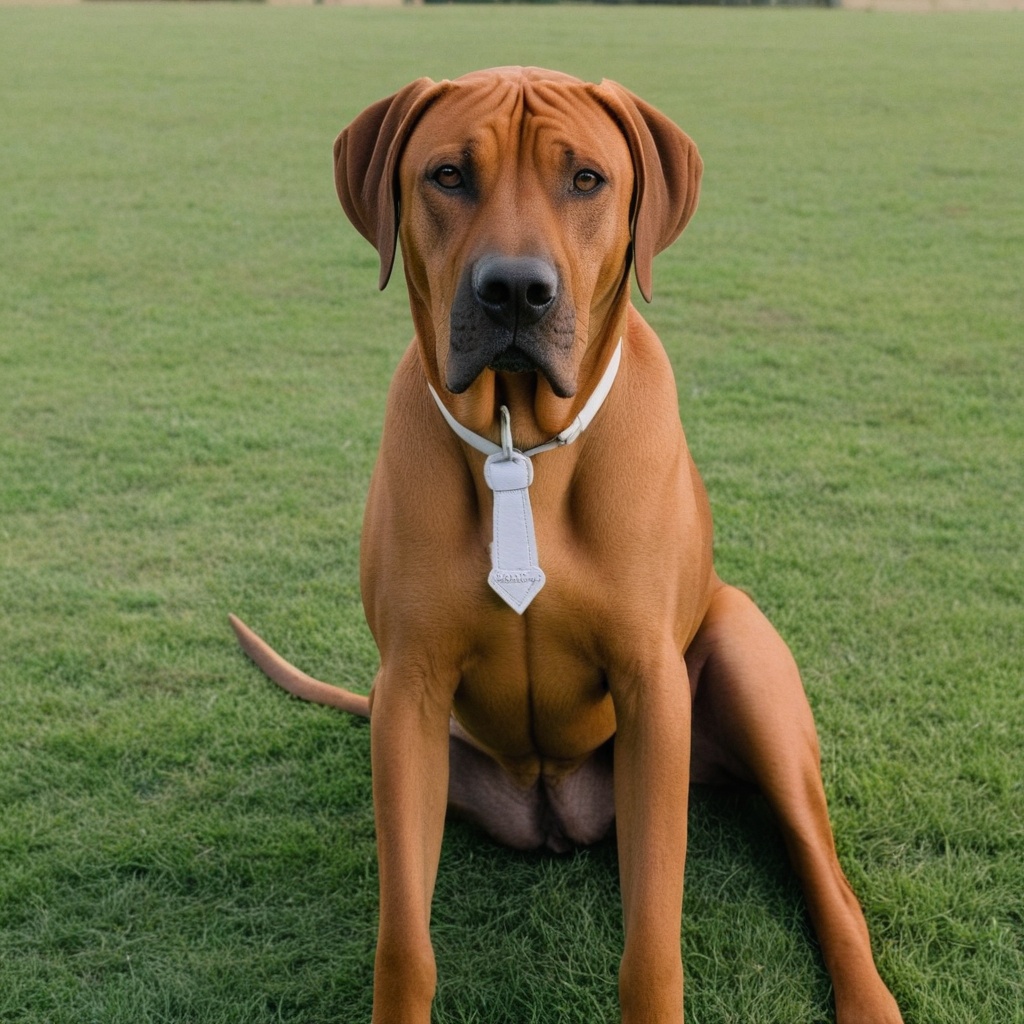} 
    \\
   \rotatebox{90}{\textbf{Front Left}} & 
    \includegraphics[width=0.15\textwidth]{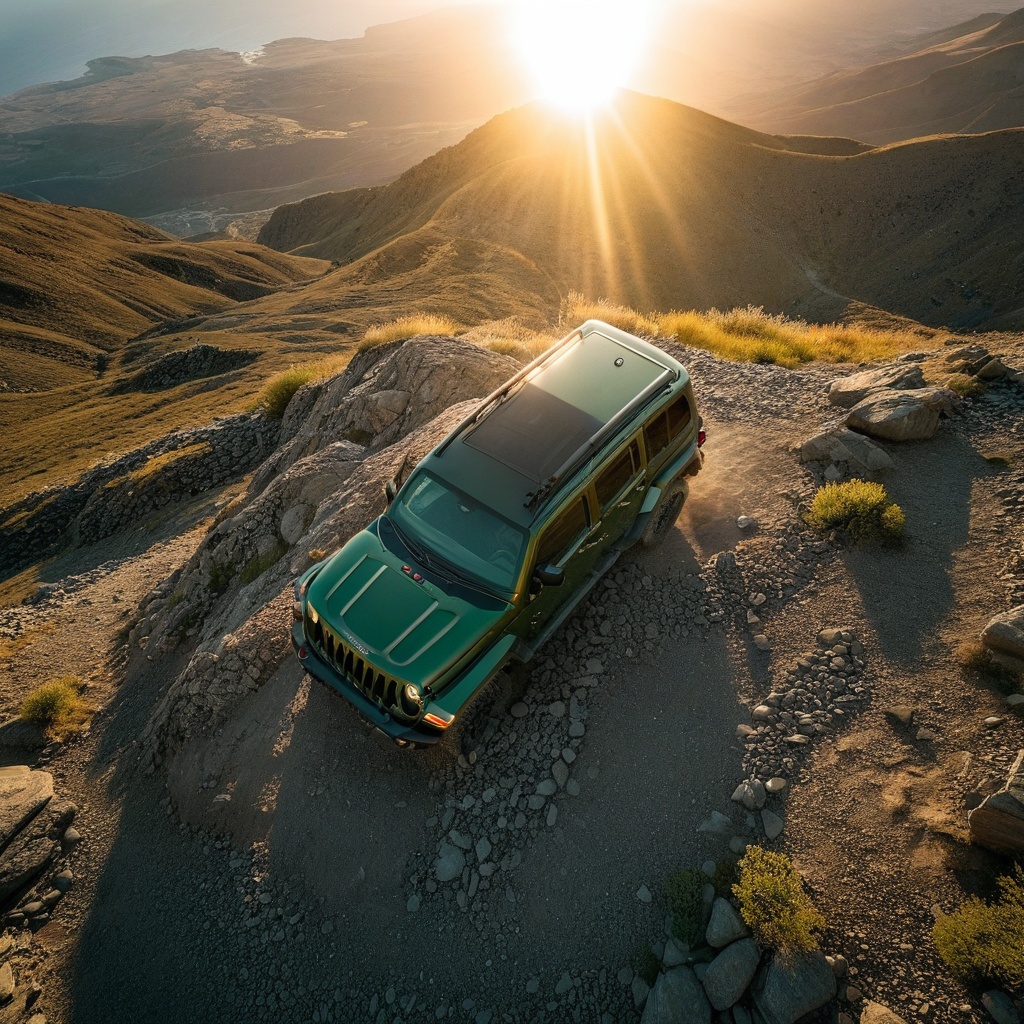} &
   \includegraphics[width=0.15\textwidth]{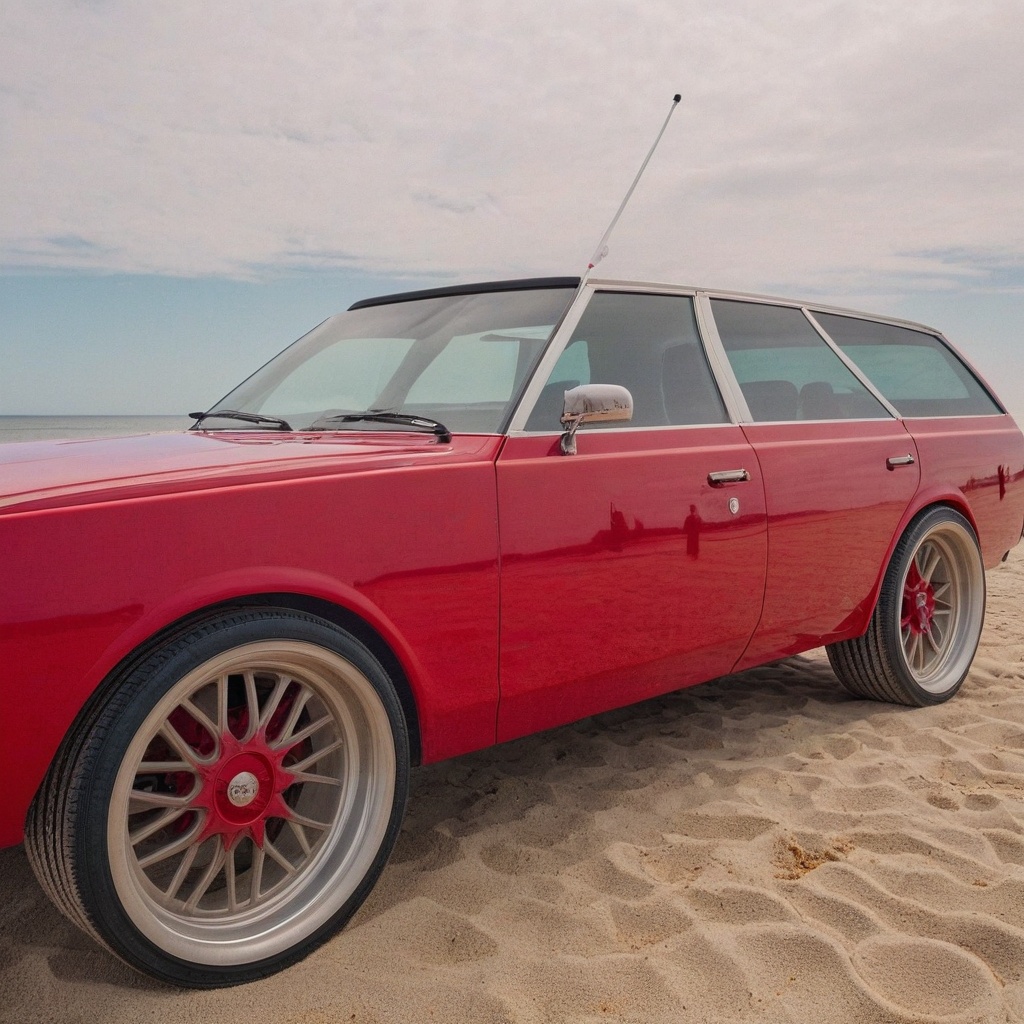} &    
   \includegraphics[width=0.15\textwidth]{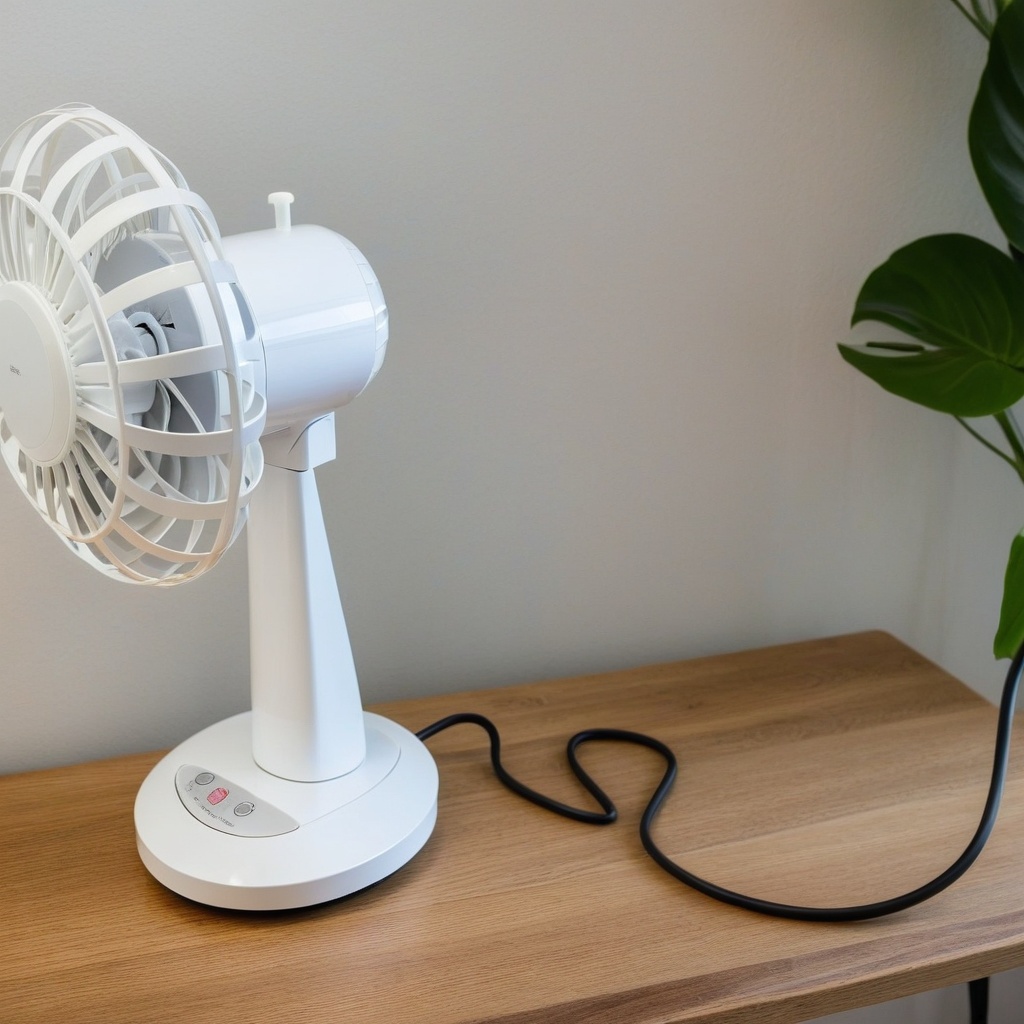} &
    \includegraphics[width=0.15\textwidth]{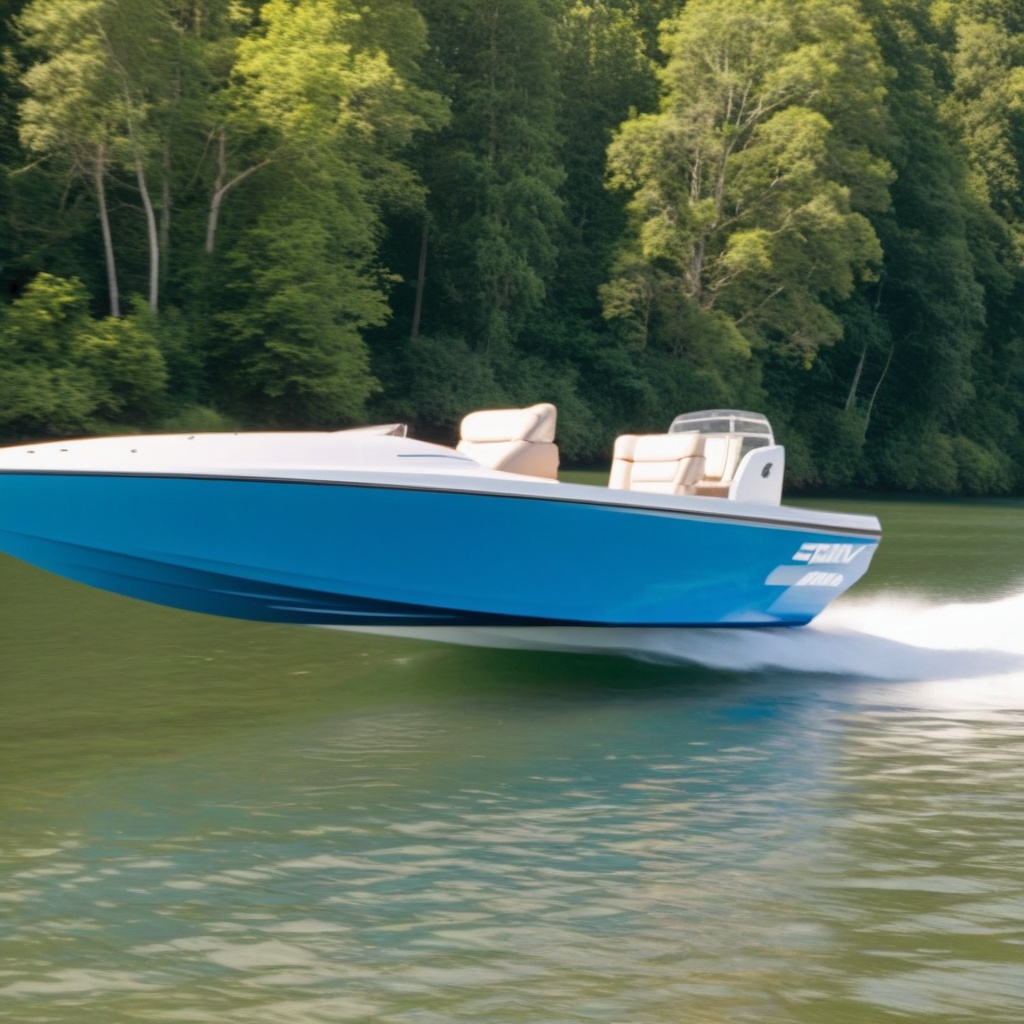} & 
    \includegraphics[width=0.15\textwidth]{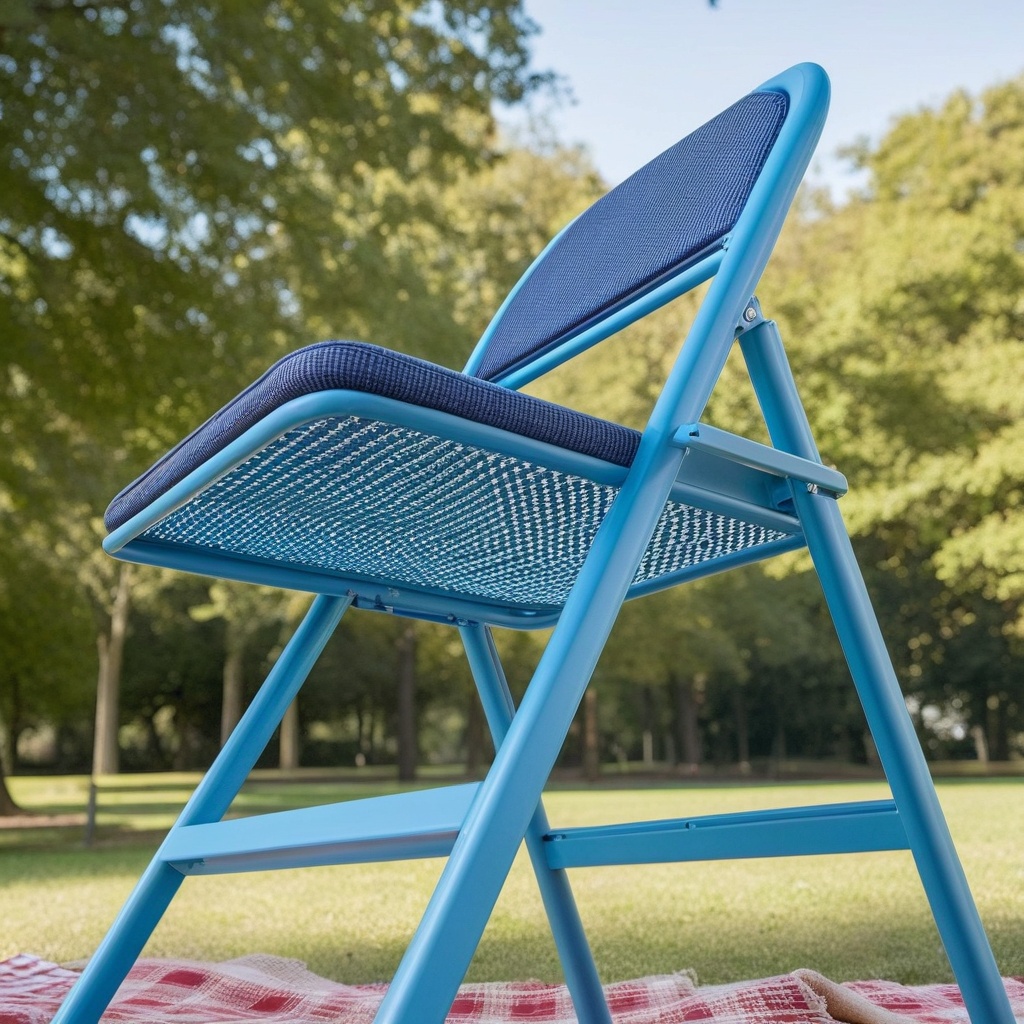} &
    \includegraphics[width=0.15\textwidth]{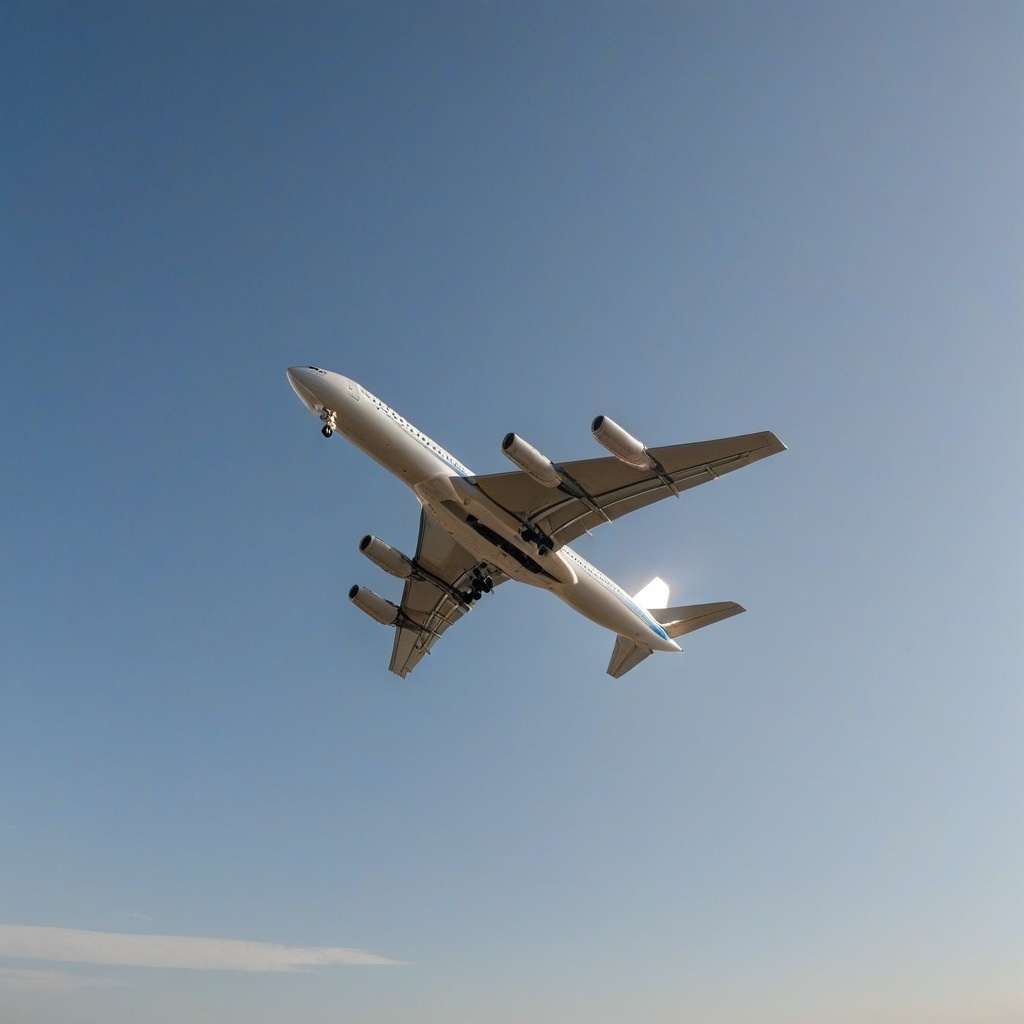} 
    \\
   \rotatebox{90}{\textbf{Left}} & 
    \includegraphics[width=0.15\textwidth]{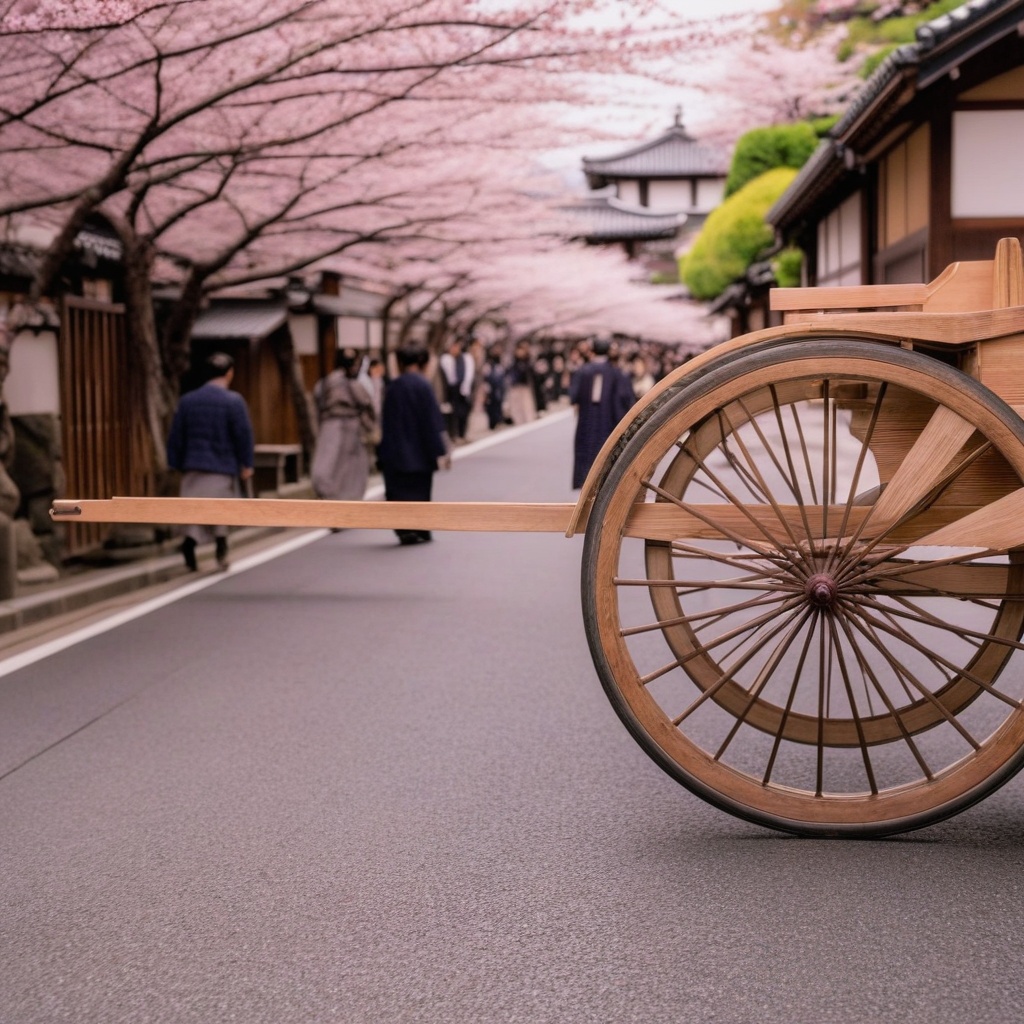} &
   \includegraphics[width=0.15\textwidth]{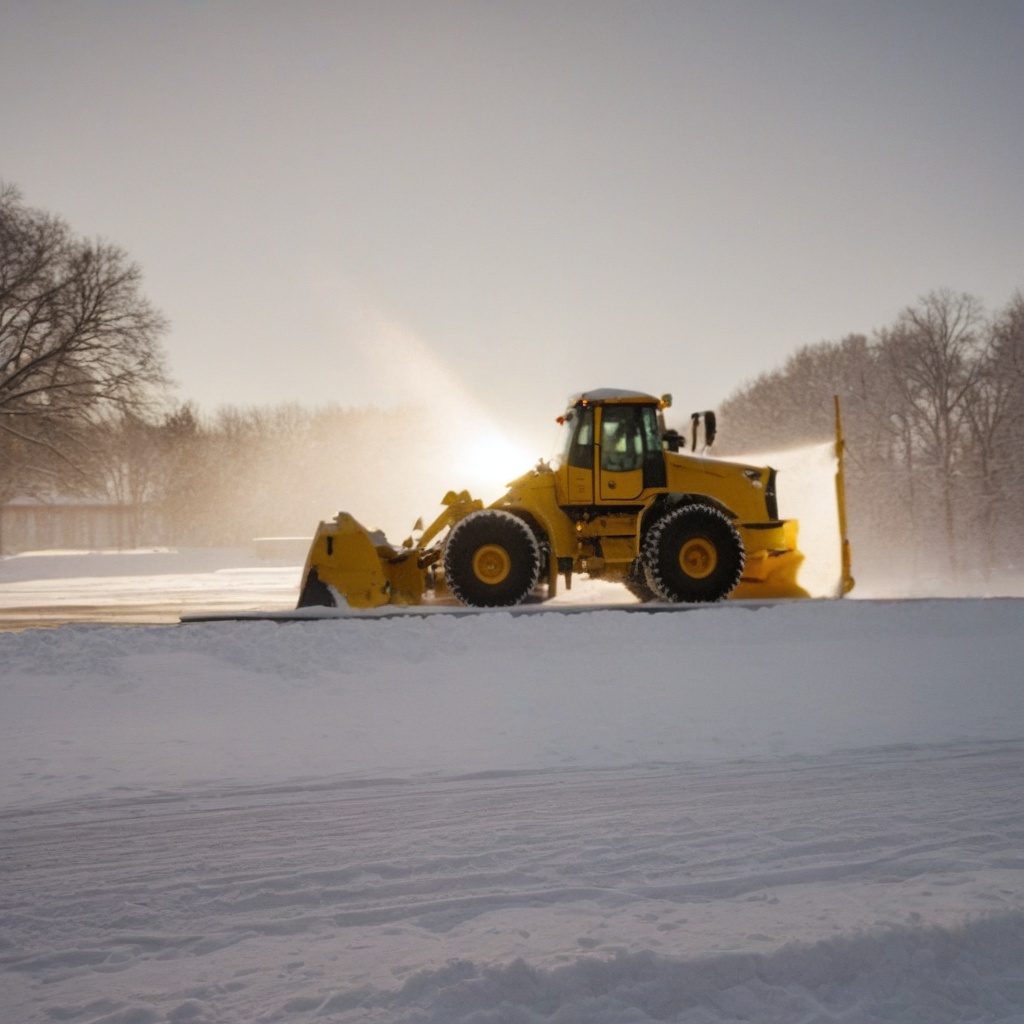} &    
   \includegraphics[width=0.15\textwidth]{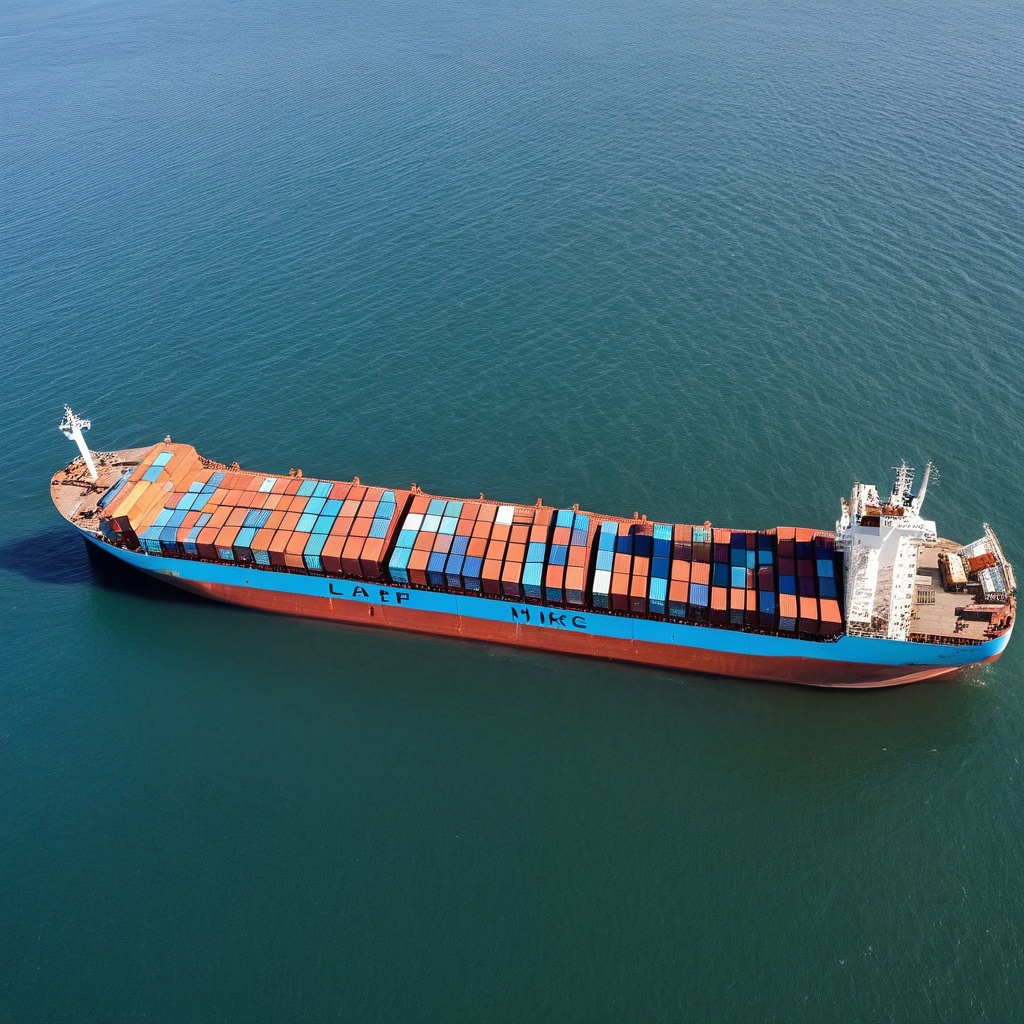} &
    \includegraphics[width=0.15\textwidth]{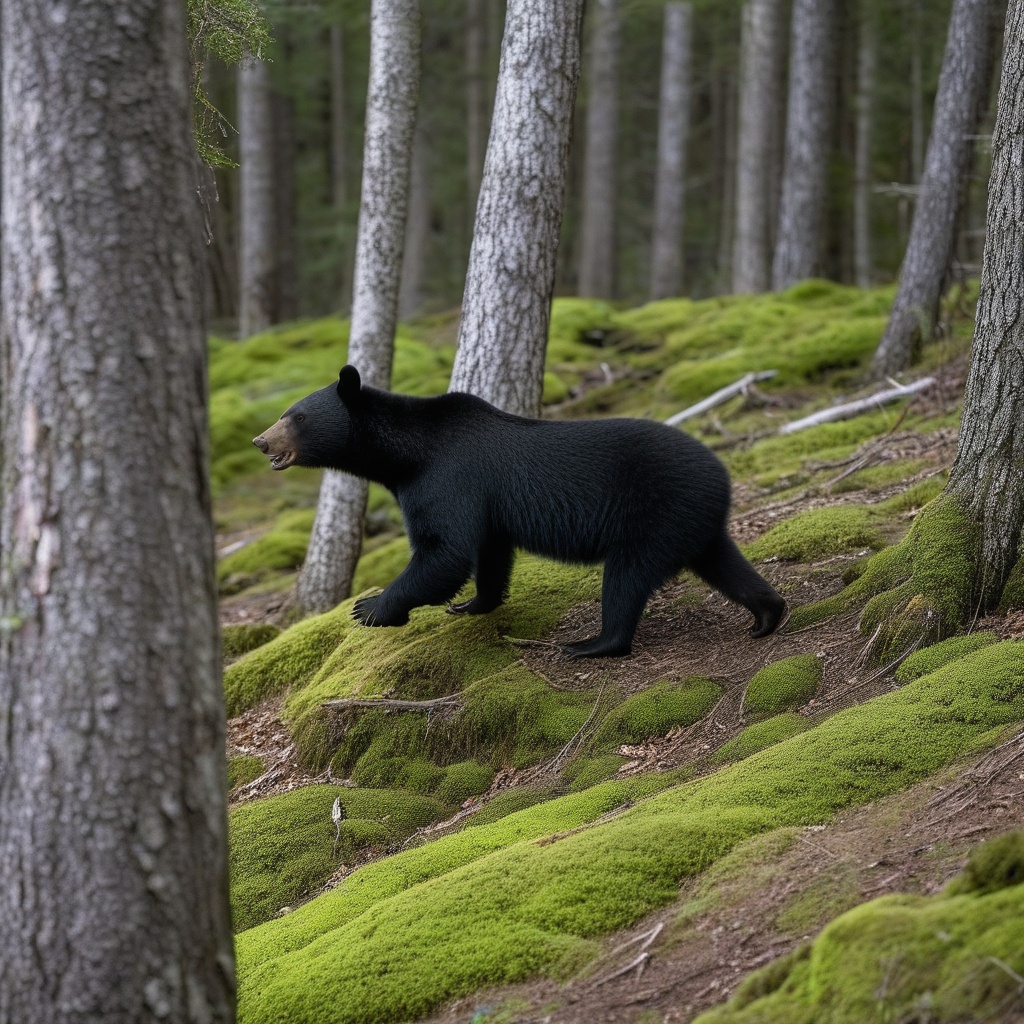} &
    \includegraphics[width=0.15\textwidth]{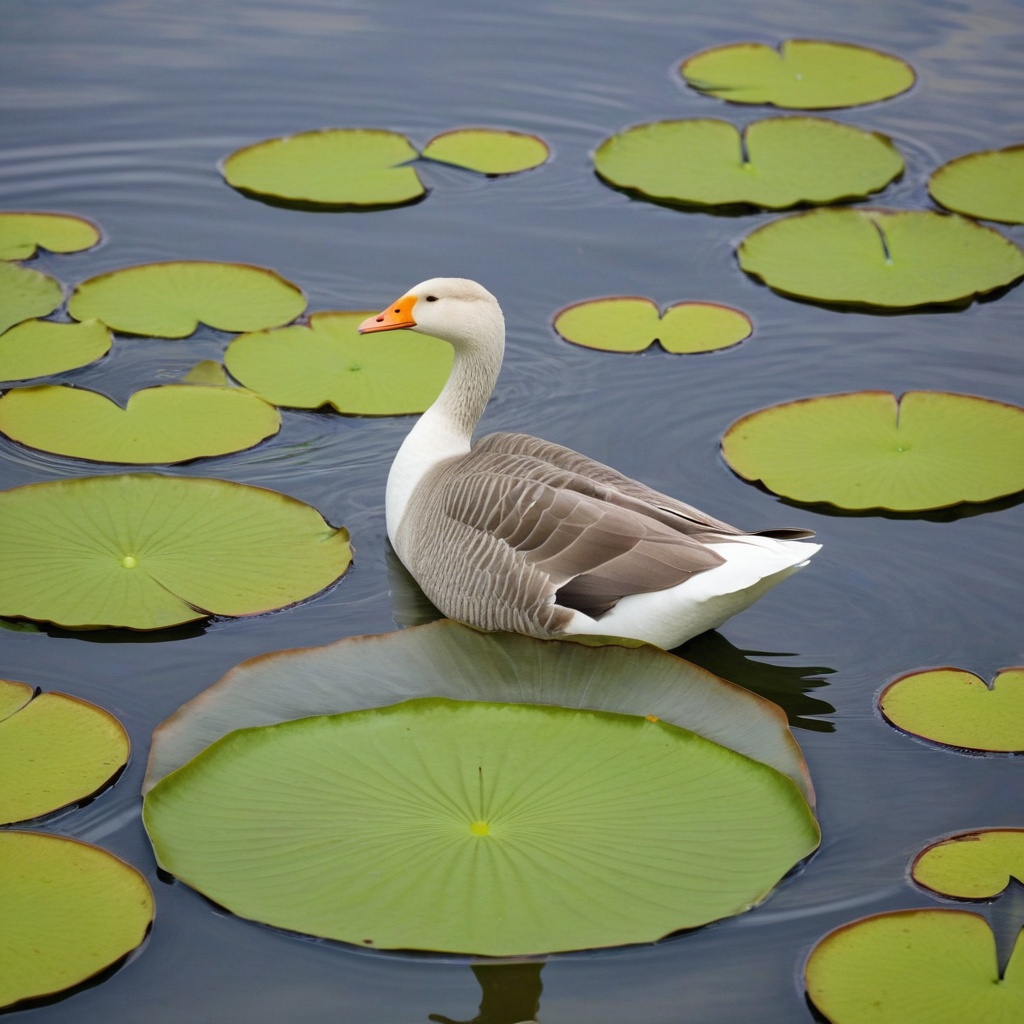} &
    \includegraphics[width=0.15\textwidth]{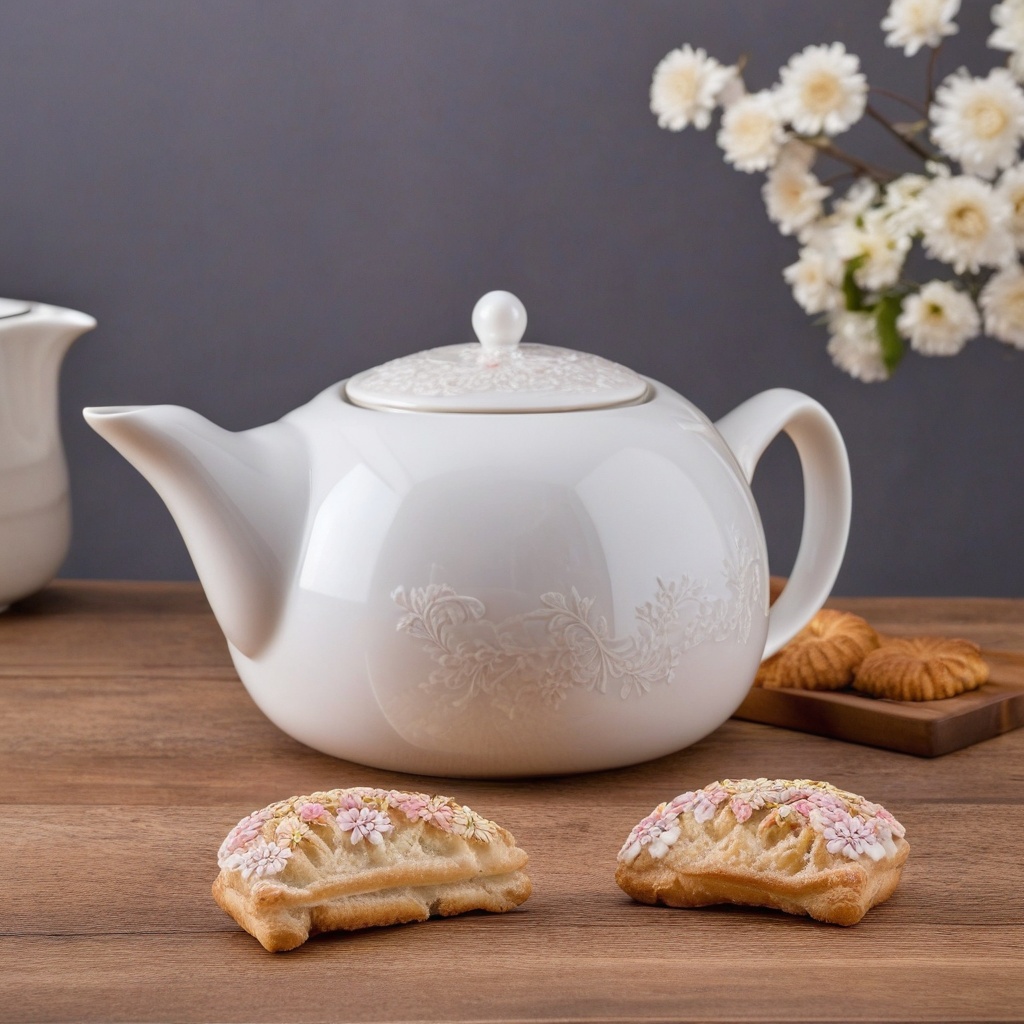} 
    \\
   \rotatebox{90}{\textbf{Back Left}} & 
    \includegraphics[width=0.15\textwidth]{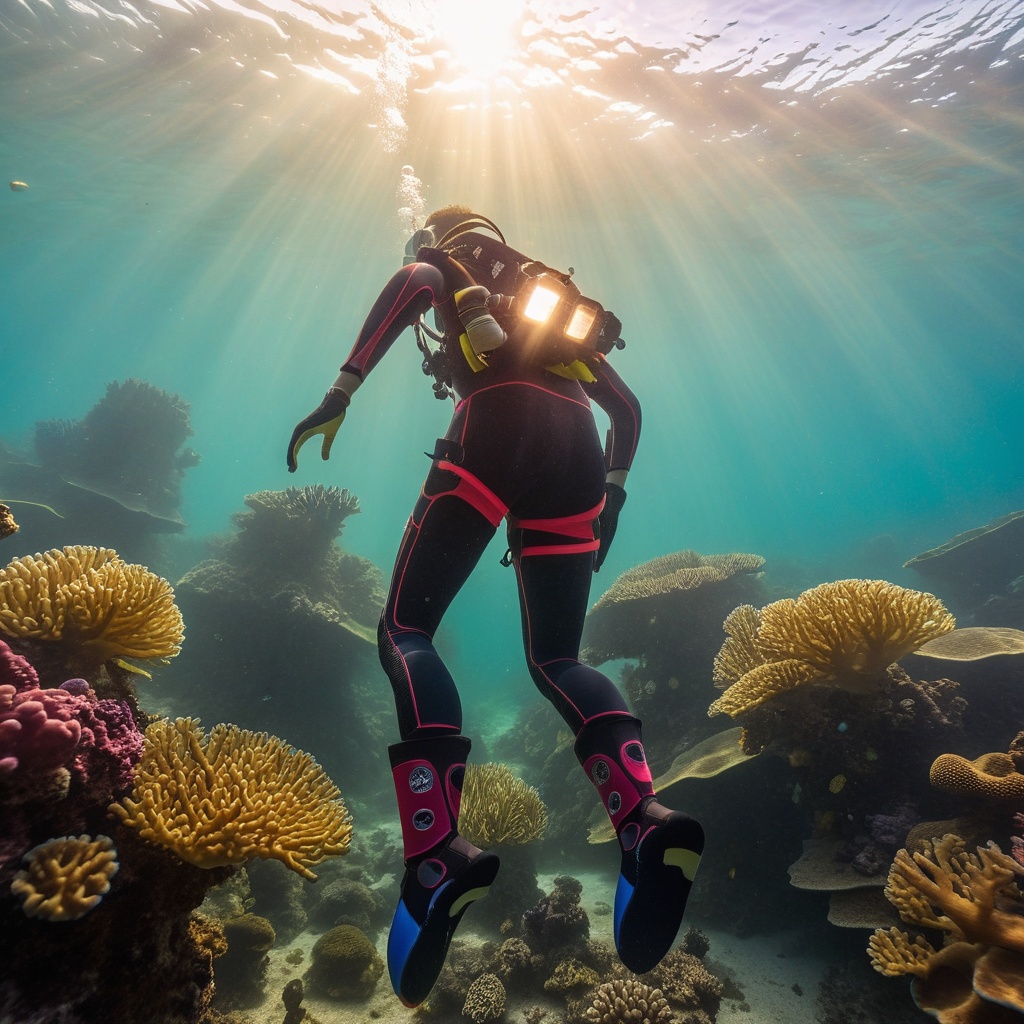} &
   \includegraphics[width=0.15\textwidth]{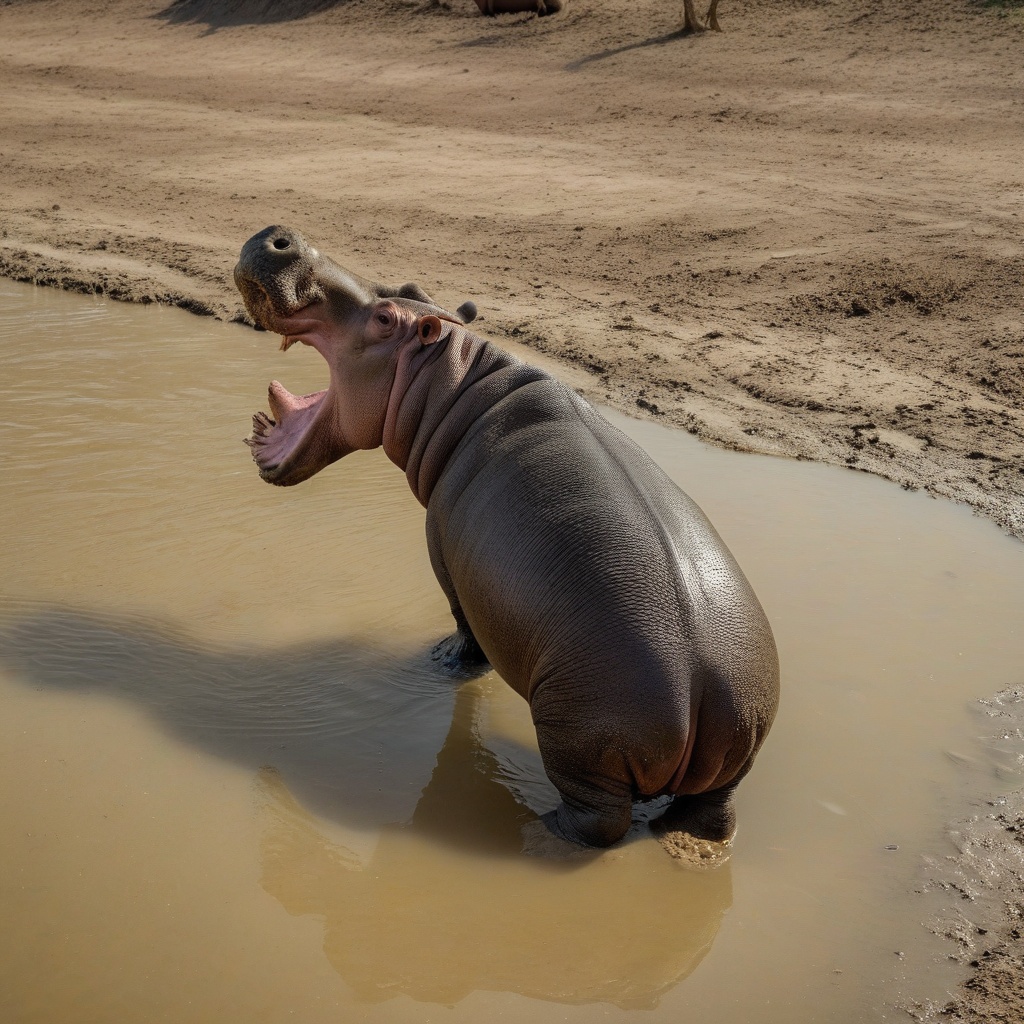} &    
   \includegraphics[width=0.15\textwidth]{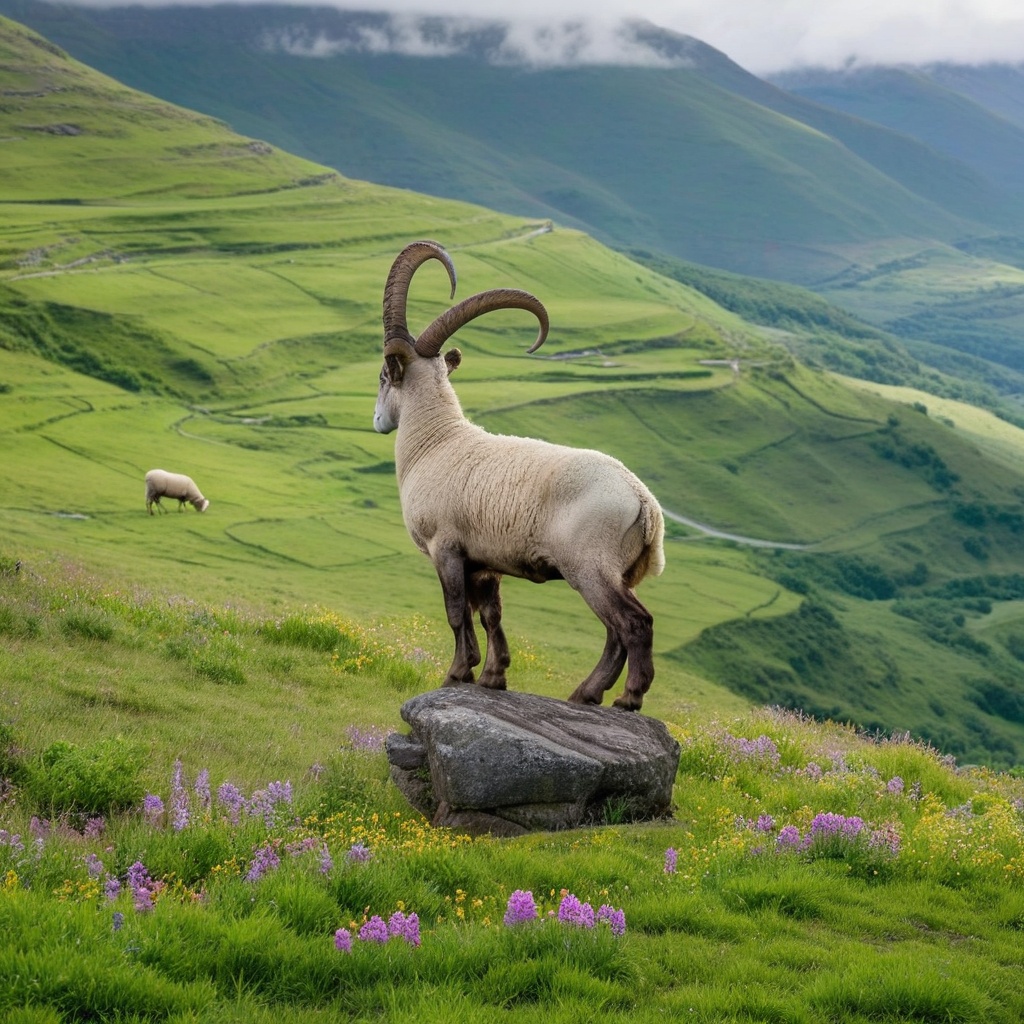} &
    \includegraphics[width=0.15\textwidth]{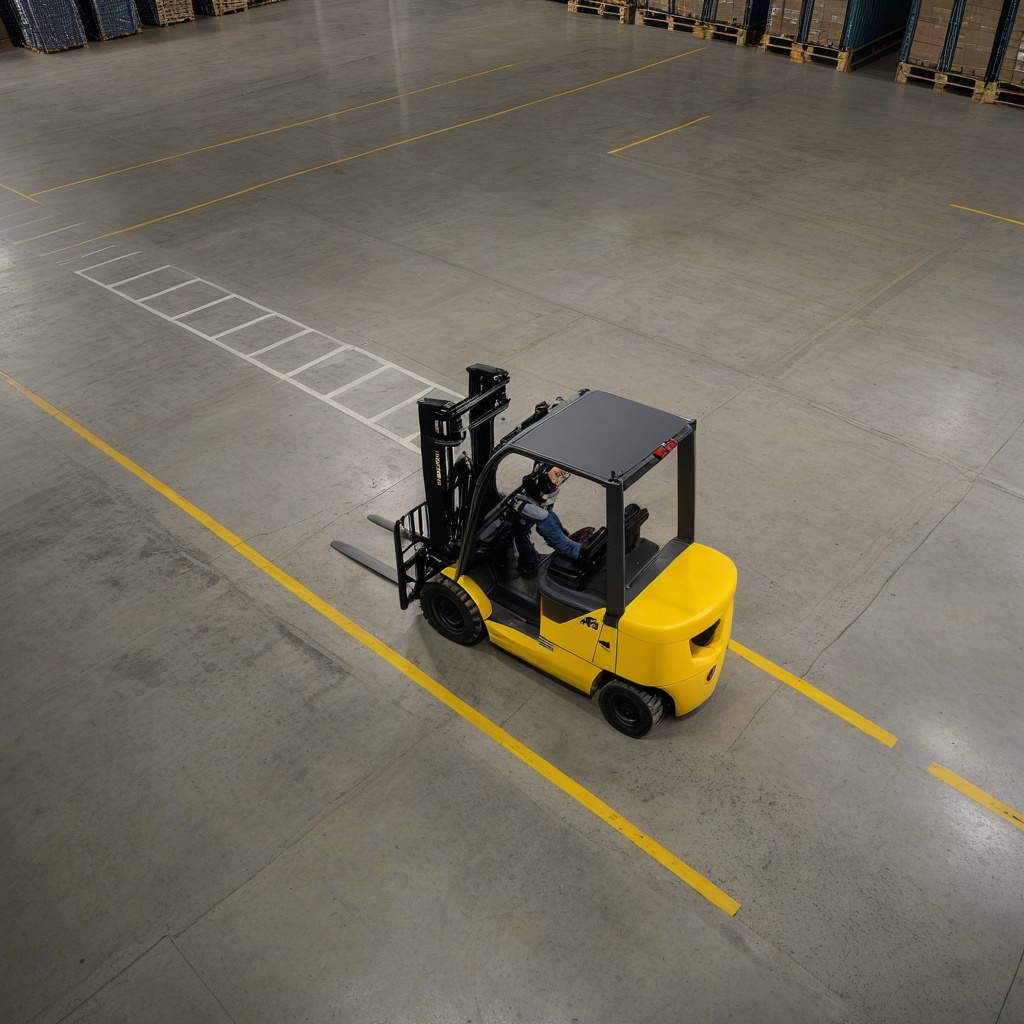} &
    \includegraphics[width=0.15\textwidth]{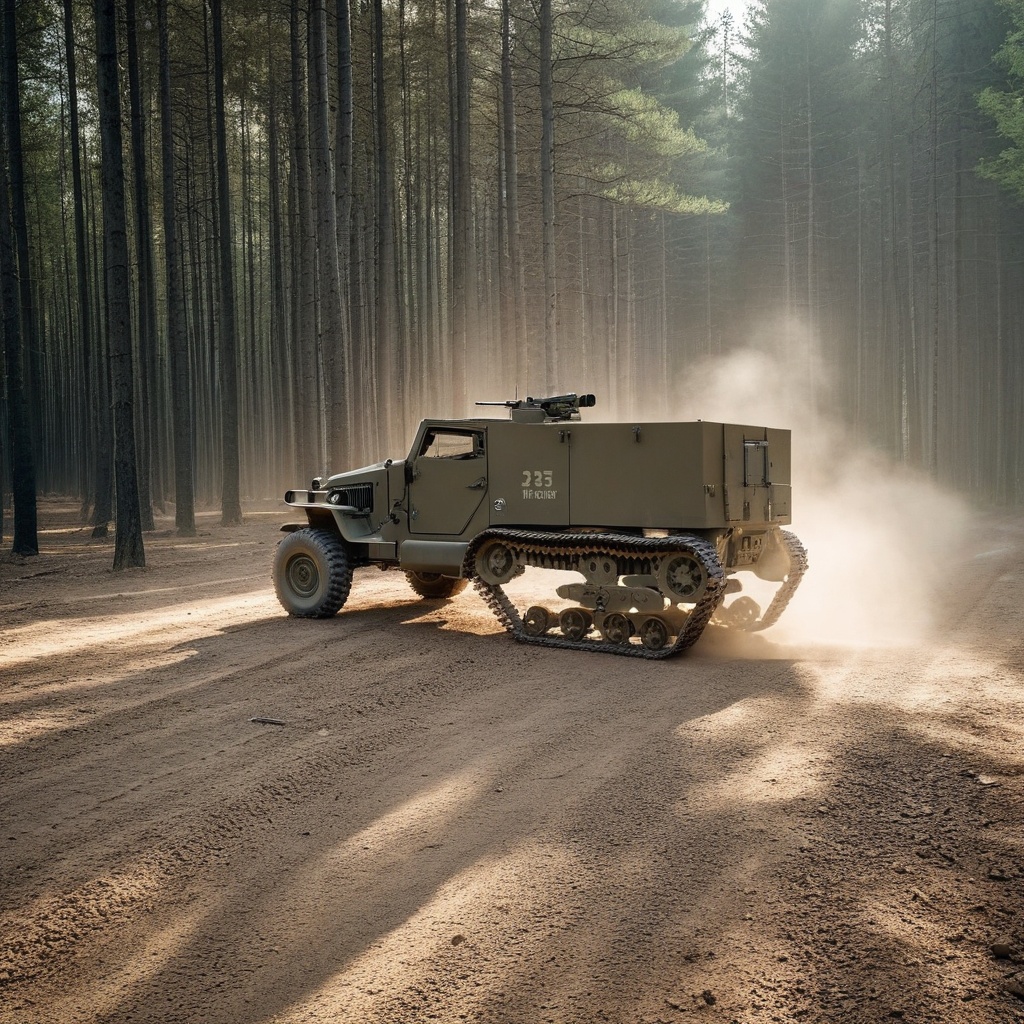} &
    \includegraphics[width=0.15\textwidth]{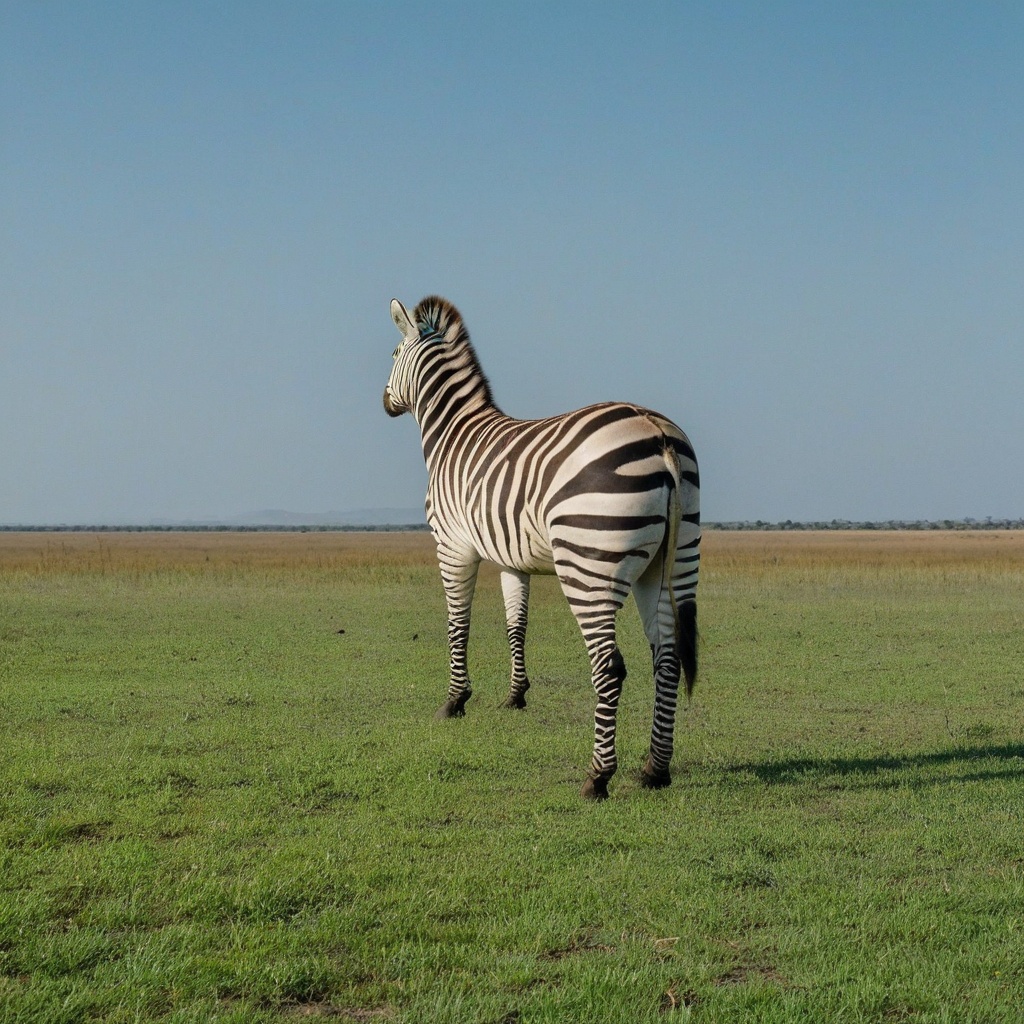} 
    \\
   \rotatebox{90}{\textbf{Back}} & 
    \includegraphics[width=0.15\textwidth]{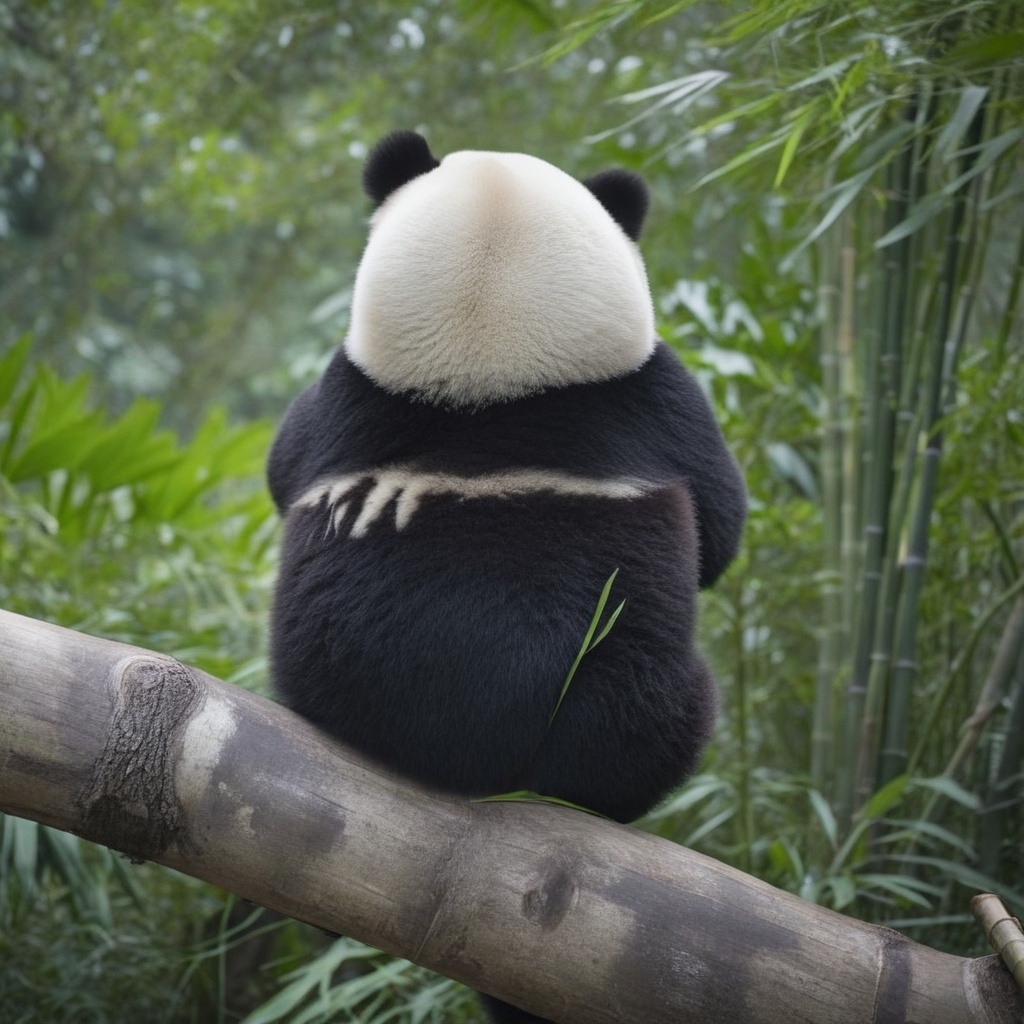} &
   \includegraphics[width=0.15\textwidth]{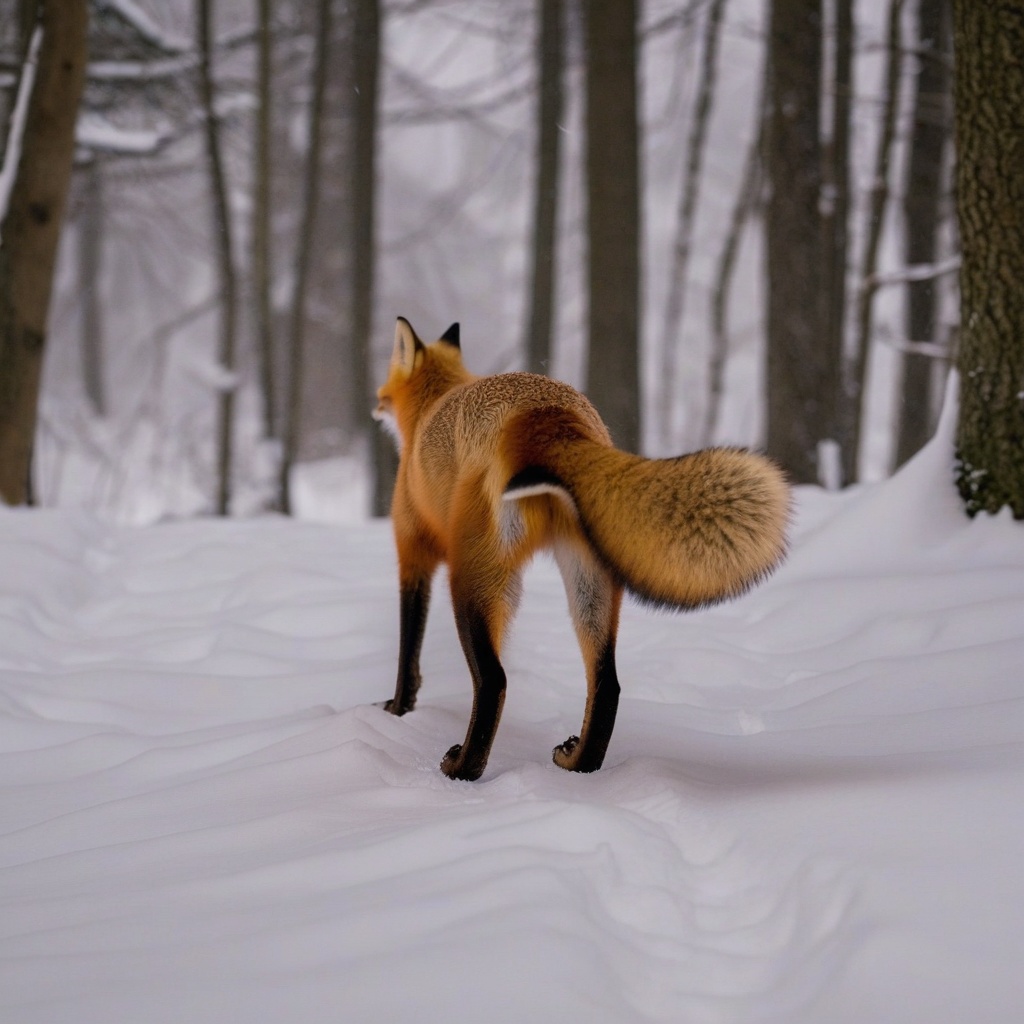} &    
    \includegraphics[width=0.15\textwidth]{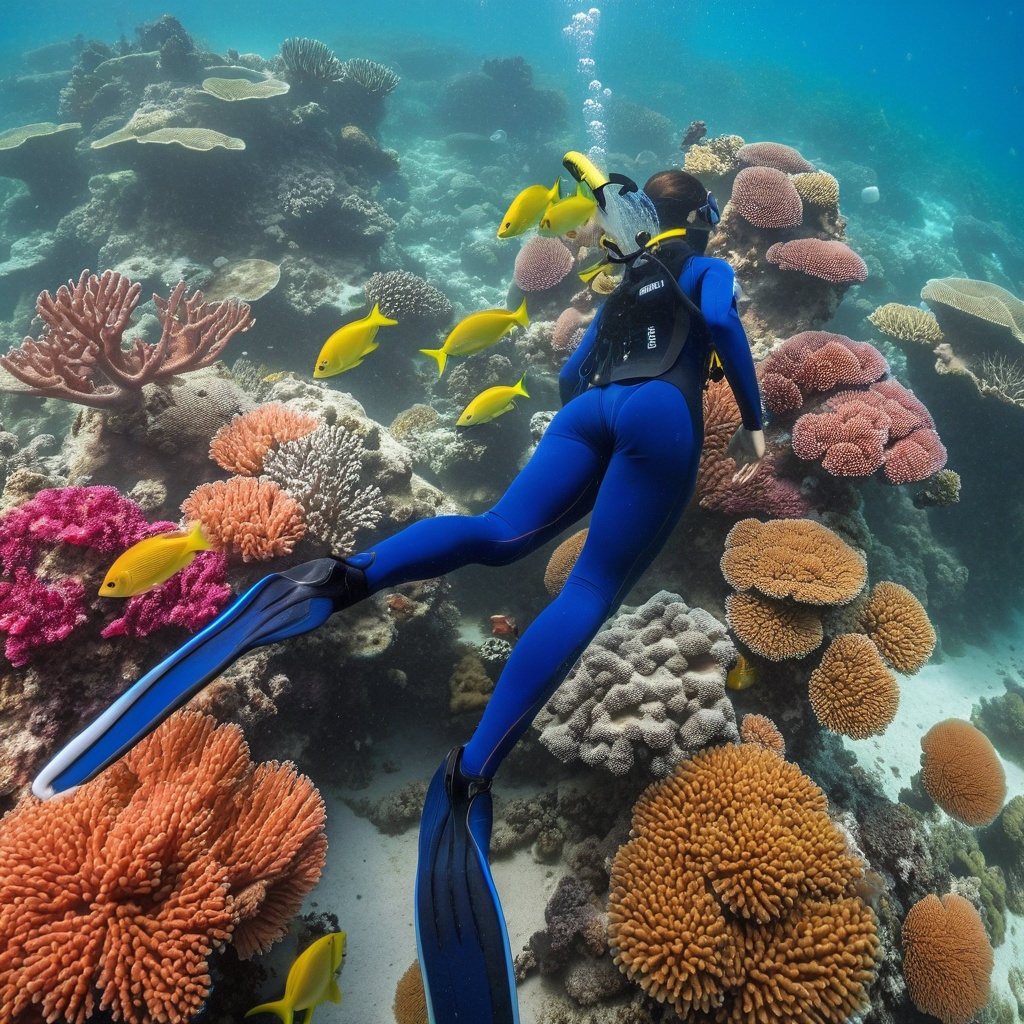} &
    \includegraphics[width=0.15\textwidth]{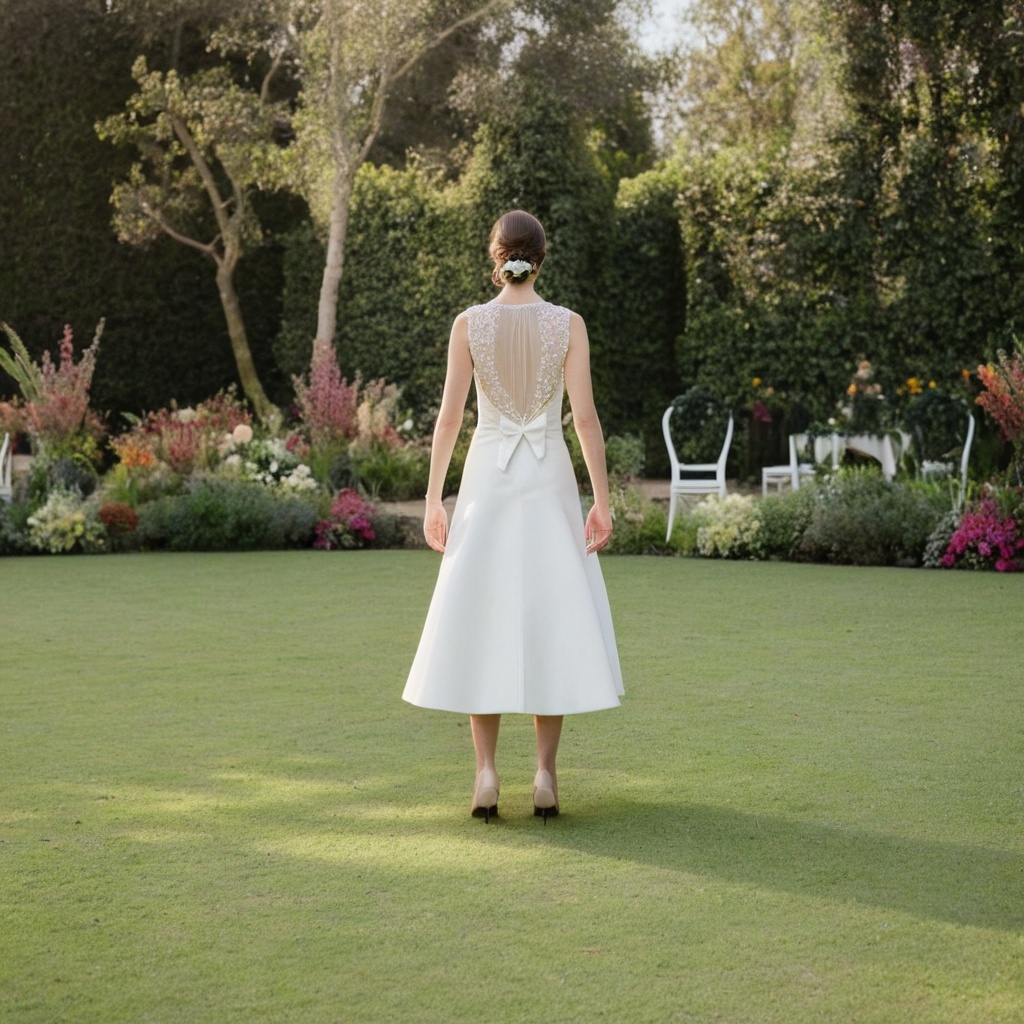} &
    \includegraphics[width=0.15\textwidth]{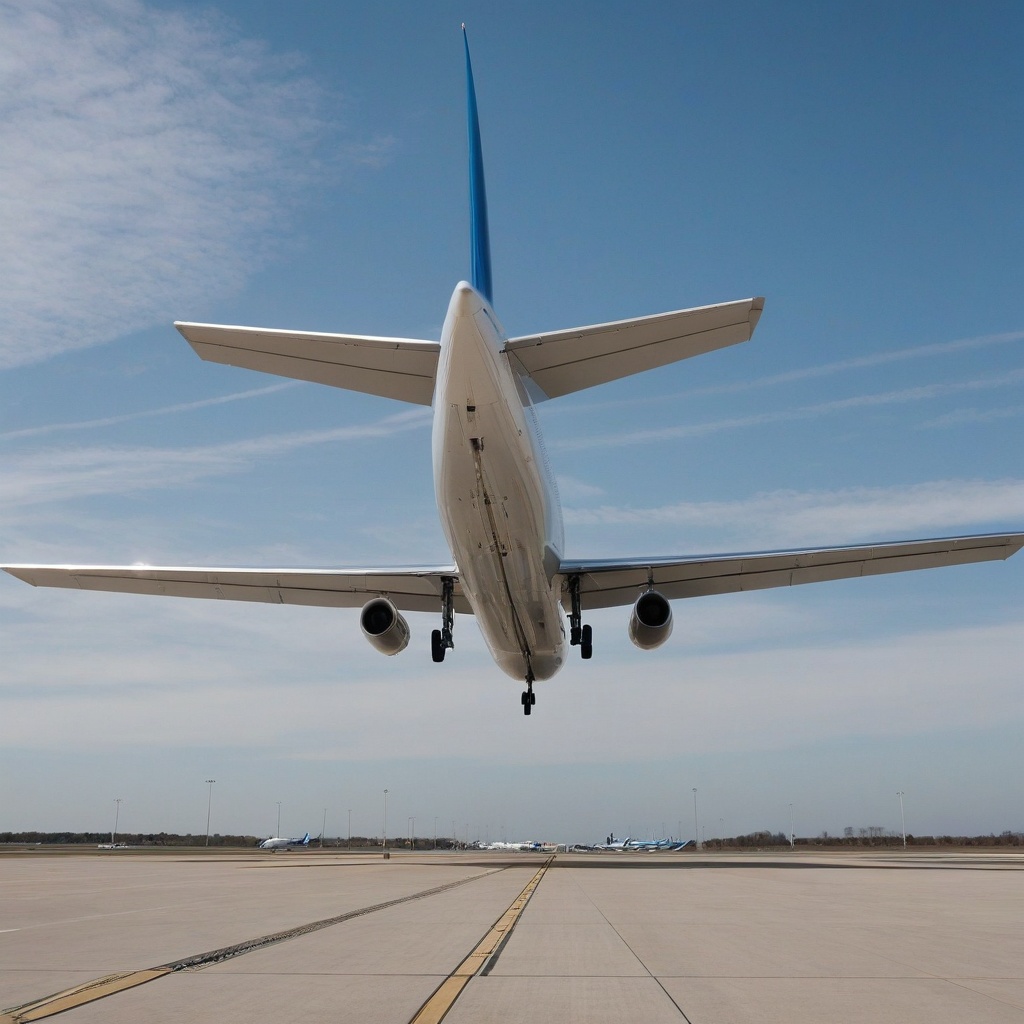} &
   \includegraphics[width=0.15\textwidth]{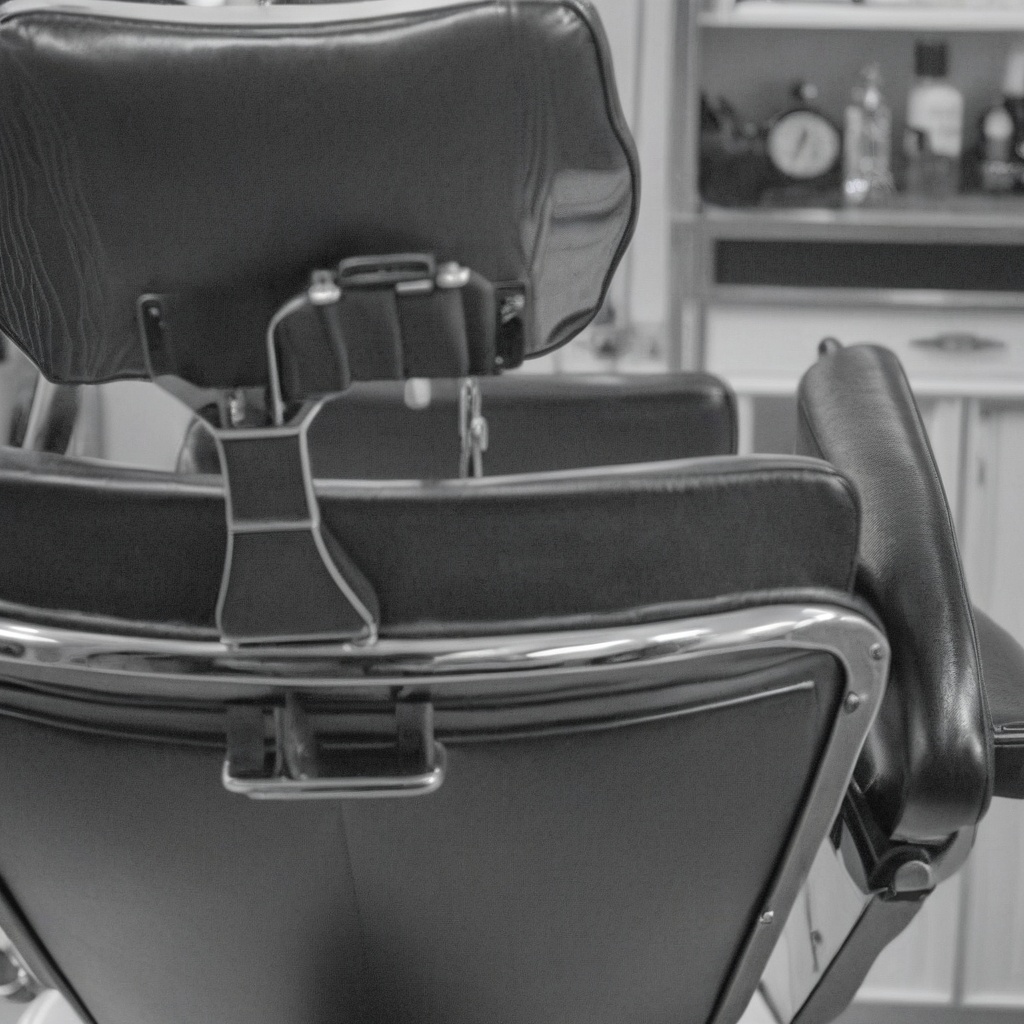} 

    \\
   \rotatebox{90}{\textbf{Back Right}} & 
    \includegraphics[width=0.15\textwidth]{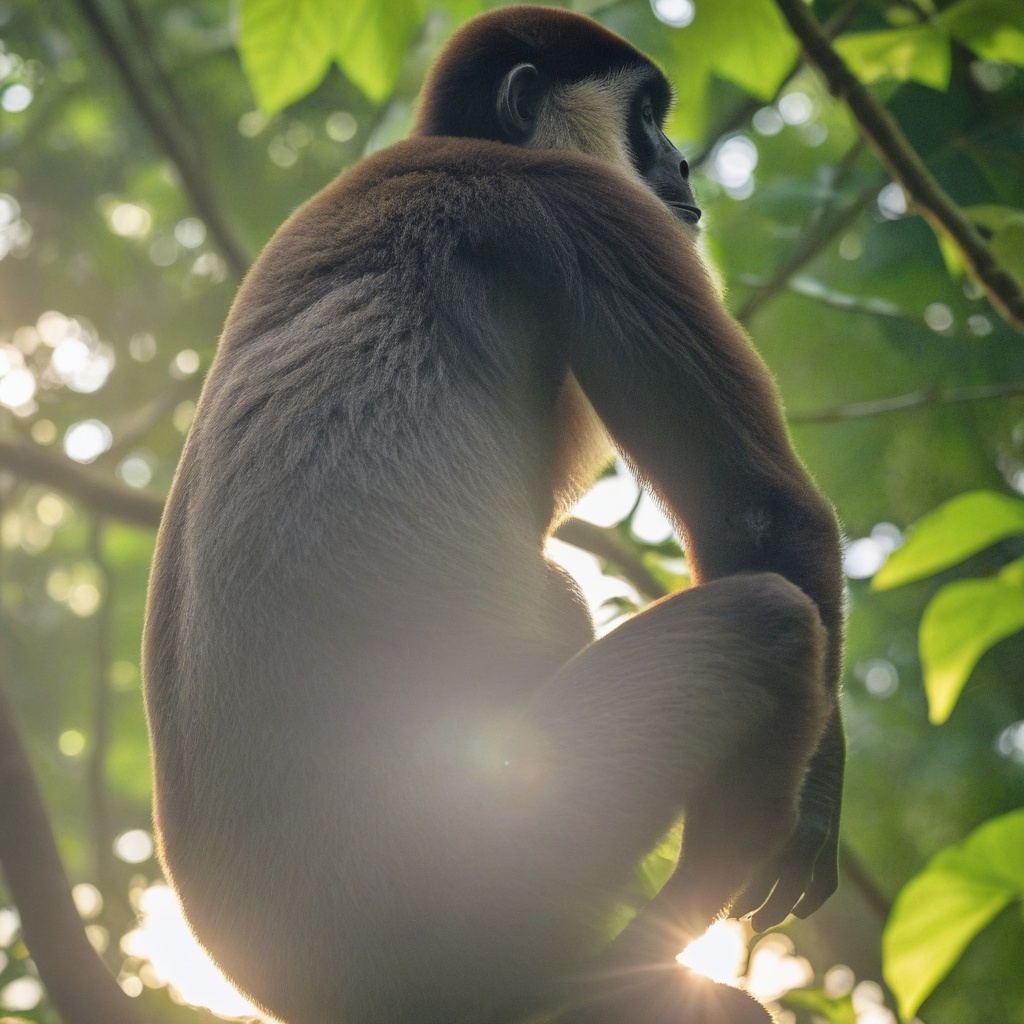} &
   \includegraphics[width=0.15\textwidth]{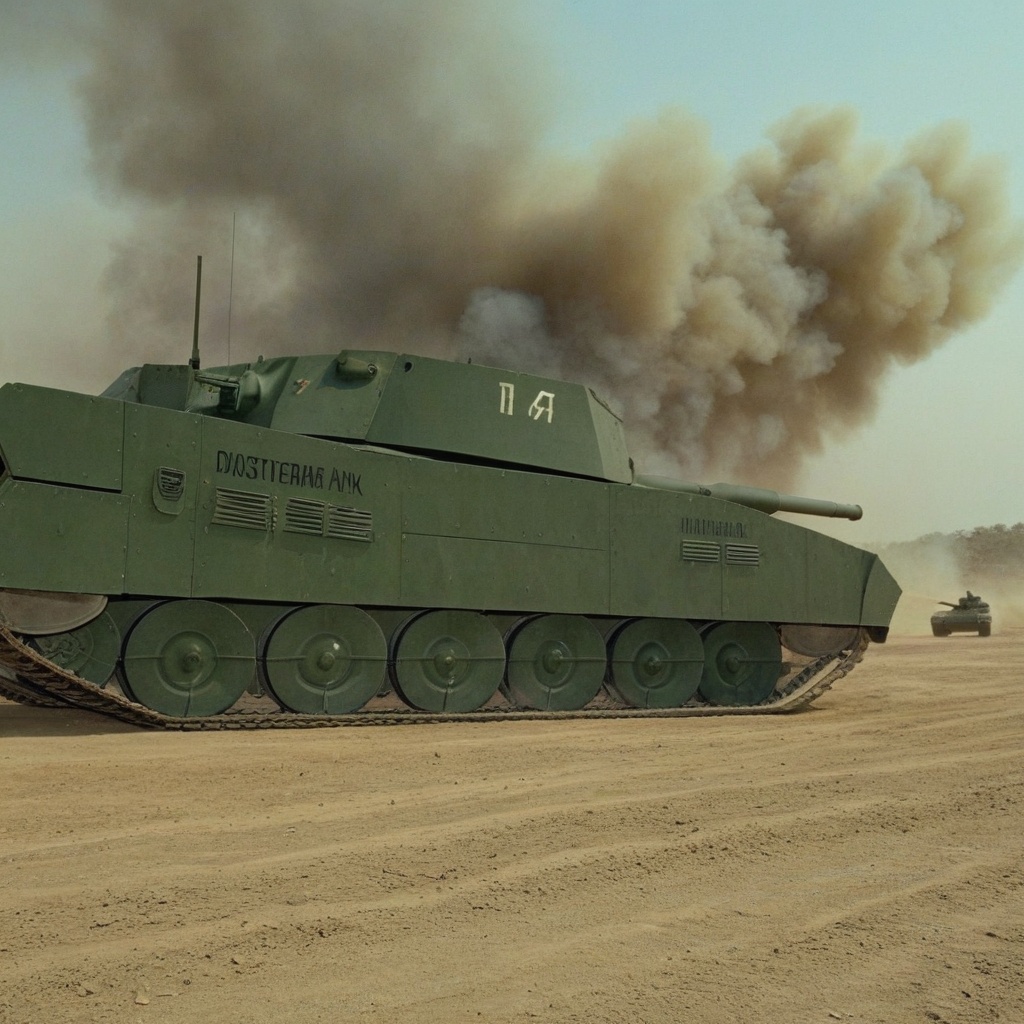} &    
   \includegraphics[width=0.15\textwidth]{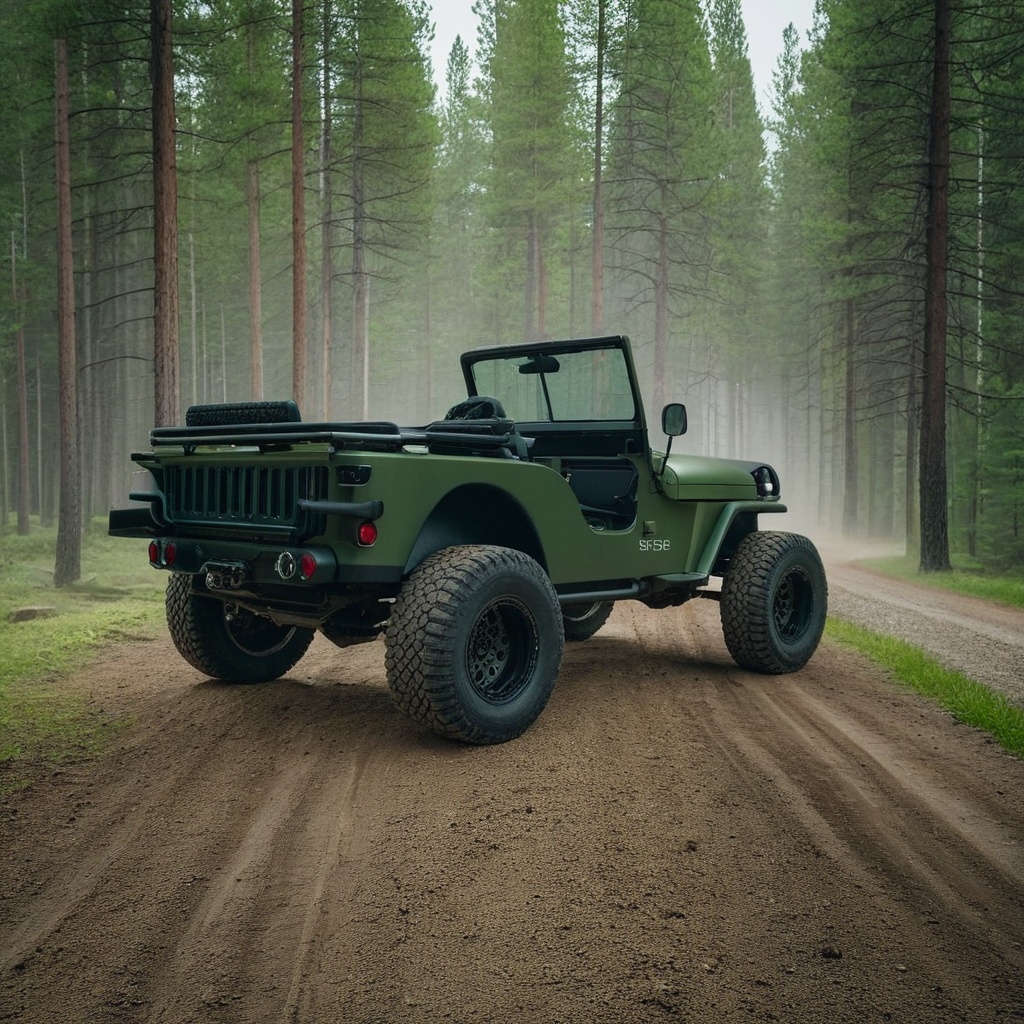} &
    \includegraphics[width=0.15\textwidth]{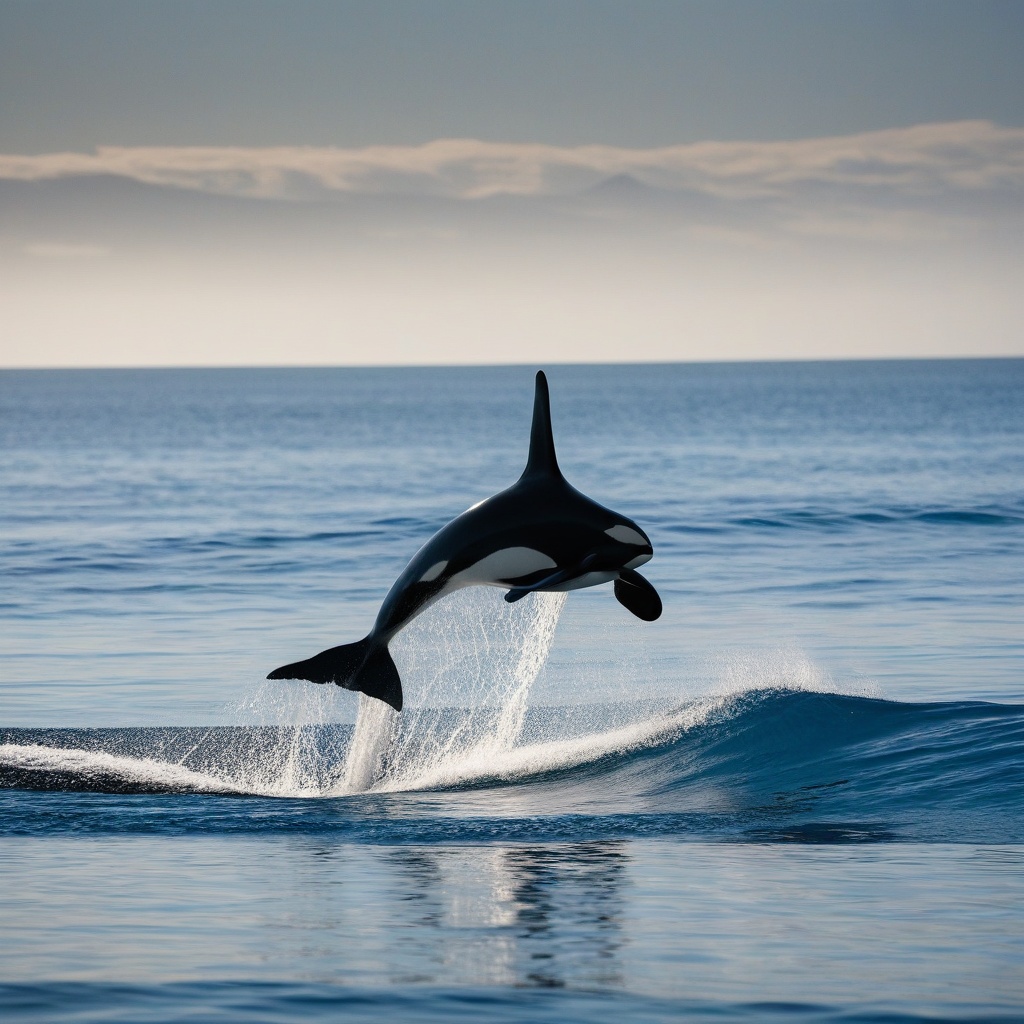} &
    \includegraphics[width=0.15\textwidth]{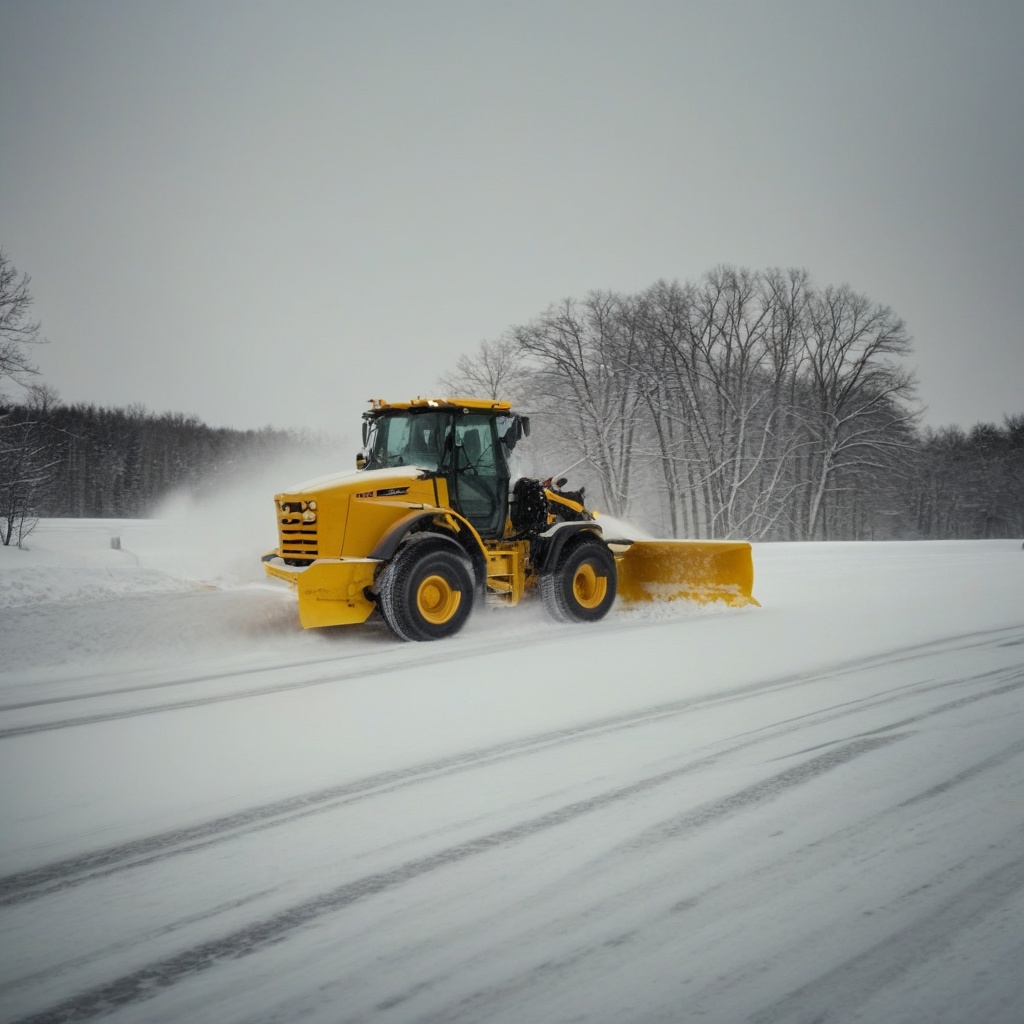} &
    \includegraphics[width=0.15\textwidth]{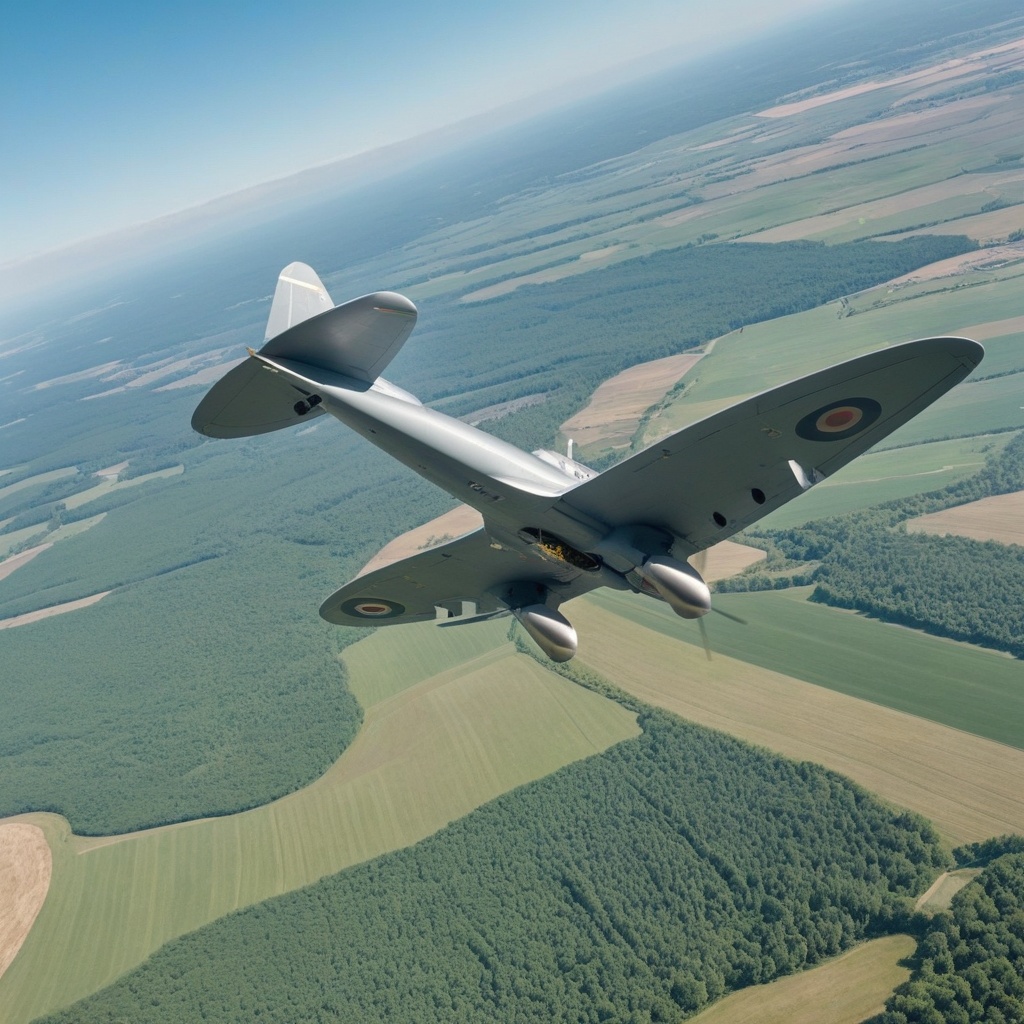} 
    \\
   \rotatebox{90}{\textbf{Right}} & 
    \includegraphics[width=0.15\textwidth]{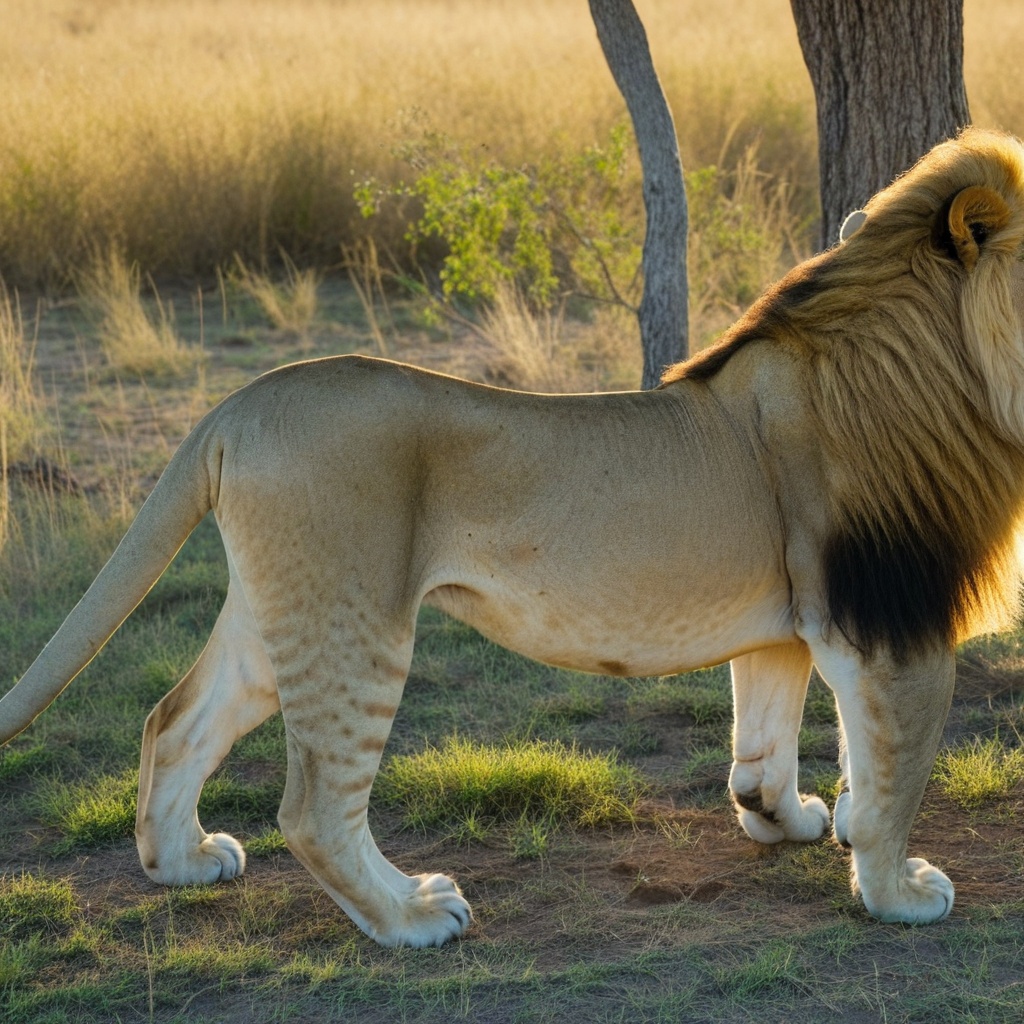} &
    \includegraphics[width=0.15\textwidth]{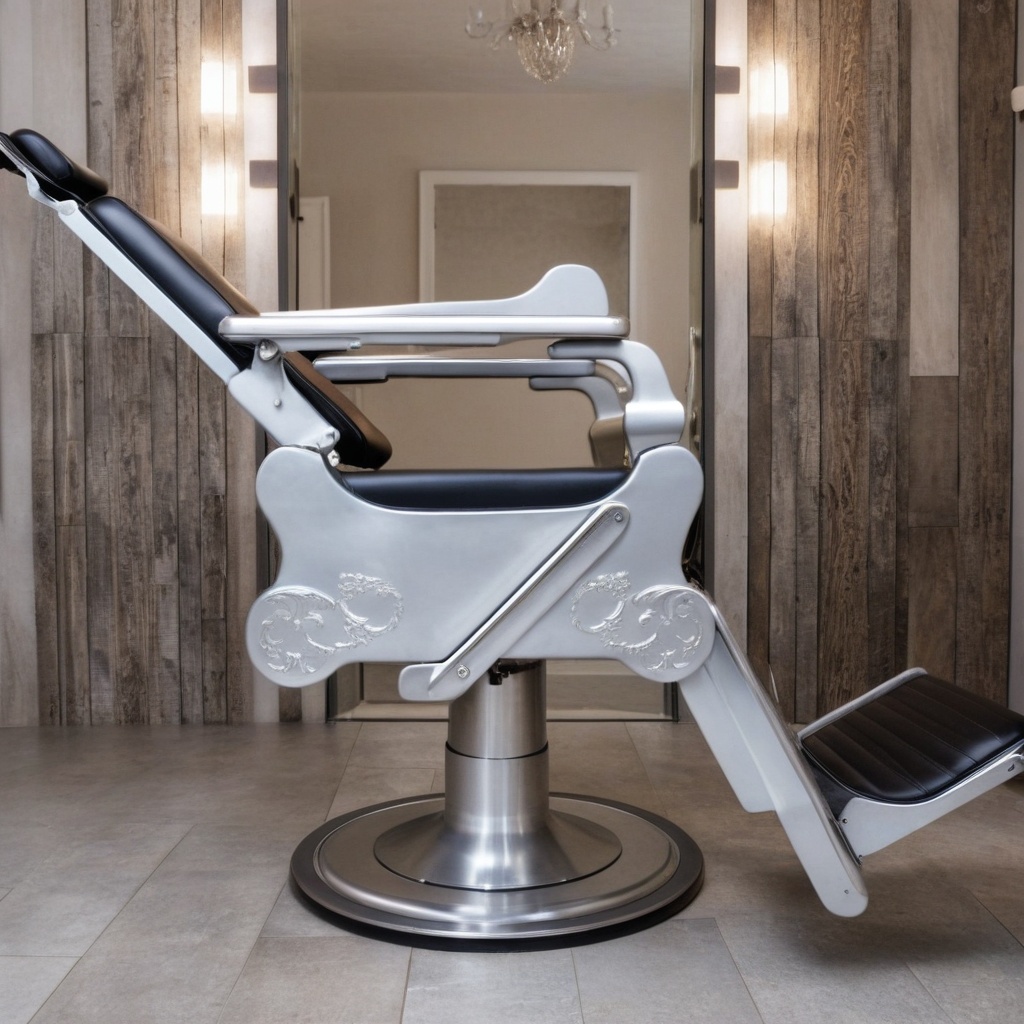} &
   \includegraphics[width=0.15\textwidth]{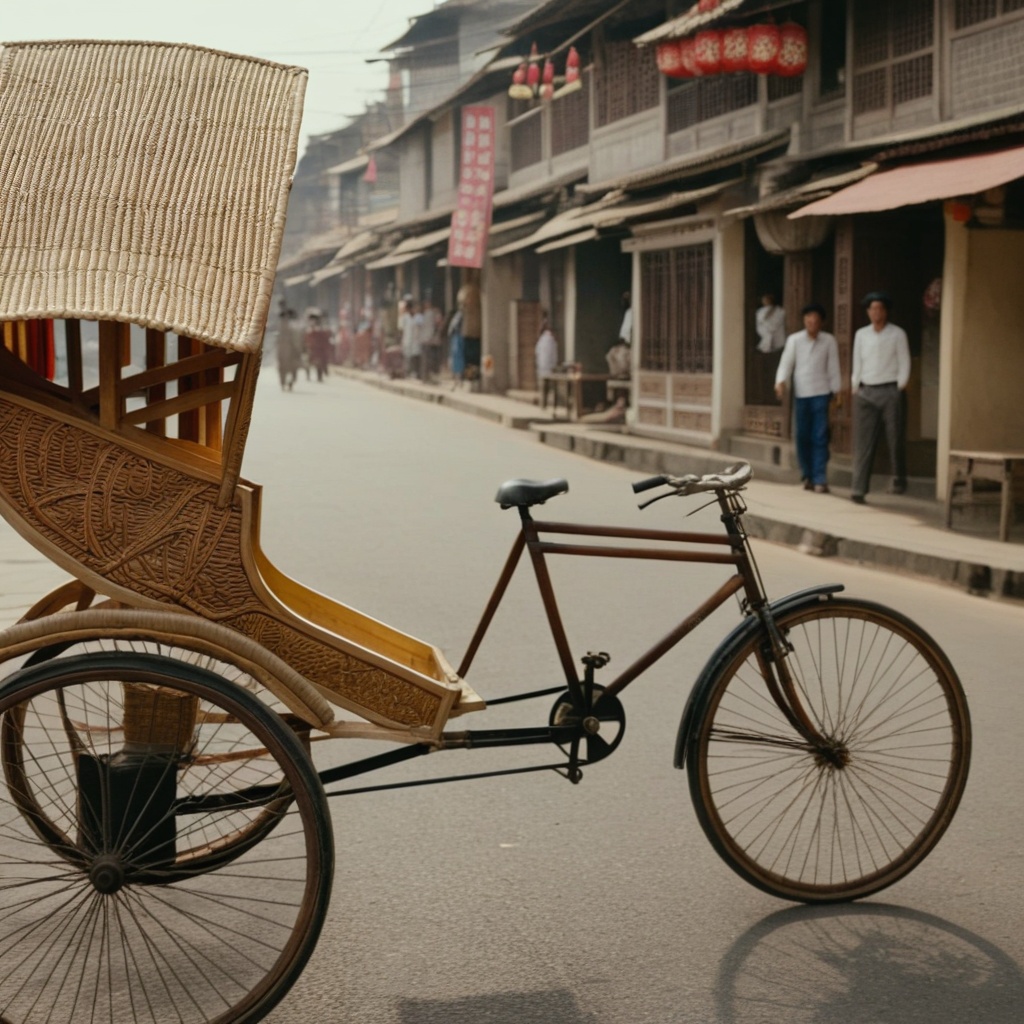} &    
   \includegraphics[width=0.15\textwidth]{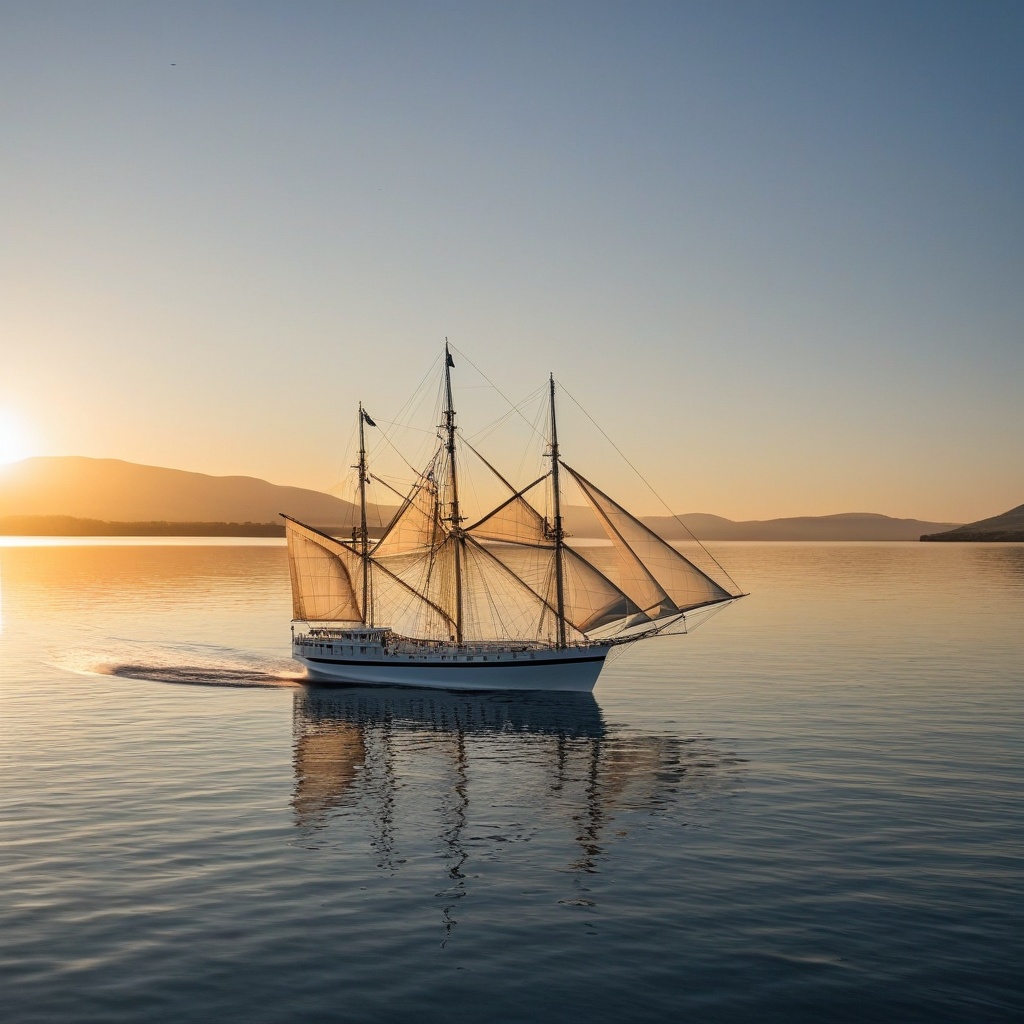} &
    \includegraphics[width=0.15\textwidth]{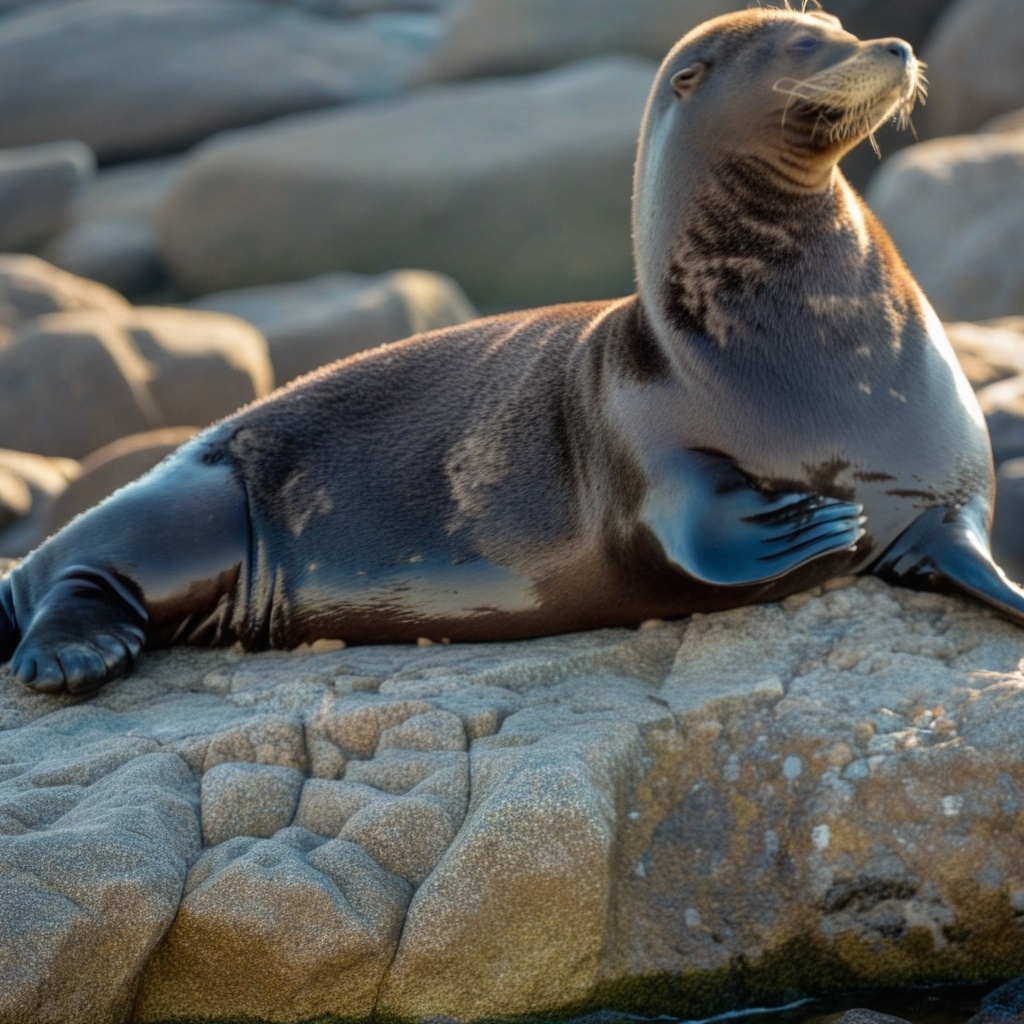} &
    \includegraphics[width=0.15\textwidth]{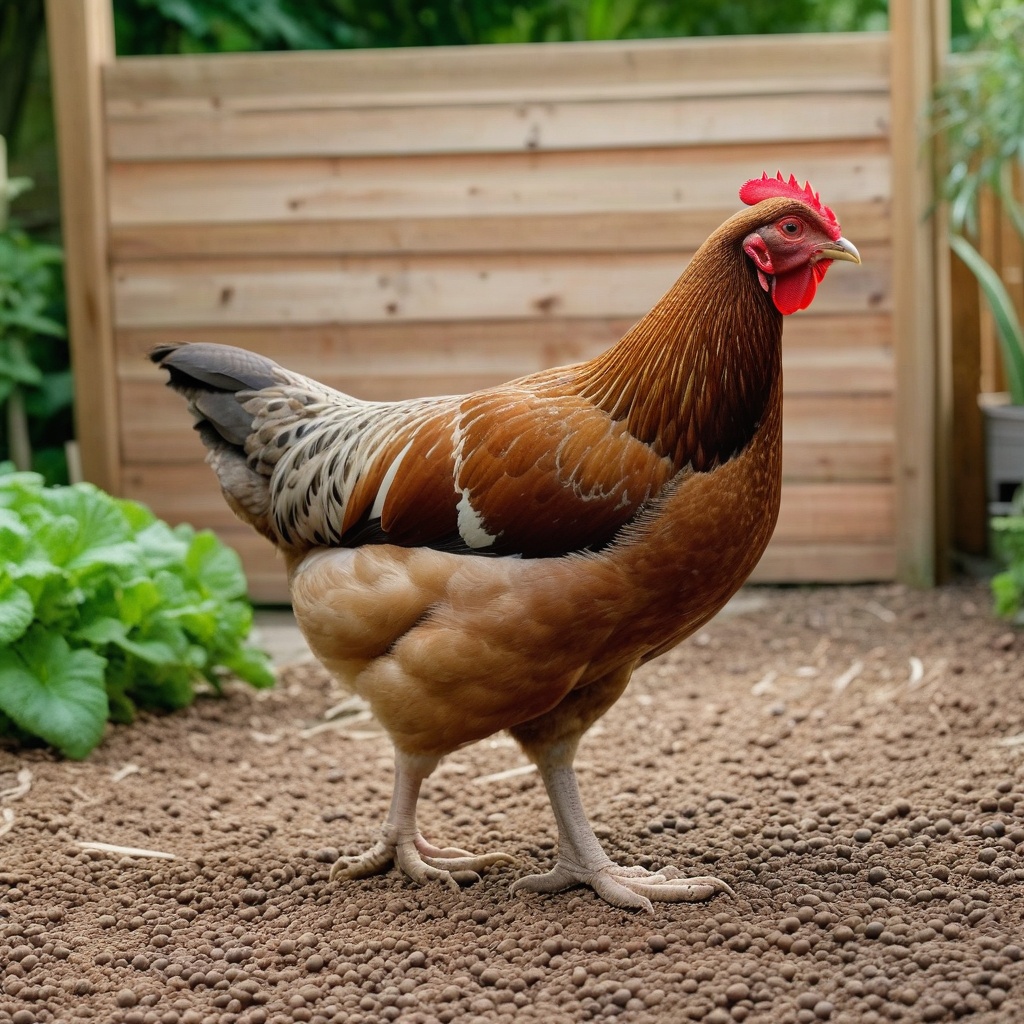} 
    \\
   \rotatebox{90}{\textbf{Front Right}} & 
   \includegraphics[width=0.15\textwidth]{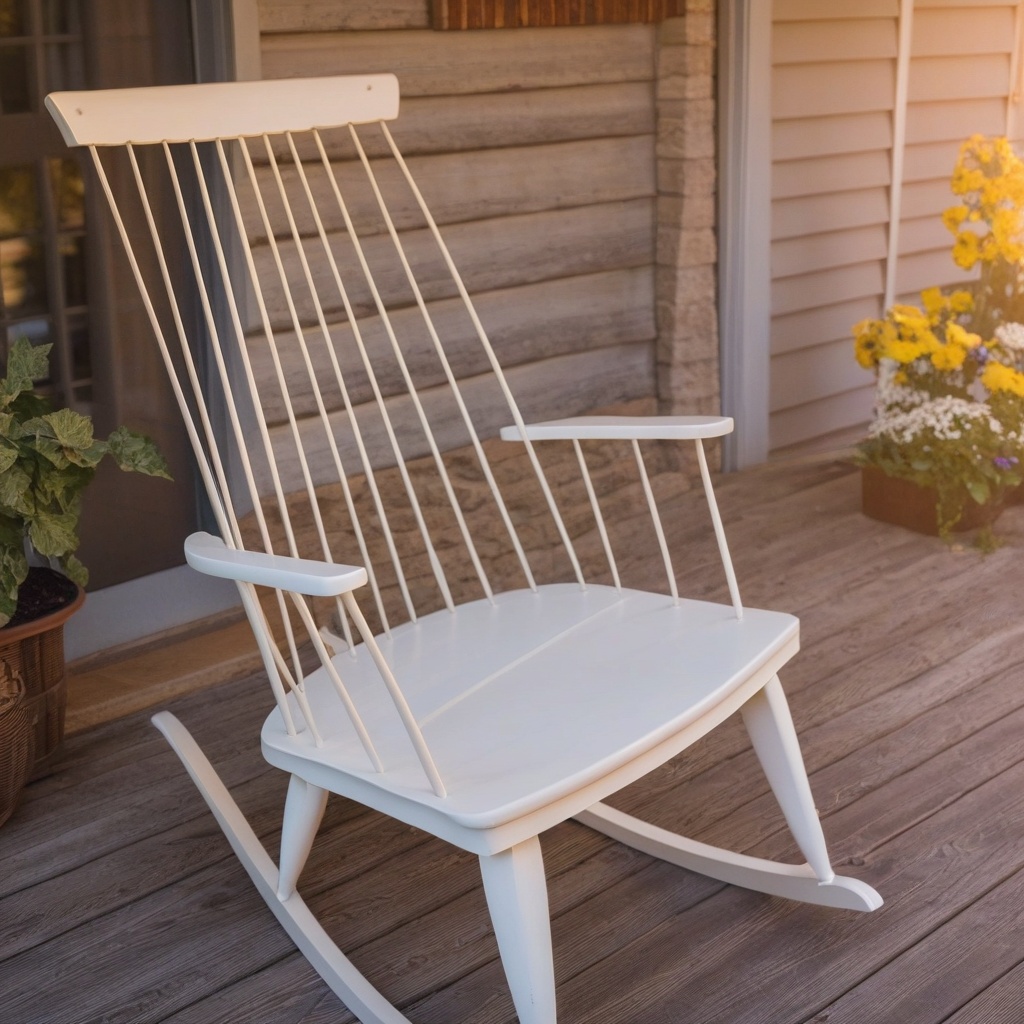} &    
   \includegraphics[width=0.15\textwidth]{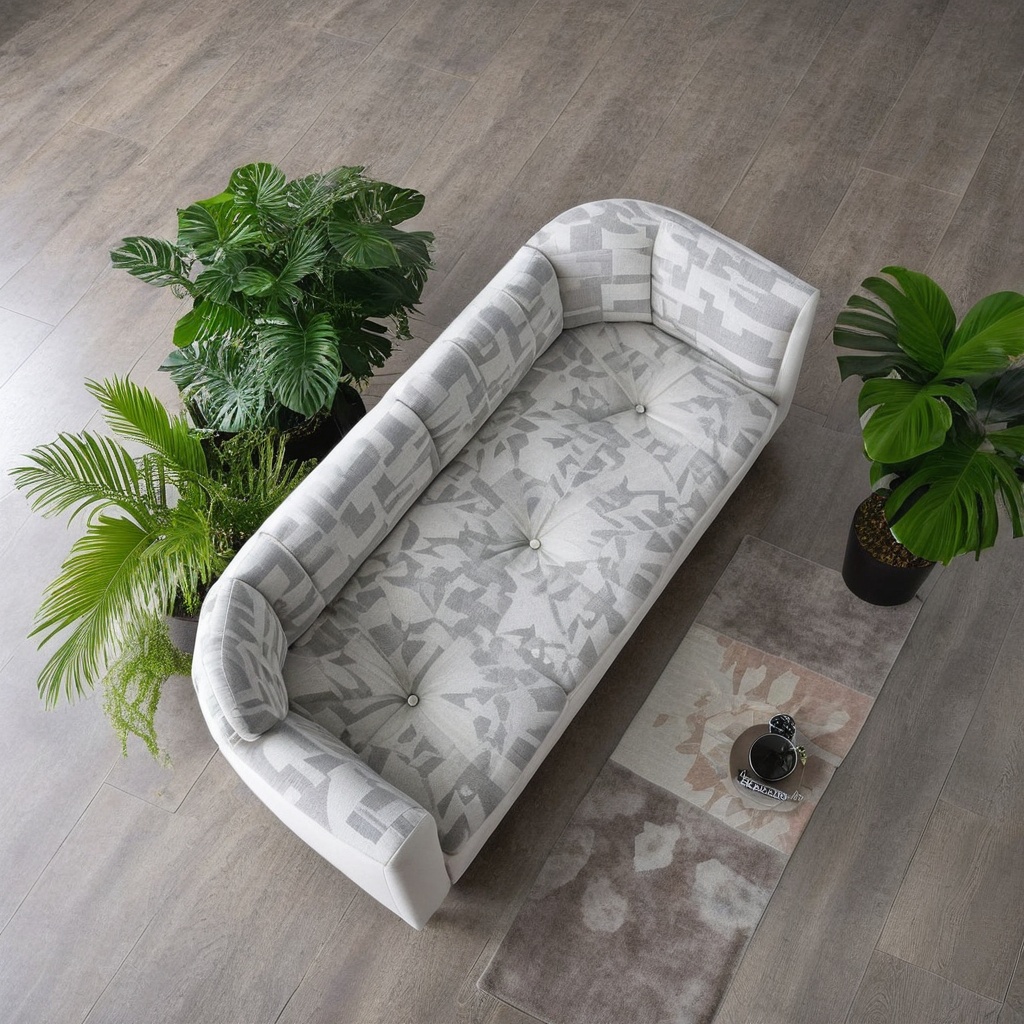} &
    \includegraphics[width=0.15\textwidth]{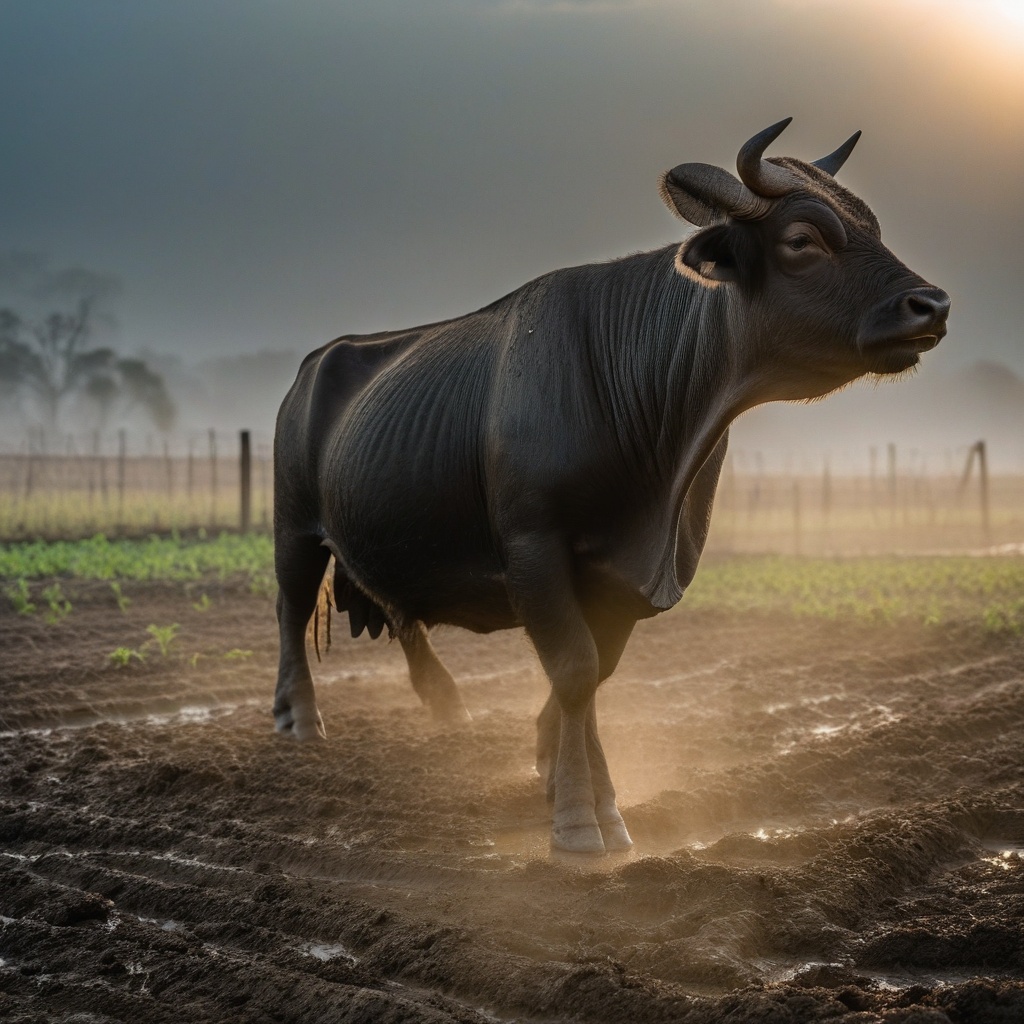} &
    \includegraphics[width=0.15\textwidth]{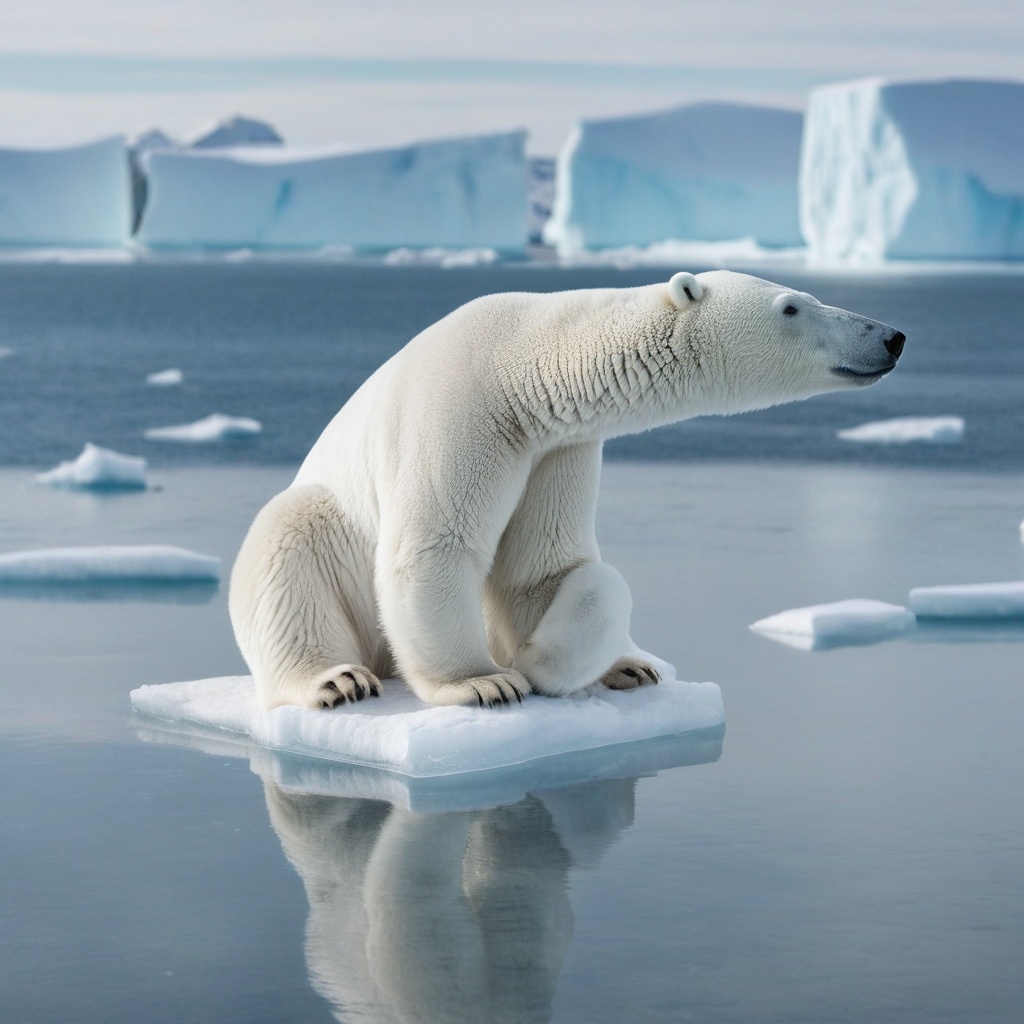} &
    \includegraphics[width=0.15\textwidth]{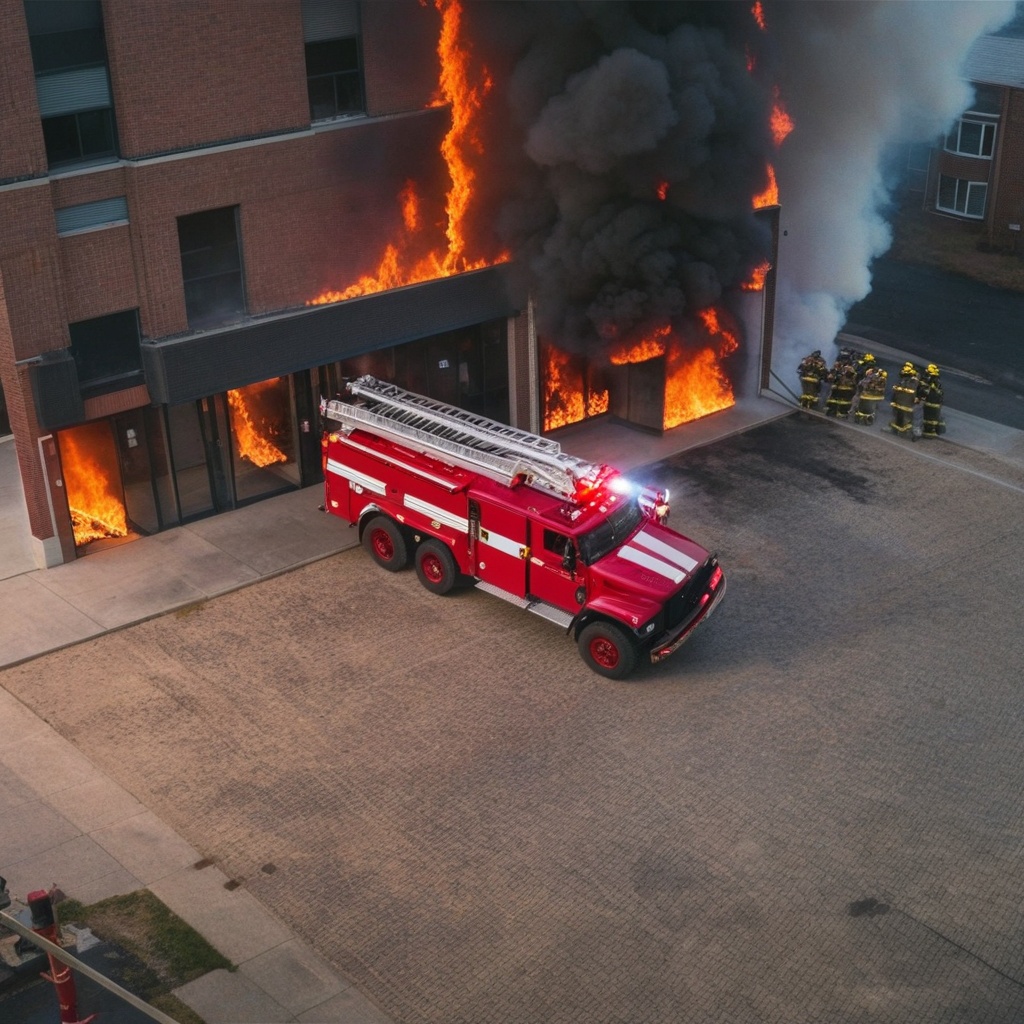} &
    \includegraphics[width=0.15\textwidth]{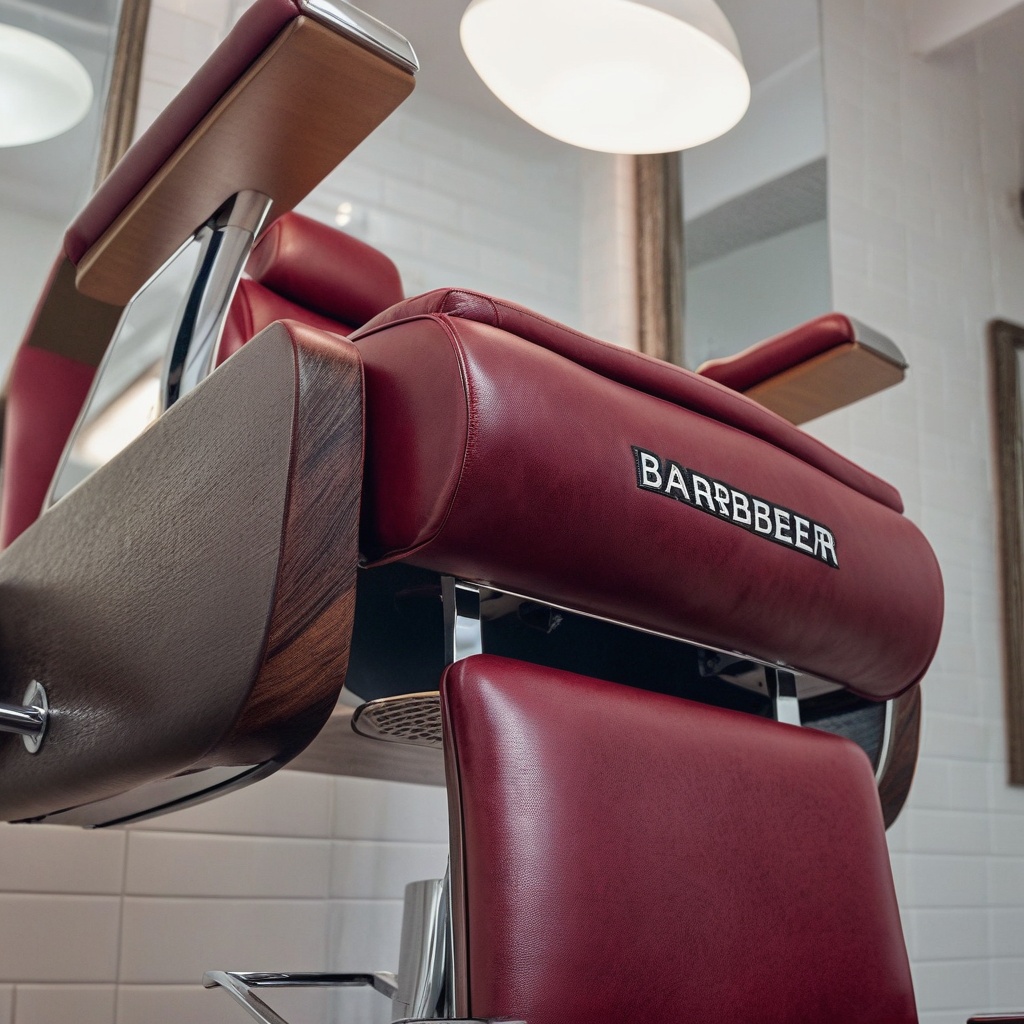}
    \\

\end{tabular}%
     \caption{\textbf{Diversity of \textit{Ultimate3D}}. Our \textit{Ultimate3D} dataset and benchmark cover 100 categories of objects, range diverse camera-object relation settings, and provide plausible image quality. (Each row shows images with the same orientation but in diverse subject and context.)}
      \label{suppfig:ultimate3D}
\end{figure*}

In Fig.~\ref{suppfig:ultimate3D}, we shows more examples of diversity on object categories, camera-object relation, and background contexts.

\section{Additional Discussions}
\label{suppsec:finetune_dicussion}
We provide some insights of using synthetic generated visual instruction dataset for MLLM finetuning. For the task of camera-object relation recognition for 100 categories (as we collected in \textit{Ultimate3D} dataset), a dataset comprising 100K to 1M VQAs will significantly improve baseline MLLMs by one-epoch finetuning. Further scaling up the dataset volume, finetuneing epochs, or model parameters may yield saturation.

Moreover, the quality of synthetic dataset is crucial to the improvement margin. Using dataset with higher success rate (ours is 93.07\% as main paper Sec. 4.5) provides better performances than using dataset generated other alternative backbones (i.e. SD-V1.5).

The coverage of arbitrary camera-object relation is also important for generalization. Our preliminary test using real images which are highly biased to front-facing direction from Pascal3D+~\cite{xiang_wacv14}, results in poor generalization on uncovered orientations and viewpoints. This illustrates the importance of using synthetic dataset to eliminate the dataset biases.

For numerical prediction of camera-object relation, our preliminary test shows plausible success using MEBOW~\cite{wu2020mebow} dataset to improve MLLMs on predicting human orientation angles. The average degree prediction accuracy is competitive to the original paper~\cite{wu2020mebow}. However, the extended numerical prediction of 100 categories and 3 types of camera-object relations may require even larger dataset volume than \textit{Ultimate3D}. Our framework is able to scale up the dataset volume. But due to the limitation of computing power, we leave it as interesting future work.

\section{Algorithm of 3D visual instruction dataset generation pipeline}
\label{suppsec:algorithm}

\begin{algorithm}[h]
\caption{Synthetic VQA generation}
\label{alg:algorithm}
\textbf{Input}: 3D asset $A$, asset category $c$, camera-object relation $\beta$.\\
\textbf{Parameter}: Renderer $\mathcal{R}$, DM-based image generator $\mathcal{G}$, image decoder $\mathcal{D}$, LLM text generator $\mathcal{L}$, system prompt given to $\mathcal{L}$ for generating image context description $p_{img}$, and for generating QA pairs $p_{qa}$, DM sampling steps $T$. \\
\textbf{Output}: Generated synthetic image $I_{syn}$, corresponding text QA pairs $\mathcal{T}_{qa}$. 
\begin{algorithmic}[1] 
    \STATE $I_{\beta} = \mathcal{R}(A, \beta)$
    \dotfill $\#$ Render 3D priors

    \STATE  $\mathcal{T}_{img} = \mathcal{L} (c, p_{img})$
    \dotfill $\#$ Generate image description

    \STATE $z_T \sim \mathcal{N}(0, I)$
    \FOR{$t=T, \cdots, 1$} 
        \STATE $z_{t-1} = \mathcal{G}(z_{t}, I_{\beta}, \mathcal{T}_{img})$ \dotfill $\#$ DM denoising 
    \ENDFOR
    \STATE $I_{syn} = \mathcal{D}(z_0)$
    
    \STATE $\mathcal{T}_{qa} = \mathcal{L} (c, \beta, p_{qa})$
    \dotfill $\#$ Generate QA pairs 

    \STATE \textbf{return} $I_{syn}, \mathcal{T}_{qa}$
    
\end{algorithmic}
\end{algorithm}


\end{document}